\documentclass[a4paper]{book}
\usepackage{makeidx,named,elsevierg}

\usepackage{epsfig}  
\usepackage{amssymb}
\usepackage{latexsym}     
\usepackage{pstricks}
\usepackage{diagrams}

\newcommand{\bpf}{\noindent {\bf Proof:} }
\def\endproof{\hfill$\blacksquare$} 

\newcommand{\cnot}{\textsc{\footnotesize CNOT}}
\newcommand{\bit}{\begin{itemize}}
\newcommand{\eit}{\end{itemize}\par\noindent}
\newcommand{\ben}{\begin{enumerate}}
\newcommand{\een}{\end{enumerate}\par\noindent}
\newcommand{\beq}{\begin{equation}}
\newcommand{\eeq}{\end{equation}\par\noindent}
\newcommand{\beqa}{\begin{eqnarray*}}
\newcommand{\eeqa}{\end{eqnarray*}\par\noindent} 
\newcommand{\beqn}{\begin{eqnarray}}  
\newcommand{\eeqn}{\end{eqnarray}\par\noindent}           

\newcommand{\dd}{\llcorner}
\newcommand{\sdot}{\bullet}
\newcommand{\ddd}{\lrcorner}
\newcommand{\uu}{\ulcorner} 
\newcommand{\uuu}{\urcorner}

\newcommand{\HH}{\mathcal{H}}

\newcommand{\ie}{\textit{i.e.}\ }
\newcommand{\Zero}{\mathbf{0}}
\newcommand{\CC}{\mathbf{C}}
\newcommand{\II}{{\rm I}}
\newcommand{\PP}{{\rm P}}

\newcommand{\FdHilb}{\mathbf{FdHilb}}
\newcommand{\FdVect}{\mathbf{FdVec}}
\newcommand{\Rel}{\mathbf{Rel}}
\newcommand{\labarrow}[1]{\stackrel{#1}{\longrightarrow}}
\newcommand{\rarr}{\rightarrow}
\newcommand{\lrarr}{\longrightarrow}
\newcommand{\Tr}{\mathsf{Tr}}
\newcommand{\linimpl}{\multimap}
\newcommand{\CPM}{\mathsf{CPM}}
\newcommand{\WProj}{\mathsf{WProj}}

\newcommand{\neoiota}{\psi}

\newarrow{Toto}<--->
\newarrow{Is}===== 
\newarrow{Dotsto}....>
\newarrow{Dots}.....
\newarrow{eq}=====

\makeindex
\collection
\begin{document}
\paper[Categorical quantum mechanics]{Categorical quantum mechanics}{Samson Abramsky and Bob Coecke}

%\tableofcontents

\section{Introduction}

Our aim is to revisit the mathematical foundations of quantum mechanics from a novel point of view. The standard axiomatic presentation of quantum mechanics in terms of Hilbert spaces, essentially due to von Neumann \shortcite{vN}, has provided the mathematical bedrock of the subject for over 70~years. Why, then, might it be worthwhile to revisit it now?

First and foremost, the advent of \emph{quantum information and computation} (QIC) as a major field of study has breathed new life into basic quantum mechanics, asking new kinds of questions and making new demands on the theory, and at the same time reawakening interest in the foundations of quantum mechanics.

As one key example, consider the changing perceptions of \emph{quantum entanglement} and its consequences. The initial realization that this phenomenon, so disturbing from the perspective of classical physics, was implicit in the quantum-mechanical formalism came with the EPR \textit{Gedanken}-experiment of the 1930's \cite{EPR}, in the guise of a ``paradox''. By the 1960's, the paradox had become a \emph{theorem} --- Bell's theorem \cite{Bell}, demonstrating that non-locality was an essential feature of quantum mechanics, and opening entanglement to experimental confirmation. By the 1990's, entanglement had become a \emph{feature}, used in quantum teleportation \cite{BBC}\index{quantum teleportation}, in protocols for quantum key distribution \cite{Ekert}, and, more generally, understood as a computational and informatic \emph{resource} \cite{BEZ}.

\subsection{The Need for High-Level Methods}
The current tools available for developing quantum algorithms and
protocols, and more broadly the whole field of quantum information and computation,  are \emph{deficient} in two main respects.
 
Firstly, they are too \emph{low-level}.  One finds a plethora of ad hoc
calculations with `bras' and `kets', normalizing constants, matrices
etc. The arguments for the benefits of a high-level, conceptual
approach to designing and reasoning about quantum
computational systems are just as compelling as for classical
computation. In particular, we have in mind the hard-learned lessons from Computer Science of the importance of \emph{compositionality}, \emph{types}, \emph{abstraction}, and the use of tools from algebra and logic in the design and analysis of complex informatic processes. 

At a more fundamental level, the standard mathematical framework
for quantum mechanics  is actually
\emph{insufficiently comprehensive} for informatic purposes.
In describing a protocol such as teleportation, or any quantum
process in which \emph{the outcome of a measurement is used to determine
subsequent actions}, the von Neumann formalism leaves feedback of
information from the classical or macroscopic level back to the quantum
\emph{implicit} and \emph{informal}, and hence not subject to
rigorous analysis and proof. As quantum protocols and computations grow
more
elaborate and complex, this point is likely to prove of increasing
importance.  

Furthermore, there are many fundamental issues in QIC which remain very much open. The current low-level methods seem unlikely to provide an adequate basis for addressing them. For example:
\begin{itemize}
\item What are the precise structural relationships between superposition, entanglement and mixedness as quantum informatic resources?  Or, more generally, 
\item  Which features of quantum mechanics account for differences in computational and informatic power as compared to classical computation? 
\item  How do quantum and classical information interact with each other, and with a spatio-temporal causal structure?  
\item  Which quantum control features (e.g.~iteration) are possible and what additional computational power can they provide?
\item  What is the precise logical status and axiomatics of No-Cloning and No-Deleting, and more generally, of the quantum mechanical formalism as a whole?
\end{itemize}
These questions gain additional force from the fact that a variety of different quantum computational architectures and information-processing scenarios are beginning to emerge.
While at first it seemed that the notions of Quantum Turing Machine \cite{Deu1} and the quantum circuit model \cite{Deu2} could supply canonical analogues of the classical computational models, recently some very different models for quantum computation have emerged, e.g.~Raussendorf and Briegel's \em one-way quantum computing \em model \cite{Briegel1,RBB} and \em measurement based quantum computing \em in general \cite{JozsaMB}, \em adiabatic quantum computing \em \cite{Adiabatic}, \em topological quantum computing \em \cite{Kitaev}, etc.  These new models have features which are both theoretically and experimentally of great interest, and the methods developed to date for the circuit model of quantum computation do not carry over straightforwardly to them. In this situation, we can have no confidence that a comprehensive paradigm  has yet been found. It is more than likely that we have overlooked many new ways of letting a quantum system compute.

Thus there is a need to design structures and develop methods and tools which apply to these \em non-standard quantum computational models \em.  We must also address the question of how the various models compare --- can they be interpreted in each other, and which computational and physical properties are preserved by such interpretations?

\subsection{High-Level Methods for Quantum Foundations}
Although our initial motivation came from quantum information and computation, in our view the development of high-level methods is potentially of great significance for the development of the foundations of quantum mechanics, and of fundamental physical theories in general.
We shall not enter into an extended discussion of this here, but simply mention some of the main points:
\begin{itemize}
\item By identifying the fundamental mathematical structures at work, at a more general and abstract level than that afforded by Hilbert spaces, we can hope to gain new structural insights, and new ideas for how various physical features can be related and combined.
\item We get a new perspective on the logical structure of quantum mechanics, radically different to the traditional approaches to quantum logic\index{quantum logic}.
\item We get a new perspective on ``No-Go'' theorems, and new tools for formulating general results applying to whole classes of physical theories.
\item Our structural tools yield an \emph{effective calculational formalism} based on a diagrammatic calculus, for which automated software tool-support is currently being developed. This is not only useful for quantum information and computation, it may also yield new ways of probing key foundational issues. 
Again, this mirrors what has become the common experience in Computer Science.
In the age of QIC, \textit{Gedanken}-experiments turn into programs!
\end{itemize}

\noindent We shall take up some of these issues again in the concluding sections.

\subsection{Outline of the Approach}
We shall use \emph{category theory}\index{category theory} as the mathematical setting for our approach.
This should be no surprise. Category theory is the language of modern structural mathematics, and the fact that it is not more widely used in current foundational studies is a regrettable consequence of the sociology of knowledge and the encumbrances of tradition. Computer Science, once again, leads the way in the applications of category theory; abstract ideas can be very practical!

We shall assume a modest familiarity with basic notions of category theory, including symmetric monoidal categories. Apart from standard references such as \cite{MacLane}, a number of introductions and tutorials specifically on the use of monoidal categories in physics are now available \cite{LNPAbramsky,LNPBaez,LNPCoecke}. More advanced textbooks in the area are \cite{Kock,StreetBook}.

We shall give an axiomatic presentation of quantum mechanics
at the abstract
level of \emph{strongly compact closed categories with biproducts}\index{strong compact closure}\index{biproducts}  ---
of which the standard von Neumann presentation in terms of Hilbert
spaces is but one example.
Remarkably enough, all the essential features of modern quantum protocols
such as \emph{quantum teleportation} \cite{BBC}, \emph{logic-gate
teleportation} \cite{Gottesman}, and \emph{entanglement swapping} 
\cite{Swap} ---which exploit quantum mechanical effects in an
essential way--- find natural counterparts at this abstract level. 
More specifically:
\begin{itemize} 
\item The basic structure of a symmetric monoidal category allows
  \emph{compound systems} to be described in a resource-sensitive
  fashion (cf.~the `no
  cloning' \cite{Dieks,WZ} and `no deleting' \cite{Pati} theorems of quantum
mechanics).
\item The compact closed structure allows \emph{preparations and 
    measurements of entangled states} to be described, and their
key properties to be proved.
%%%INC
\item The strong compact closed structure brings in the central
notions of adjoint, unitarity and sesquilinear inner product ---allowing an involution such as complex conjugation
to play a role--- and it gives rise to a two-dimensional generalization of Dirac's \emph{bra-ket} calculus \cite{Dirac}, in which the structure of compound systems is fully articulated, rather than merely implicitly encoded by labelling of basis states.
%%%
\item Biproducts allow \emph{probabilistic branching} due to
measurements, \emph{classical
communication} and \emph{superpositions} to be captured.   
Moreover, from the combination  of the---apparently purely qualitative---structures
of strong compact closure and biproducts  there emerge \em scalars \em and a  \em
Born rule\em.
\end{itemize} 
We are then able to use this abstract
setting to give precise formulations of quantum teleportation, logic
gate teleportation, and entanglement swapping, and
to prove correctness of these protocols --- for example, proving correctness of
teleportation means showing that the final state of Bob's qubit equals the
initial state of Alice's qubit. 

\subsection{Development of the Ideas}
A first step in the development of these ideas was taken in \cite{AC}, where it was recognized that compact-closed structure could be expressed in terms of bipartite projectors in Hilbert space, thus in principle enabling the structural description of  information flows in entangled quantum systems. In \cite{Coe1} an extensive  analysis of a range of quantum protocols was carried out concretely,  in terms of Hilbert spaces, with a highly suggestive but  informal  graphical notation of information-flow paths through networks of projectors.
The decisive step in the development of the categorical approach was taken in \cite{AC2}, with \cite{AC3} as a supplement improving the definition of strongly compact closed category.
The present article is essentially an extended and revised version of \cite{AC2}. There have been numerous subsequent developments in the programme of categorical quantum mechanics since \cite{AC2}. We shall provide an overview of the main developments in Section~7, but  the underlying programme as set out in \cite{AC2} still stands, and we hope that the present article will serve as a useful record of this approach in its original conception.

\subsection{Related Work}
To set our approach in context, we compare and contrast it with some related approaches.

\subsubsection{Quantum Logic}
Firstly, we discuss the relationship with quantum logic\index{quantum logic} as traditionally conceived, \ie the study of lattices abstracted from the lattice of closed linear subspaces of Hilbert space \cite{BvN}.

We shall not emphasize the connections to logic in the present article, but in fact our categorical axiomatics can be seen as the algebraic or semantic counterpart to a \emph{logical type theory} for quantum processes. This type theory has a resource-sensitive character, in the same sense as Linear logic \cite{Girard} --- and this is directly motivated by the no-cloning and no-deleting principles of quantum information. The correspondence of our formalism to a logical system, in which a notion of \emph{proof-net} (a graphical representation of multiple-conclusion proofs) gives a diagrammatics for morphisms in the free  strongly compact closed category with biproducts, and simplification of diagrams corresponds to \emph{cut-elimination}, is developed in detail in \cite{AD}.

This kind of connection with logic belongs to the proof-theory side of logic, and more specifically to the Curry-Howard correspondence, and the three-way connection between logic, computation and categories which has been a staple of categorical logic, and of logical methods in computer science, for the past three decades \cite{LS,LNPAbramsky}.

The key point is that we are concerned with the direct mathematical representation of \emph{quantum processes}.
By contrast, traditional quantum logic is concerned with \emph{quantum propositions}, which express \emph{properties} of quantum systems. There are many other differences. For example, compound systems and the tensor product\index{tensor product} are central to our approach, while quantum logic has struggled to accommodate these key features of quantum mechanics in a mathematically satisfactory fashion. However, connections between our approach and the traditional setting of orthomodular posets\index{orthomodular posets} and lattices have been made by John Harding \shortcite{Harding,HardingBis}.

\subsubsection{Categories in Physics}
There are by now several approaches to using category theory in physics.\index{categories in physics} For comparison, we mention the following:
\begin{itemize}
\item \cite{BaezDolan,Crane}. Higher-dimensional categories, TQFT's, categorification, etc.
\item \cite{IshBut,IshamII}. The topos-theoretic approach.
\end{itemize}

\paragraph{Comparison with the topos approach}
The topos approach aims ambitiously at providing a general framework for the formulation of physical theories. It is still in an early stage of development. Nevertheless, we can make some clear comparisons.
\begin{center}
\begin{tabular}{lcl}
Our approach & & Topos approach \\ \hline
monoidal & \textit{vs.} & cartesian \\
linear & \textit{vs.} & intuitionistic \\
processes & \textit{vs.} & propositions \\
geometry of proofs & \textit{vs.} & geometric logic 
\end{tabular}
\end{center}
Rather as in our comparison with quantum logic, the topos approach is primarily concerned with quantum propositions, whereas we are concerned directly with the representation of quantum processes. Our underlying logical setting is linear, theirs is cartesian, supporting the intuitionistic logic of toposes. It is an interesting topic for future work to relate, and perhaps even usefully combine, these approaches.

\paragraph{Comparison with the $n$-categories approach}
The $n$-categories approach is mainly motivated by the quest for quantum gravity.
In our approach, we emphasize the following key features which are essentially absent from the $n$-categories work:
\begin{itemize}
\item operational aspects
\item the interplay of quantum and classical
\item compositionality
\item open \textit{vs.}~closed systems.
\end{itemize}
These are important for applications to quantum informatics, but also of foundational significance.

There are nevertheless some intriguing similarities and possible connections, notably in the r\^ole played by  \emph{Frobenius algebras}, which we will mention briefly in the context of our approach in Section~7.

\subsection{Outline of the Article}
In Section~2, we shall give a rapid review of quantum mechanics and some quantum protocols such as teleportation. In Sections~3, 4 and 5, we shall present the main ingredients of the formalism: compact and strongly compact categories, and biproducts. In Section~6, we shall show how quantum mechanics can be axiomatized in this setting, and how the formalism can be applied to the complete specification and verification of a number of important quantum protocols. In Section~7 we shall review some of the main developments and advances made within the categorical quantum mechanics programme since \cite{AC2}, thus giving a picture of the current state of the art.

\section{Review of Quantum Mechanics and Teleportation}
\label{qmtelepsec}

In this paper, we shall only consider \emph{finitary} quantum
mechanics, in which all Hilbert spaces are finite-dimensional. This is 
standard in most current discussions of quantum computation and
information \cite{Nielsen},
and corresponds physically to considering only observables with finite
spectra, such as \emph{spin}. (We  refer briefly to the extension 
of our approach to the infinite-dimensional case in the
Conclusion.)    

Finitary quantum theory has the following basic ingredients (for
more details, consult standard texts such as \cite{Isham}).
\ben
\item[{\bf 1.}] The \em state space \em of the system is represented as a
  finite-dimensional Hilbert space $\HH$, \ie a finite-dimensional complex
  vector space with a `sesquilinear' inner-product written $\langle \phi \mid \psi
  \rangle$, which is conjugate-linear in the first argument and linear 
  in the second.  A \emph{state} of a quantum system
corresponds to a one-dimensional subspace ${\cal A}$ of $\HH$,
 and is standardly
represented by a vector $\psi\in {\cal A}$ of unit norm. 
\item[{\bf 2.}] For informatic purposes, the basic type is that of
  \emph{qubits}, namely $2$-dimensional Hilbert space, equipped with a
  \em computational basis \em $\{|0\rangle,|1\rangle\}$.
\item[{\bf 3.}] \em Compound systems \em are described by tensor products of the
  component systems. It is here that the key phenomenon of
 \em entanglement \em arises, since the general form of a vector in $\HH_1
  \otimes \HH_2$ is
\[ 
\sum_{i=1}^n \alpha_i\cdot \phi_i \otimes \psi_i
\] 
Such a vector may encode \emph{correlations} between the 
first and second components of the system, and cannot simply be
resolved into a pair of vectors in the component spaces.
\een
The \em adjoint \em to a linear map
$f:{\cal H}_1\to{\cal H}_2$ is the  linear map $f^\dagger:{\cal
H}_2\to{\cal H}_1$ such that, for all $\phi\in {\cal H}_2$ and  
$\psi\in {\cal H}_1$,
\[
\langle\phi \mid f(\psi)\rangle_{{\cal H}_2} = \langle f^\dagger(\phi) \mid
\psi\rangle_{{\cal H}_1}\,.
\]
\emph{Unitary
  transformations} are linear isomorphisms 
$U:{\cal H}_1\to{\cal H}_2$
such that
\[
U^{-1}\!\!=U^\dagger:{\cal H}_2\to{\cal H}_1\,.
\]
Note that all such transformations 
\em preserve the inner product \em since,
for all $\phi,\psi\in{\cal H}_1$,
\[
\langle U(\phi) \mid U(\psi)\rangle_{{\cal H}_2}=\langle (U^\dagger U)(\phi) \mid
\psi\rangle_{{\cal H}_1}=\langle \phi \mid \psi\rangle_{{\cal H}_1}\,.
\]
\em Self-adjoint operators \em are linear transformations
$M :{\cal H}\to{\cal H}$
such that $M=M^\dagger$.

\ben
\item[{\bf 4.}] The \em basic data transformations \em are represented by unitary
transformations.  Note that all such data transformations are necessarily
\emph{reversible}.
\item[{\bf 5.}] The \em measurements \em which can be performed on the system are
represented by  self-adjoint operators.  
\een
The act of measurement itself consists of two parts:
\bit
\item[{\bf 5a.}] The observer is informed about the measurement outcome,
  which is a value $x_i$ in the spectrum $\sigma(M)$ of
the corresponding self-adjoint operator $M$. For convenience we assume
$\sigma(M)$ to be \em non-degenerate \em (linearly independent eigenvectors
have distinct eigenvalues). 
\item[{\bf 5b.}] The state of the system undergoes a change, represented by
  the action of the \em projector \em ${\rm P}_i$ arising from  the \em spectral decomposition \em 
\[ 
M=x_1\cdot {\rm P}_1+\ldots+x_n\cdot {\rm P}_n
\]
\eit
In this spectral decomposition the projectors ${\rm P}_i:{\cal H}\to{\cal
H}$ are  idempotent, self-adjoint, and
mutually orthogonal
\[ 
{\rm P}_i\circ{\rm P}_i={\rm P}_i\qquad\quad\quad{\rm P}_i={\rm
P}_i^\dagger \qquad\quad\quad {\rm P}_i\circ{\rm P}_j=0, \ \
i\not=j . 
\]

This spectral decomposition always exists and is unique by the
\em spectral theorem \em for self-adjoint operators.  By our assumption
that $\sigma(M)$ was non-degenerate each projector ${\rm P}_i$ has a
one-dimensional subspace of
${\cal H}$ as its fixpoint set (which equals its image).

The probability of $x_i\in\sigma(M)$ being the actual outcome is given
by the \em Born rule \em which does not depend on
the value of $x_i$ but on ${\rm P}_i$ and the system state $\psi$, explicitly 
\[
{\sf Prob}({\rm
P}_i,\psi)=\langle\psi\mid {\rm P}_i(\psi)\rangle\,. 
\]
The
status of the Born rule within our abstract setting will emerge in
Section~8.
The derivable notions of \em mixed states \em and \em non-projective
measurements
\em will not play a significant r\^ole in this paper.

The values $x_1,\ldots,x_n$ are in effect merely labels distinguishing  
the projectors ${\rm P}_1,\ldots,{\rm P}_n$ in the
above sum. Hence we can abstract over them and think of
a measurement as a list of $n$ mutually orthogonal
projectors
$({\rm P}_1,\ldots,{\rm P}_n)$
where $n$ is the dimension of the Hilbert space. 

Although real-life experiments in many cases destroy the
system (e.g.~any measurement of a photon's location destroys it) measurements always have the same shape in
the quantum formalism.  When distinguishing between `measurements which preserve the system' and `measurements which
destroy the system' it would make sense to decompose a measurement explicitly in two components:
\bit
\item \em Observation \em consists of receiving the information on the outcome of the measurement, to be thought of
as specification of the index $i$ of the outcome-projector ${\rm P}_i$ in the above list. Measurements which destroy
the system can be seen as `observation only'$\!$. 
\item \em Preparation \em consists of producing the state ${\rm P}_i(\psi)$.
\eit
In our abstract setting these arise naturally as the two `building blocks' which are used to construct
projectors and measurements. 

We now discuss some important quantum protocols which we chose because 
of the key r\^ole entanglement plays in
them --- they involve both initially entangled states, and measurements against a basis of entangled states.

\subsection{Quantum teleportation}

%%%I blended the introduction description with the one here

The quantum teleportation \index{quantum teleportation} protocol \cite{BBC} (see also
\cite{Coe1}\S2.3\&\S3.3) involves three qubits $a$, $b$ and $c$ and two spatial
regions
$A$ (for ``Alice'') and $B$ (for ``Bob''). 

\vspace{2.5mm}\noindent{
\hspace{-15pt}\begin{minipage}[b]{1\linewidth}
\centering{\epsfig{figure=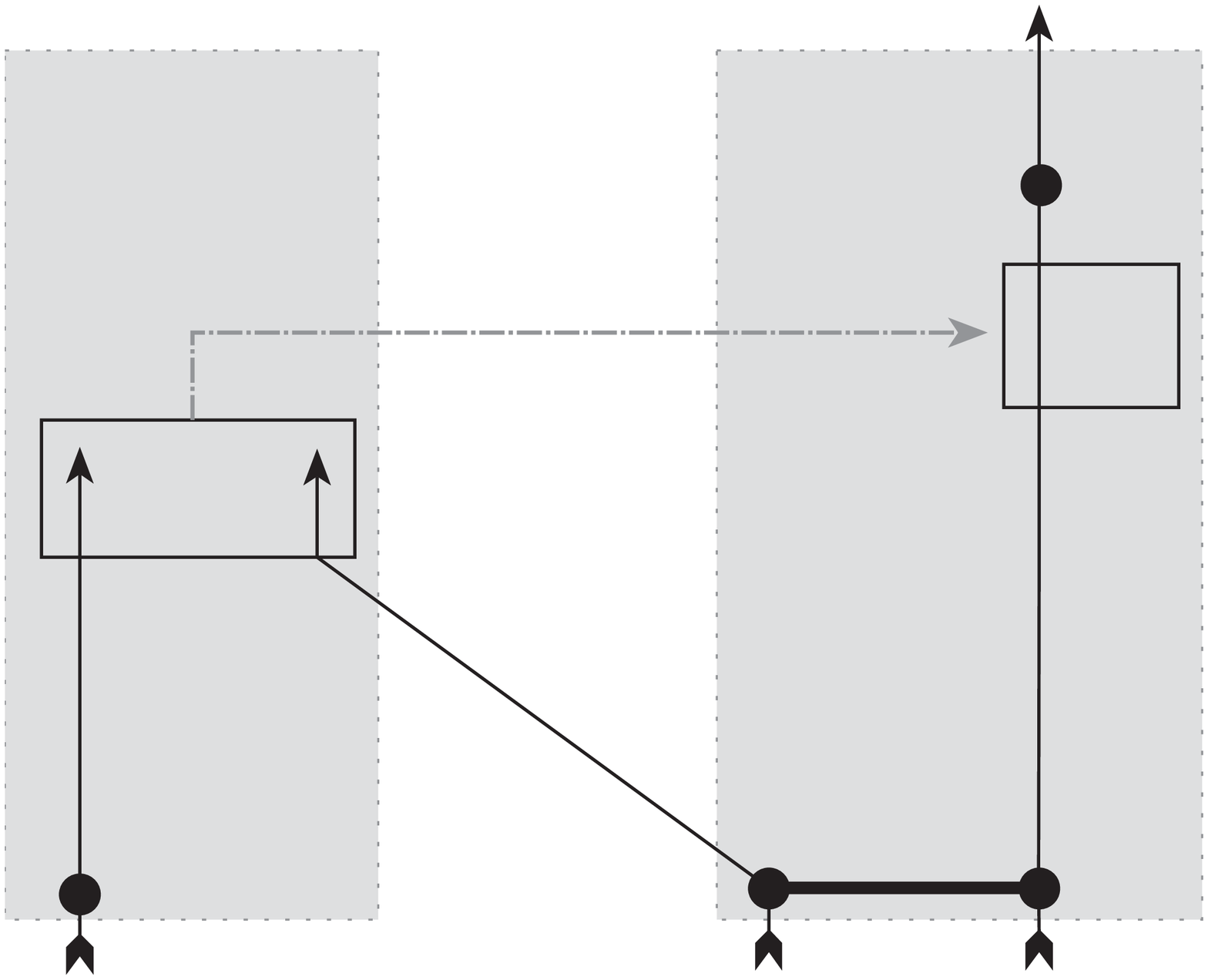,width=245pt}}

\hspace{3mm}\begin{picture}(245,0)
\put(155.5,37){\large$|00\rangle\!+\!|
\hspace{-0.5pt}1\hspace{-0.5pt}1\hspace{-0.5pt}\rangle$}
\put(20,108){\Large$M_{\!Bell}$}
\put(214,139){\Large$U_{\!x}$}
\put(91.5,149){\large${x\in\mathbb{B}^2}$}
\put(209.1,182.2){\large$|\phi\rangle$}
\put(14,37){\large$|\phi\rangle$}
\put(255,78){\vector(0,1){60}}
\put(257,104){${\rm time}$}
\put(-1,189){\Large$A$}
\put(144,189){\Large$B$}
\put(7.4,1){\Large$a$}
\put(149,0){\Large$b$}
\put(203.4,1){\Large$c$}
\end{picture}
\end{minipage}}

\vspace{1.5mm}\noindent
Qubit $a$ is in a state
$|\phi\rangle$ and located in $A$.
Qubits $b$ and $c$ form an `EPR-pair', that is, their joint state is
$|00\rangle+|11\rangle$. We assume that these qubits are
initially in $B$ e.g.~Bob created them. 
After \emph{spatial relocation} so that $a$ and $b$ are located in $A$, while
$c$ is positioned in $B$, or in other words, ``Bob sends qubit $b$ to
$Alice$'', we can start the actual teleportation of qubit $a$.
Alice performs a \em Bell-base measurement
\em on $a$ and $b$ at $A$, that is, a measurement such that each ${\rm P}_i$ projects on one of the
one-dimensional subspaces spanned by a vector in the \em
Bell basis\em: 
\[
b_1:={|00\rangle\!+\!|11\rangle\over \sqrt{2}}\quad
b_2:={|01\rangle\!+\!|10\rangle\over \sqrt{2}}\quad
b_3:={|00\rangle\!-\!|11\rangle\over \sqrt{2}}\quad
b_4:={|01\rangle\!-\!|10\rangle\over \sqrt{2}}\,.\!\!\!
\]
This measurement can be of the type `observation only'. Alice observes
the outcome of the measurement and ``sends these two classical bits
($x\in\mathbb{B}^2$) to
Bob''. Depending on which classical bits he receives Bob then
performs one of the unitary
transformations
\[
\beta_1:=\left(\begin{array}{rr} 
1&0\\
0&1
\end{array}\right)
\quad
\beta_2:=\left(\begin{array}{rr}
0&1\\
1&0
\end{array}\right)
\quad
\beta_3:=\left(\begin{array}{rr}
1&0\\
0&\!\!\!\!\!-\!1
\end{array}\right)
\quad
\beta_4:=\left(\begin{array}{rr}
0&\!\!\!\!\!-\!1\\
1&0
\end{array}\right) 
\]
on $c$ --- $\beta_1,\beta_2,\beta_3$
are all self-inverse while
$\beta_4^{-1}=-\beta_4$. 
The final state of $c$ proves to be $|\phi\rangle$ as well. (Because of the measurement, $a$ no longer
has this state --- the information in the source has been `destroyed' in 
transferring it to the target).  Note that the state of $a$
constitutes continuous data ---an arbitrary pair of complex
numbers $(\alpha ,\beta)$ satisfying $|\alpha|^2 + |\beta|^2 = 1$--- while the actual physical data transmission only
involved two classical bits. We will be able to derive this fact in
our abstract setting. 
Teleportation is simply the most basic of a family of quantum
protocols, and already illustrates the basic ideas, in particular the
use of \emph{preparations of entangled states} as carriers for information
flow, performing \emph{measurements} to propagate information, using
\emph{classical information} to control branching behaviour to ensure the
required behaviour despite quantum indeterminacy, and performing local 
data transformations using \emph{unitary operations}. (Local here means
that we apply these operations only at $A$ or at $B$, which are
assumed to be spatially separated, and not simultaneously at both).

Since in quantum teleportation a continuous variable has been 
transmitted while the actual \em
classical communication \em involved only two bits, besides this \em classical
information flow \em there has to exist 
%%%INC
some kind of
%%% 
\em quantum information flow\em.  The
nature of this quantum flow has been analyzed by one of the authors in
\cite{Coe1,Coe2}, building on the joint work in \cite{AC}. We recover those results
in our abstract setting (see Subsection 
\ref{sec:ABSETNNETW}), which also reveals additional
`fine structure'. To identify it we have to separate it from
the classical information flow. Therefore we decompose the
protocol into:
\ben
\item a {\it tree\,} with the  operations as nodes, and
  with \emph{branching} caused by the indeterminism of
measurements;
\item  a \em network \em of the operations in terms of 
the order they are applied and the subsystem
to which they apply. 
\een 

\vspace{0.8mm}\noindent{\footnotesize  
\begin{minipage}[b]{1\linewidth}   
\centering{\epsfig{figure=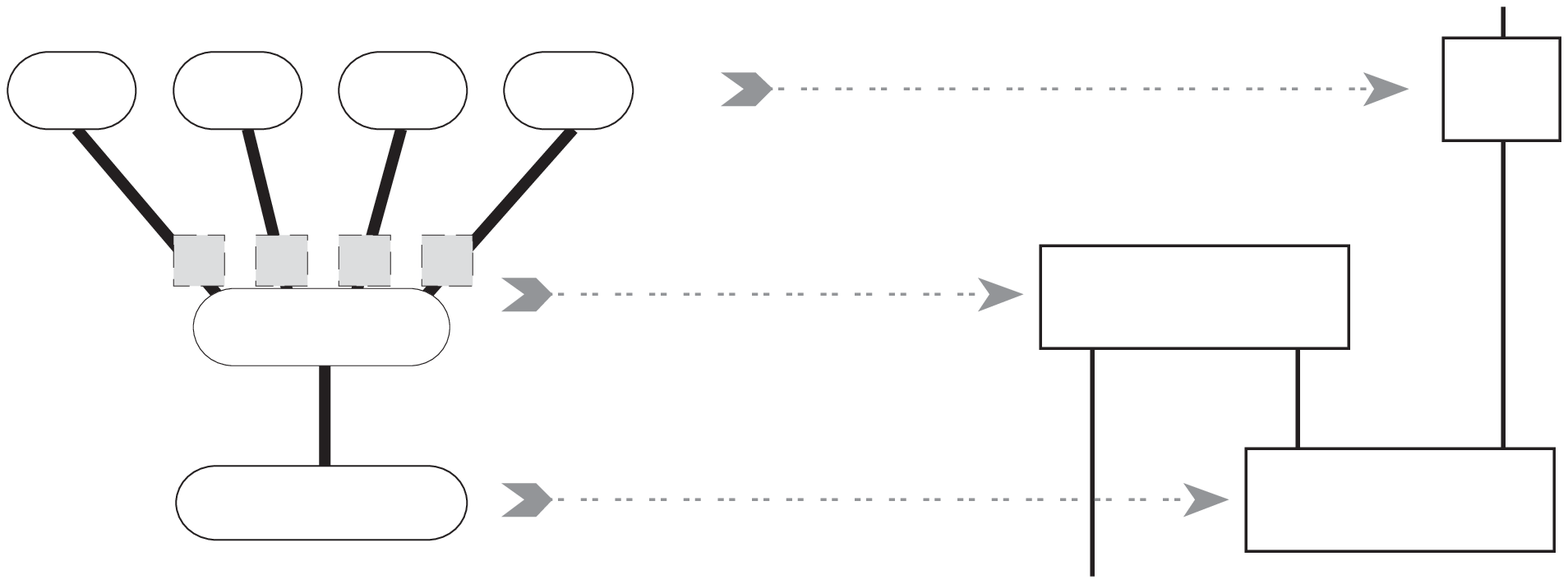,width=205pt}}       
  
\begin{picture}(205,0)   
\put(26.5,24.0){\scriptsize$|00\rangle\!+\!|\hspace{-0.5pt}1\hspace{-0.5pt}1\hspace{-0.5pt}\rangle$}     
\put(31,46.2){$M_{\!Bell}$}  
\put(3,76.9){$U_{\!00}$}   
\put(25,76.9){$U_{\!01}$}    
\put(47,76.9){$U_{\!10}$}   
\put(69,76.9){$U_{\!11}$}   
\put(22.8,56.6){\scriptsize${}_{0\hspace{-0.5pt}0}$}     
\put(33.6,56.6){\scriptsize${}_{0\hspace{-0.5pt}1}$}      
\put(44.2,56.6){\scriptsize${}_{1\hspace{-0.5pt}0}$}    
\put(55.55,56.6){\scriptsize${}_{1\hspace{-0.7pt}1}$} 
\put(194.5,75.8){$...$}   
\put(153.2,48.8){$...$}   
\put(180.7,21.5){$...$}     
\put(141,9){\normalsize$a$}  
\put(168.5,9){\normalsize$b$}  
\put(196,9){\normalsize$c$}
\end{picture}  
\end{minipage}}

\vspace{-1.8mm}\noindent
The nodes in the tree are connected to the boxes in the network by
their temporal coincidence. 
Classical communication is encoded in the tree as the dependency
of operations on the branch they are in. For each path from the
root of the tree to a leaf, by `filling in the
operations on the included nodes in the corresponding boxes of the
network', we obtain an \em entanglement network\em, that is,
a network
  
\vspace{3.4mm}\noindent{\footnotesize  
\begin{minipage}[b]{1\linewidth}  
\centering{\epsfig{figure=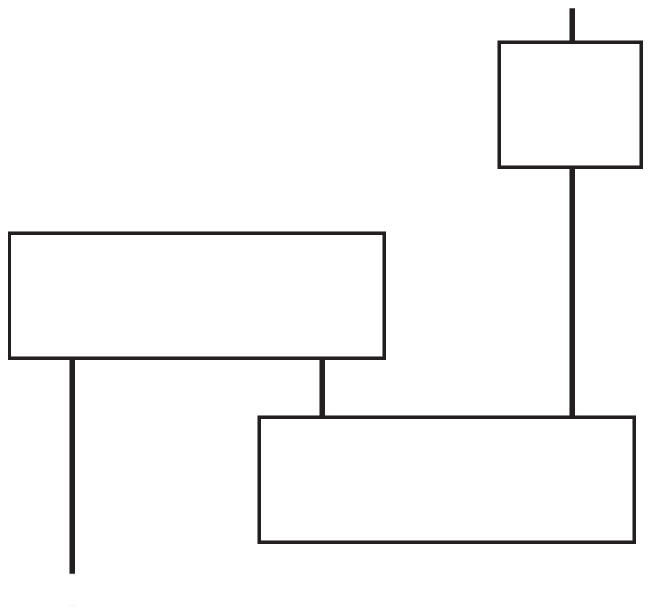,width=70pt}}      
  
\begin{picture}(70,0)   
\put(33,22.2){\scriptsize$|00\rangle\!+\!|\hspace{-0.5pt}1\hspace{-0.5pt}1
\hspace{-0.5pt}\rangle$}      
\put(16.8,42.5){${\rm P}_{\!x}$}   
\put(57.5,63.5){$U_{\!x}$}    
\put(5,8){\normalsize$a$}
\put(32.5,8){\normalsize$b$} 
\put(60,8){\normalsize$c$}
\put(95,23){\vector(0,1){40}}  
\put(97,39){${\rm time}$}  
\end{picture}  \vspace{-1mm}
\end{minipage}}
  
\vspace{0mm}\noindent
for each of the four values $x$ takes.
A component ${\rm P}_{\!x}$ of an 
observation will be referred to as an
\em observational branch\em. It will
be these networks, from which we have
removed the classical
information flow, that we will
study in Subsection
\ref{sec:ABSETNNETW}.  (There
is a clear analogy with the idea of 
unfolding a Petri net into its set
of `processes' \cite{Petri}).
The
classical information flow will be
reintroduced in Section~\ref{sec:biprods}.

\subsection{Logic gate teleportation}

Logic gate teleportation \cite{Gottesman} (see also \cite{Coe1}\S3.3) generalizes
the above protocol in that $b$ and $c$ are initially not necessarily an EPR-pair but may
be in some other (not arbitrary) entangled state $|\Psi\rangle$.  Due to this
modification the final state of $c$ is not
$|\phi\rangle$ but
$|f_\Psi(\phi)\rangle$ where $f_\Psi$ is a linear map which depends on $\Psi$.  As
shown in \cite{Gottesman}, when  this construction is applied to the
situation where  $a$, $b$ and $c$ are  each a pair of qubits rather
than a single
qubit, it provides a universal quantum computational
primitive which is moreover fault-tolerant \cite{Shor} and enables
the construction of a quantum computer based on single qubit unitary operations,
Bell-base measurements and only one kind of prepared state (so-called GHZ states).
The connection between $\Psi$, $f_\Psi$ and the unitary corrections $U_{\Psi\!,x}$ will
emerge straightforwardly in our abstract setting.

\subsection{Entanglement swapping}

Entanglement swapping \cite{Swap} (see also \cite{Coe1}\S6.2) is another
modification of the
teleportation protocol where $a$ is not in a state $|\phi\rangle$ but is a
qubit in an
EPR-pair together with an ancillary qubit $d$.  The result is that after
the protocol $c$ forms an EPR-pair
with $d$. If the measurement on $a$ and $b$ is non-destructive, we can also
perform a
unitary operation on $a$, resulting in  $a$ and $b$ also constituting
an EPR-pair.
Hence we have `swapped' entanglement:
 
\vspace{5.2mm}\noindent{\footnotesize  
\begin{minipage}[b]{1\linewidth}  
\centering{\epsfig{figure=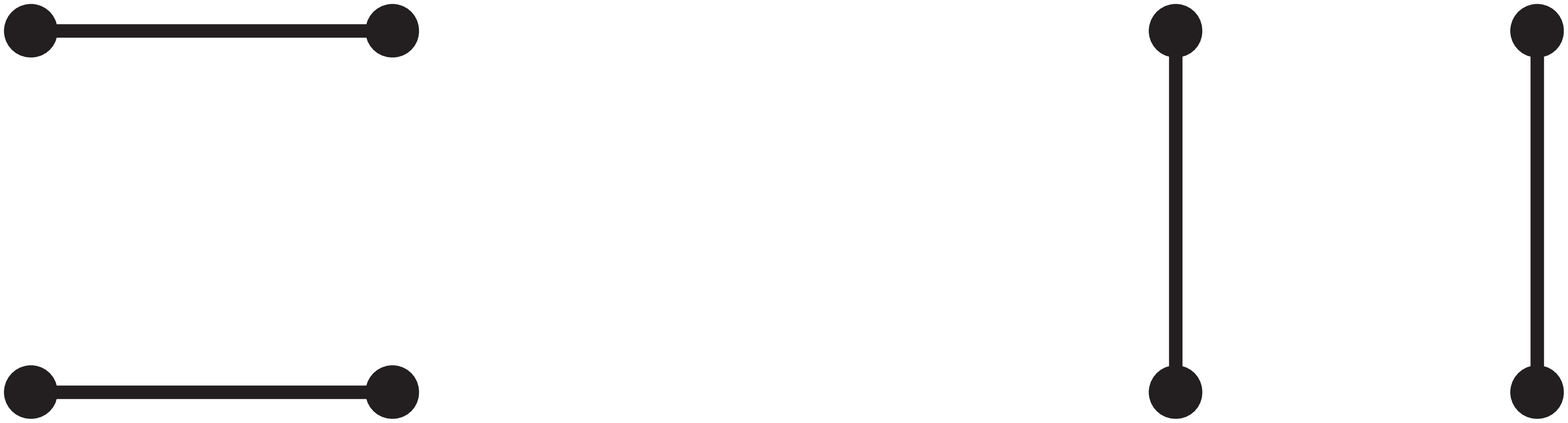,width=141.81pt}}       
   
\begin{picture}(141.81,0)   
\put(06,15.2){\tiny$|00\rangle\!\!+\!\!|\hspace{-0.5pt}1\hspace{-0.5pt}1\hspace{-0.5pt}\rangle$}      
\put(06,48.2){\tiny$|00\rangle\!\!+\!\!|\hspace{-0.5pt}1\hspace{-0.5pt}1\hspace{-0.5pt}\rangle$}      
\put(107.5,28.2){\tiny$|00\rangle\!\!+\!\!|\hspace{-0.5pt}1\hspace{-0.5pt}1\hspace{-0.5pt}\rangle$}      
\put(140.2,28.2){\tiny$|00\rangle\!\!+\!\!|\hspace{-0.5pt}1\hspace{-0.5pt}1\hspace{-0.5pt}\rangle$}      
\put(66.0,23.5){\large$\leadsto$}       
\put(-5,6){$b$}       
\put(-5,47){$a$}       
\put(38,47){$d$}       
\put(38,6){$c$}       
\put(99,6){$b$}       
\put(99,47){$a$}        
\put(142,47){$d$}       
\put(142,6){$c$}       
\end{picture}  
\end{minipage}}  
 
\vspace{-0.5mm}\noindent
In this case the entanglement networks have the shape: 

\vspace{1.8mm}\noindent{\footnotesize  
\begin{minipage}[b]{1\linewidth}  
\centering{\epsfig{figure=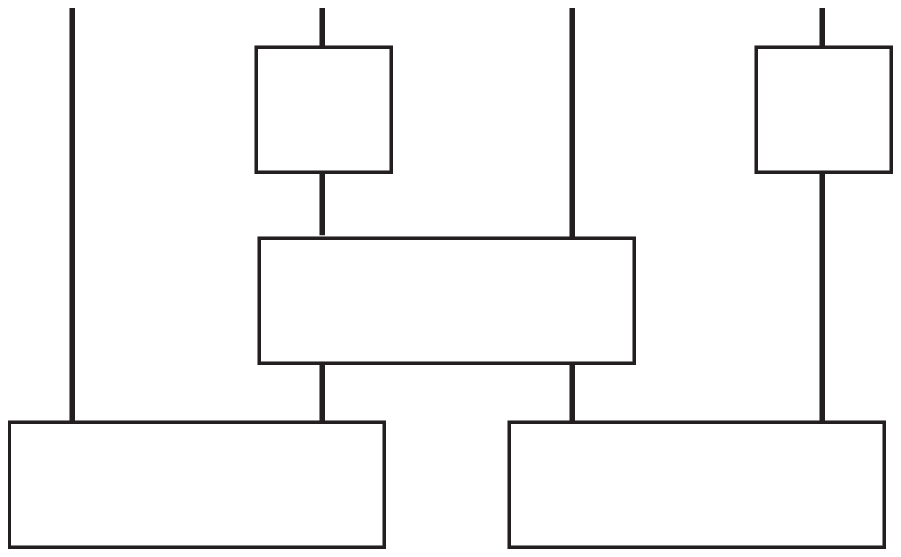,width=98pt}}       
   
\begin{picture}(98,0)   
\put(05,15.2){\scriptsize$|00\rangle\!+\!|\hspace{-0.5pt}1\hspace{-0.5pt}1\hspace{-0.5pt}\rangle$}      
\put(60.5,15.2){\scriptsize$|00\rangle\!+\!|\hspace{-0.5pt}1\hspace{-0.5pt}1\hspace{-0.5pt}\rangle$}      
\put(45.8,35.5){${\rm P}_{\!x}$}   
\put(31.5,56.5){$U_{\!x}\hspace{-1.1mm}\mbox{\rm'}$}     
\put(86.5,56.5){$U_{\!x}$}     
\put(5,2.2){\normalsize$d$}
\put(32.5,2.5){\normalsize$a$} 
\put(60,2.2){\normalsize$b$}
\put(87.5,2.5){\normalsize$c$}
\put(125,19){\vector(0,1){40}}  
\put(127,35){${\rm time}$}    
\end{picture}  
\end{minipage}}
 
\vspace{0.6mm}\noindent
Why this protocol works will again  emerge straightforwardly from our
abstract setting, as will 
generalizations of this protocol which have a much more sophisticated
compositional content (see Subsection
\ref{sec:ABSETNNETW}).

\section{Compact Closed Categories and the Logic of Entanglement}
\label{sec:CCC}
\subsection{Monoidal Categories}

Recall that a \em symmetric monoidal category \em consists of a category {\bf C},
a bifunctorial\index{monoidal category}\index{symmetric monoidal category} \em tensor \em 
\[
-\otimes-:{\bf C}\times{\bf C}\to{\bf C}\,,
\]
a \em unit \em object
${\rm I}$, and natural isomorphisms 
\[
\lambda_A : A \simeq {\rm I}\otimes A\quad\quad\quad\quad\quad\ \ \rho_A: A \simeq
A\otimes{\rm I}
\]
\[
\alpha_{A,B,C}:A\otimes(B\otimes C)\simeq (A\otimes B)\otimes C\vspace{0.5mm} 
\] 
\[
\sigma_{A,B}:A\otimes B\simeq B\otimes A
\]
which satisfy certain coherence conditions \cite{MacLane}.

\paragraph{Examples}
The following two examples are of particular importance and will recur through this section.
\begin{enumerate}
\item  The category $\FdVect_{\mathbb{K}}$,  of finite-dimensional vector spaces over a field $\mathbb{K}$ and linear maps. The tensor product\index{tensor product} is the usual construction on vector spaces. The unit of the tensor is $\mathbb{K}$, considered as a one-dimensional vector space over itself.
\item The category $\Rel$\index{category of relations} of sets and relations, with cartesian product as the `tensor', and a one-element set as the unit. Note that cartesian product is \emph{not} the categorical product in $\Rel$.
\end{enumerate}

\paragraph{The Logic of Tensor Product}
Tensor can express \emph{independent} or \emph{concurrent} actions (mathematically: bifunctoriality):
\[ \begin{diagram}
A_1 \otimes A_2 & \rTo^{f_1 \otimes 1} & B_1 \otimes A_2 \\
\dTo^{1 \otimes f_2} & & \dTo_{1 \otimes f_2} \\
A_1 \otimes B_2 & \rTo_{f_1 \otimes 1} & B_1 \otimes B_2 \\
\end{diagram}
\]
But tensor is \emph{not} a categorical product, in the sense that \emph{we cannot reconstruct an `element' of the tensor from its components}.

This turns out to comprise the \emph{absence} of \emph{diagonals} and \emph{projections}:
\[ \begin{array}{lcc}
& A \labarrow{\Delta} A \otimes A & A_1 \otimes A_2 \labarrow{\pi_i} A_i \\
\mbox{Cf.} & A \vdash A \wedge A & A_1 \wedge A_2 \vdash A_i 
\end{array} 
\]
Hence monoidal categories provide a setting for \emph{resource-sensitive} logics such as Linear Logic \cite{Girard}. 
No-Cloning and No-Deleting are built in!
\emph{Any} symmetric monoidal category can be viewed as \emph{a setting for describing  processes in a resource sensitive way, closed under sequential and parallel composition}

\subsection{The `miracle' of scalars}

A key step in the development of the categorical axiomatics for Quantum Mechanics  was the recognition that the notion of \emph{scalar} is meaningful in great generality --- in fact, in any monoidal (not necessarily symmetric) category.

Let $(\CC , \otimes , \II, \lambda, \alpha, \sigma )$ be a  monoidal category .
We define a \emph{scalar} in $\CC$ to be a morphism $s : \II \rightarrow \II$, \ie an endomorphism of the tensor unit.

\begin{example}
In $\mathbf{FdVec}_{\mathbb{K}}$, linear maps $\mathbb{K} \to \mathbb{K}$ are uniquely
determined by the image of $1$, and hence correspond biuniquely to
elements of $\mathbb{K}\,$; composition corresponds to multiplication of 
scalars. In $\mathbf{Rel}$\index{category of relations}, there are just two scalars, corresponding
to the Boolean values $0$, $1$.
\end{example}

\noindent The (multiplicative) monoid of scalars is then just the endomorphism monoid $\CC (\II , \II )$.
The first key point is the elementary but beautiful observation by Kelly and Laplaza \cite{KL} that this monoid is always commutative.
\begin{lemma}
\label{scprop}
$\CC (\II , \II )$ is a commutative monoid
\end{lemma}
\bpf
\[ \begin{diagram}
\II & \rTo^{\rho_{\II}} & \II \otimes \II &  \rEq & \II \otimes \II & \rTo^{\lambda_{\II}^{-1}} & \II \\
\uTo^{s} & & \uTo^{s \otimes 1} & & \dTo_{1 \otimes t} & & \dTo_{t} \\
\II & \rTo^{\rho_{\II}} & \II \otimes \II & \rTo^{s \otimes t} & \II \otimes \II & \rTo^{\lambda_{\II}^{-1}} & \II \\
\dTo^{t} & & \dTo^{1 \otimes t} & & \uTo_{s \otimes 1} & & \uTo_{s} \\
\II & \rTo_{\lambda_{\II}} & \II \otimes \II &  \rEq & \II \otimes \II & \rTo_{\rho_{\II}^{-1}} & \II \\
\end{diagram}
\]
using the coherence equation $\lambda_{\II} = \rho_{\II}$.
\hfill\endproof\newline

The second point is that a good notion of \emph{scalar multiplication} exists at this level of generality.
That is, each scalar
$s:\II\to\II$ induces a natural transformation
\[ \begin{diagram}
s_A :A & \rTo^{\simeq} & \II \otimes \!A & \rTo^{s \otimes 1_A} & \II
\otimes\! A & \rTo^{\!\!\simeq\ } & A\,.\ \ \ \ \
\end{diagram}
\]
with the naturality square
\[ \begin{diagram}
A & \rTo^{s_A} & A \\
\dTo^{f} & & \dTo_{f} \\
B & \rTo_{s_B} & B \\
\end{diagram}
\]
We write $s \sdot f$ for $f \circ s_A=s_B\circ f$.
Note that
\[ \begin{array}{lcl}
1 \sdot f & = & f \label{sdotident} \\
s \sdot (t \sdot f) & = &  (s \circ t) \sdot f \label{sdotact}\\
(s \sdot g)\circ(t \sdot f) & =  & (s\circ t)\sdot(g\circ f) \label{sdotcomp}\\
(s \sdot f) \otimes (t \sdot g) & = &  (s \circ t) \sdot (f \otimes g) \label{sdotten}
\end{array}
\]
which exactly generalizes the multiplicative part of the usual properties of scalar multiplication.
Thus scalars act globally on the whole category.

\subsection{Compact Closure}
A category {\bf C} is \em $*$-autonomous \em \cite{Barr} if
it is symmetric monoidal, and comes equipped with a full and
faithful functor 
\[
(\ )^*:{\bf C}^{op}\to{\bf C}
\]
such that a bijection
\[ 
{\bf C}(A\otimes B,C^*)\simeq {\bf C}(A,(B\otimes C)^*)
\]
exists which is natural in all variables. 
Hence a $*$-autonomous category is closed, with
\[
{A\multimap B:=(A\otimes B^*)^*}\,.
\]
These $*$-autonomous categories
provide a categorical semantics for the multiplicative
fragment of linear logic \cite{Seely}.

A \em compact closed category \em \cite{Kelly} is a
$*$-autonomous category with a self-dual tensor\index{compact closure},
i.e.~with natural isomorphisms
\[
u_{A,B}:(A\otimes B)^*\simeq A^*\otimes B^* \qquad u_{\II} : \II^* \simeq
\II\,.
\] 
It follows that 
\[
A\multimap B\simeq A^*\otimes B\,.
\]

\noindent A very different definition arises when one considers a symmetric monoidal
category as a one-object bicategory. In this context,
compact closure simply means that every object $A$, qua 1-cell of the
bicategory, has a specified adjoint \cite{KL}. 
\begin{definition}[Kelly-Laplaza]\label{def:compclos}\em 
A \em compact closed category \em is a symmetric monoidal
category in which to each object $A$ a \em dual object \em  
$A^*$, a \em unit \em 
\[
\eta_A:{\rm I}\to A^*\otimes A
\]
 and a \em  
counit \em 
\[
\epsilon_A:A\otimes A^*\to {\rm I}
\]
are assigned, in
such a way that the diagram
\begin{diagram}  
A&\rTo^{\rho_A}&A\otimes{\rm
I}&\rTo^{1_A\otimes\eta_A}&A\otimes(A^*\otimes
A)\\ 
\dTo^{1_A}&&&&\dTo^{\alpha_{A,A^*\!\!,A}}\\ 
A&\lTo_{\lambda_A^{-1}}&{\rm I}\otimes A&\lTo_{\epsilon_A\otimes
1_A}&(A\otimes A^*)\otimes A
\end{diagram}
and the dual one for $A^*$ both commute.  
\end{definition} 

\paragraph{Examples}The symmetric monoidal categories $({\bf Rel},\times)$ of sets, relations
and cartesian product\index{category of relations} and $({\bf FdVec}_\mathbb{K},\otimes)$ of finite-dimensional vector spaces over a field
$\mathbb{K}$, linear maps and tensor product
are both compact closed.  In $({\bf Rel},\times)$, we simply set $X^{*} = X$.  Taking a one-point set $\{ * \}$ as the unit for $\times$, and writing $R^{\cup}$ for the converse of a relation $R$:
\[
\eta_X=\epsilon_X^{\cup}=\{(*,(x,x))\mid x\in X\}\,.
\] 
For $({\bf FdVec}_\mathbb{K},\otimes)$, we take $V^{*}$ to be the dual space of linear functionals on $V$.
The unit and counit in $({\bf FdVec}_\mathbb{K},\otimes)$ are
\[
\eta_V:\mathbb{K}\to V^*\otimes V::1\mapsto\sum_{i=1}^{i=n}\bar{e}_i\otimes
e_i
\qquad
{\rm and}
\qquad
\epsilon_V:V\otimes
V^*\to\mathbb{K}::e_i\otimes\bar{e}_j\mapsto \bar{e}_{j}( e_{i})
\] 
where $n$ is the dimension of $V$, $\{e_i\}_{i=1}^{i=n}$ is a basis of
$V$ and $\bar{e}_i$ is the linear functional in $V^*$ determined by
$\bar{e}_{j}( e_{i}) = \delta_{ij}$.

\begin{definition}\label{def:name}\em
The \em name \em $\uu f\uuu$ and the \em coname \em $\dd f\ddd$ of a morphism $f:A\to B$ in a compact
closed category are
\begin{diagram} 
A^*\!\!\otimes\! A&\rTo^{1_{A^*}\!\!\otimes\! f}&A^*\!\otimes\! B&&&&&{\rm I}\\
\uTo^{\eta_A}&\ruTo_{\uu f\uuu}&&&&&\ruTo^{\dd f\ddd}&\uTo_{\epsilon_B} \\     
{\rm I}&&&&&A\!\otimes\! B^*&\rTo_{f\!\otimes\! 1_{B^*}}&B\!\otimes\! B^*&&
\end{diagram}
\end{definition}

For $R\in{\bf Rel}(X,Y)$ we have
\[
\uu R\uuu=\{(*,(x,y))\mid xRy,x\in X, y\in Y\}\quad{\rm and}\quad
\dd R\ddd=\{((x, y),*)\mid xRy,x\in X, y\in Y\}
\]

\noindent 
and for $f\in {\bf FdVec}_\mathbb{K}(V,W)$ with $(m_{ij})$ the matrix of
$f$ in bases
$\{e_i^V\}_{i=1}^{i=n}$ and $\{e_j^W\}_{j=1}^{j=m}$ of $V$ and $W$
respectively
\[
\uu f\uuu:\mathbb{K}\to V^*\otimes
W::1\mapsto\!\!\sum_{i,j=1}^{\!i,j=n,m\!}\!\!m_{ij}\cdot
\bar{e}_i^V\otimes e_j^W
\]
and
\[ 
\dd f\ddd:V\otimes
W^*\to\mathbb{K}::e_i^V\otimes\bar{e}_j^W\mapsto m_{ij}.
\]

\noindent Given $f:A\to B$ in any compact closed category ${\bf C}$  we can define $f^*:B^*\to A^*$ as
\begin{diagram}  
B^*&\rTo^{\lambda_{B^*}}&{\rm I}\otimes B^*&\rTo^{\eta_A\otimes 1_{B^*}}&A^*\otimes A\otimes
B^*\\ 
\dTo^{f^*}&&&&\dTo_{1_{A^*}\!\otimes f\otimes 1_{B^*}}\\   
A^*&\lTo_{\rho_{A^*}^{-1}}&A^*\otimes {\rm I}&\lTo_{1_{A^*}\otimes \epsilon_B}&A^*\otimes
B\otimes B^*
\end{diagram}
This operation $(\ )^*$ is functorial and makes Definition \ref{def:compclos} coincide
with the one given at the beginning of this section. It then follows by 
\[
{\bf C}(A\otimes B^*,{\rm I})\simeq{\bf C}(A,B)\simeq{\bf C}(I,A^*\otimes B)
\]
that every morphism
of type $\II\!\to\!   A^*\!\otimes B$ is the name of some morphism of type ${A\to B}$ and every
morphism of type ${A\otimes B^*\!\to{\rm I}}$ is the coname of some morphism of
type ${A\to B}$.  In the case of the unit and the counit we have
\[
\eta_A={\uu 1_A\uuu}\quad\quad{\rm and}
\quad\quad\epsilon_A={\dd 1_A\ddd}\,.  
\]
For $R\in{\bf Rel}(X,Y)$ the dual is the converse,
$R^*=R^{\cup}\in{\bf Rel}(Y,X)$, and for $f\in{\bf FdVec}_\mathbb{K}(V,W)$, the dual is
\[
f^*:W^*\to V^*::\,\phi \mapsto\,\phi\circ f\,.
\]
The following holds by general properties of adjoints and symmetry of 
the tensor \cite{KL}\S6.

\begin{proposition}\label{prop:triqcc}
In a compact closed category ${\bf C}$ there is a natural isomorphism
${d_A:A^{**}\simeq A}$ and the diagrams
\begin{diagram}
A^*\otimes
A&\rTo^{\sigma_{A^*\!\!,A}}&A\otimes A^*&&&\II&\rTo^{\eta_{A^*}}&A^{**}\otimes A^*\\
\dTo^{1_{A^*}\otimes d^{-1}_A}&&\dTo_{\epsilon_A}&&&\dTo^{\eta_A}&&\dTo_{d_A\otimes 1_{A^*}}\\
A^*\otimes
A^{**}&\rTo_{\epsilon_{A^*}}&\II&&&A^*\otimes A&\rTo_{\sigma_{A^*\!\!,A}}&A\otimes A^*
\end{diagram} 
commute for all objects $A$ of ${\bf C}$.
\end{proposition}

\paragraph{Graphical representation.}   Complex algebraic expressions for morphisms in symmetric monoidal categories can rapidly become hard to read.  Graphical representations exploit two-dimensionality, with the vertical dimension  corresponding to composition and the horizontal to the monoidal tensor,
and provide more intuitive presentations of morphisms.  We depict  objects by wires, morphisms by boxes with input and output wires, composition by connecting outputs to inputs, and the monoidal tensor by locating boxes side-by-side.  We distinguish between an object and its dual in terms of directions of the wires. In particular,
$g\circ f$, $f\otimes g$, $\uu f\uuu$ and $\dd f\ddd$ will respectively be depicted by
  \par\vspace{3mm}\par\noindent
\begin{minipage}[b]{1\linewidth}
\centering{\epsfig{figure=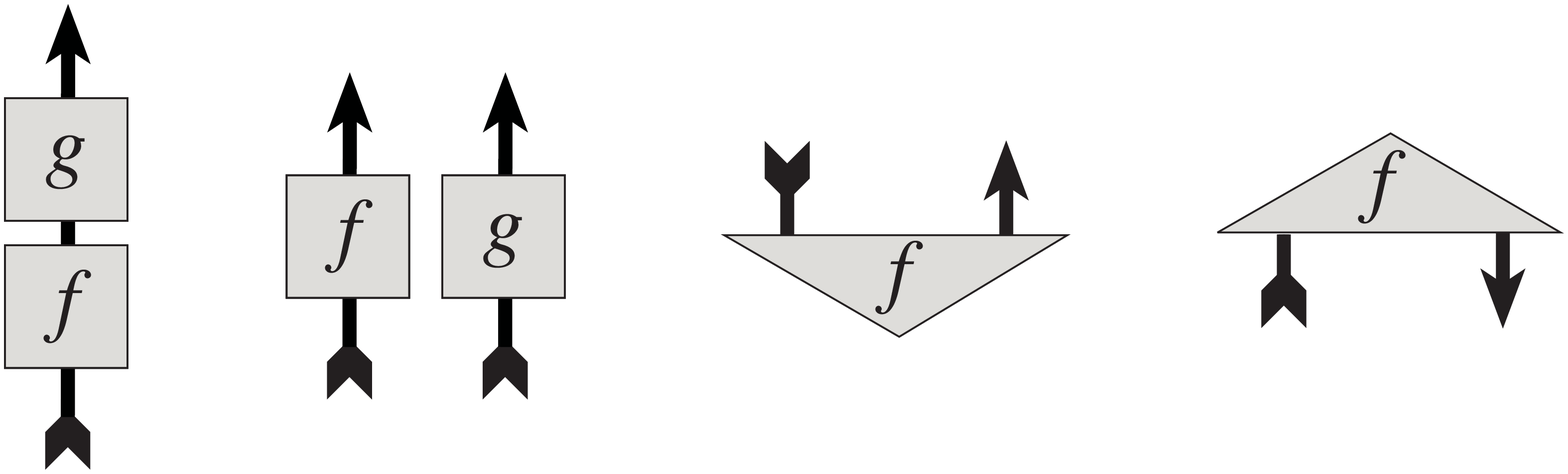,width=255pt}}
\end{minipage}
\par\vspace{3mm}\par\noindent
Implicit in the use of this graphical notation is that we assume we are working in a \emph{strict} monoidal category, in which the unit and associativity isomorphisms are identities. We can always do this because of the coherence theorem for monoidal categories \cite{MacLane}. Similarly, strictness is assumed for the duality in compact closed categories:
\[ A^{**} = A, \qquad (A \otimes B)^{*} = A^{*} \otimes B^{*}, \qquad \II^{*} = \II \, . \]
Pointers to references on diagrammatic representations and corresponding calculi are in Section \ref{sec:Diagrammatic}.

\subsection{Key lemmas}

The following Lemmas constitute the core of our
interpretation of entanglement in compact closed categories.
It was however observed by Radha Jagadeesan \shortcite{Radha} that they can
be shown in arbitrary $*$-autonomous categories using some of the results in
\cite{CS}.
 
\begin{lemma}[absorption]\label{lm:precompos} For
$A\rTo^{f}B\rTo^{g}C$
we have that
\[
(1_{A^*}\!\!\otimes g)\circ \uu f\uuu=\uu g\circ f\uuu.
\]
\end{lemma}
\bpf
Straightforward by Definition \ref{def:name}.
\hfill\endproof\newline

In a picture,
\par\vspace{3mm}\par\noindent
\begin{minipage}[b]{1\linewidth}
\centering{\epsfig{figure=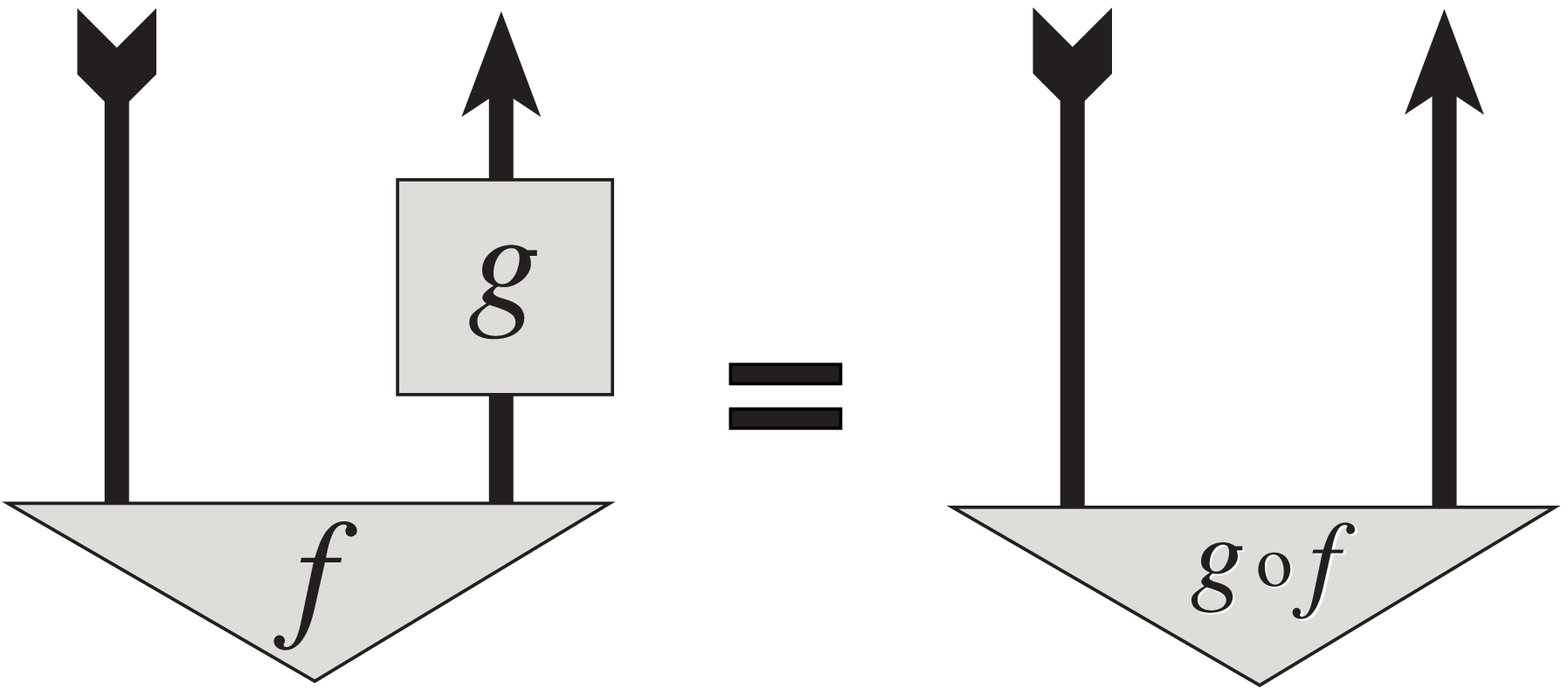,width=145pt}}
\end{minipage}
\par\vspace{3mm}\par\noindent

\begin{lemma}[Compositionality]\label{lm:compos} For
$A\rTo^{f}B\rTo^{g}C$
we have that
\[
\lambda^{-1}_C\circ (\dd f\ddd\otimes 1_C)\circ(1_A\otimes\uu
g\uuu)\circ\rho_A=g\circ f\,.
\]
\end{lemma}
\bpf
{\small\begin{diagram}
A&&&&&&\rTo^{g\circ f}&&&&&&C\\ %1
&\rdTo^{\rho_A}&&&&&\mbox{\bf Lemma
\ref{lm:compos}}&&&&&\ruTo^{\lambda^{-1}_C}&\\ %2
&&A\otimes {\rm I}&&\rTo^{1_A\otimes\uu g\uuu}&&A\otimes B^*\!\otimes C&&\rTo^{\dd f\ddd\otimes 1_C}&&{\rm I}\otimes C&&\\ %3
&&&\rdTo~{1_A\otimes\eta_B}&&\ruTo~{1_{A\otimes B^*}\!\!\otimes g}&&\rdTo~{f\otimes 1_{B^*\!\otimes C}}&&\ruTo~{\ \epsilon_B\otimes 1_C}&&&\\ %4
\dTo^f&&\dTo^{f\otimes 1_{\rm I}}&&A\otimes B^*\!\otimes B&&&&B\otimes B^*\!\otimes C&&\uTo_{1_{\rm I}\otimes g}&&\uTo_g\\
&&&&&\rdTo~{f\otimes 1_{B^*\!\otimes B}}&&\ruTo~{1_{B^*\!\otimes B}\otimes g}&&&&&\\
&&B\otimes{\rm I}&&\rTo_{1_B\otimes\eta_B}&&B\otimes B^*\otimes B&&\rTo_{\epsilon_B\otimes 1_B}&&{\rm I}\otimes B&&\\
&\ruTo_{\rho_B}&&&&&\hspace{-2cm}\mbox{\bf Compact
closedness}\hspace{-2cm}&&&&&\rdTo_{\lambda^{-1}_B}&\\
B&&&&&&\rTo_{1_B}&&&&&&B
\end{diagram}

}

\noindent
The top trapezoid
is the statement of  the Lemma. 
The diagram uses bifunctoriality and naturality of
$\rho$ and
$\lambda$.
\hfill\endproof\newline

In a picture,
\par\vspace{3mm}\par\noindent
\begin{minipage}[b]{1\linewidth}
\centering{\epsfig{figure=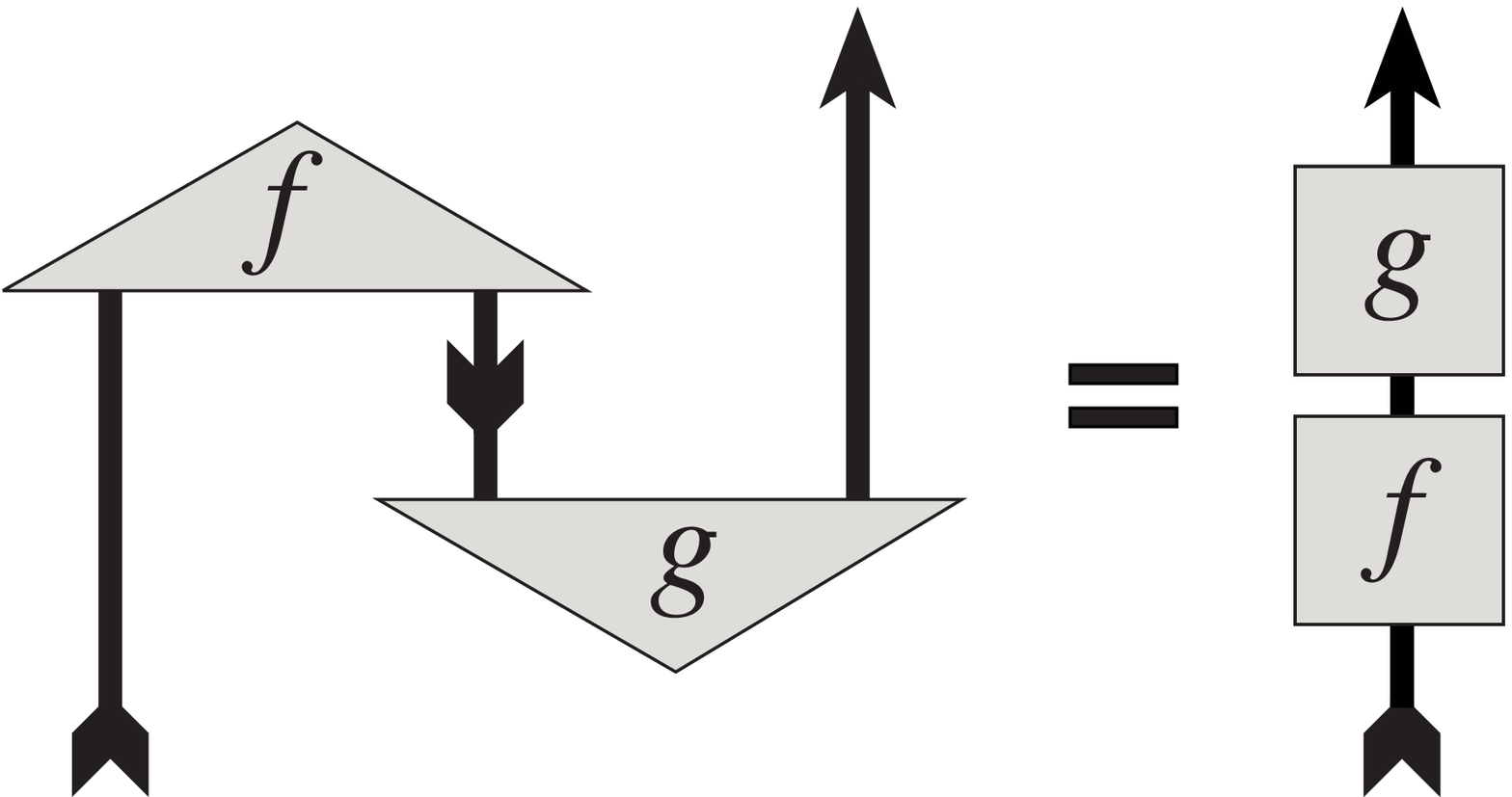,width=145pt}}
\end{minipage}
\par\vspace{3mm}\par\noindent

\begin{lemma}[Compositional CUT]\label{lm:CUT}   
For
$A\rTo^{f}B\rTo^{g}C\rTo^{h}D$
we have that
\[ 
\hspace{-1.5mm}(\rho^{-1}_A\!\otimes 1_{D^*}\!)\circ(1_{A^*}\!\otimes\dd g\ddd\otimes\!
1_D)\circ({\uu f\uuu}\otimes{\uu  h\uuu})\circ\rho_I =\uu h\circ g\circ f\uuu.
\hspace{-1.5mm}
\]
\end{lemma}
\bpf\par\noindent\hspace{-4.5mm}
{\small\mbox{\begin{diagram}
{\rm I}&&&&\rTo^{\uu h\circ g\circ f\uuu}&&&&A^*\!\!\otimes\! D\\ %1
&\rdTo^{\rho_{\rm I}}&&&\mbox{\bf Lemma
\ref{lm:CUT}}\hspace{-0.8cm}&&&\ruTo^{\rho^{-1}_A\!\otimes\!1_{D^*}}_{1_A\!\otimes\!
\lambda^{-1}_{D^*}}\ruTo(2,4)_{1_{A^*}\!\!\otimes\!(h\!\circ\! g)\qquad\quad}&\\  %2
&&{\rm I}\otimes {\rm I}&\rTo^{\!\!\!{\uu f\uuu}\!\otimes\!{\uu
h\uuu}}&A^*\!\!\otimes\! B\!\otimes\! C^*\!\!\otimes\! D&\rTo^{1_{A^*}\!\otimes\!
\dd g\ddd\!\otimes\! 1_D}&A^*\!\!\otimes\! {\rm I}\!\otimes\! D&&\\ %3
&&\dTo^{\!\!\eta_A\!\otimes\! 1_{\rm I}}&\rdTo~{{\uu f\uuu}\!\otimes\! 1_{\rm I}}&\uTo~{1_{A^*}\!\!\otimes\!{\uu h\uuu}}&\hspace{-0.9cm}\mbox{\bf Lemma
\ref{lm:compos}}&&&\\ %4
\dTo^{\ \ \ \ \eta_A}&&A^*\!\!\otimes\! A\otimes\!{\rm I}&\rTo^{\!\!\!1_{A^*}\!\!\otimes\! f\!\otimes\! 1_{\rm I}\!\!\!\!\!}&A^*\!\!\otimes\! B\!\otimes\!{\rm I}&\lTo^{\rho_{A^*\otimes B}}&A^*\!\!\otimes\! B&&\uTo~{\hspace{-7mm}1_{A^*}\!\!\otimes\!(h\!\circ\! g\!\circ\! f)}\\ %5
&\ruTo_{\rho_{A^*\!\otimes A}}&&&&&&\luTo_{1_{A^*}\!\!\otimes\! f}&\\ %6
A^*\!\!\otimes\!
A&&&&\rTo_{1_{A^*\!\otimes A}}&&&&A^*\!\!\otimes\! A
\end{diagram}}\bigskip

}

\noindent
The top trapezoid
is the statement of  the Lemma.
The diagram uses Lemma \ref{lm:compos} and naturality of $\rho$ and $\lambda$.
\hfill\endproof\newline

In a picture,
\par\vspace{3mm}\par\noindent
\begin{minipage}[b]{1\linewidth}
\centering{\epsfig{figure=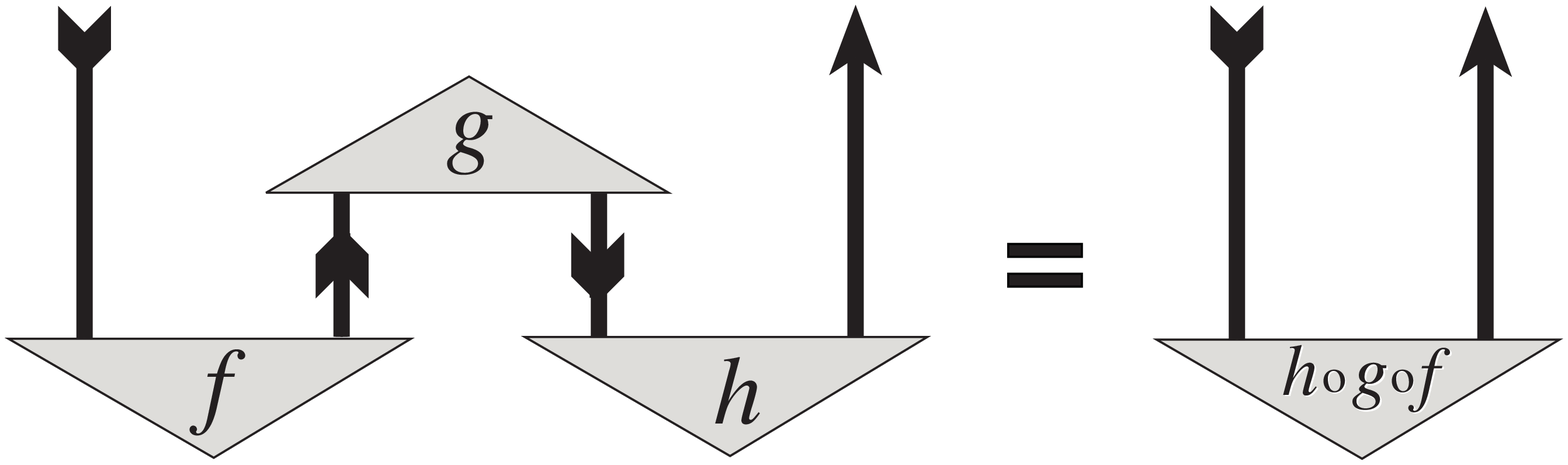,width=220pt}}
\end{minipage}
\par\vspace{3mm}\par\noindent

\paragraph{Discussion.} On the right hand side of Lemma
\ref{lm:compos} we have
$g\circ f$, that is, we first apply $f$ and then $g$, while on the left hand
side we first apply  the coname of $g$, and then the coname of $f$.  In
Lemma~\ref{lm:CUT} there is a similar, seemingly `acausal' inversion of the
order of application, as $g$ gets inserted between $h$ and $f$.

\medskip
For completeness we  add the following `backward' absorption lemma, which again involves a
reversal of the composition order.

\begin{lemma}[backward absorption]\label{lm:precompos2} For 
$C\rTo^{g}A\rTo^{f}B$
we have that
\[
(g^*\otimes 1_{A^*}\!)\circ \uu f\uuu=\uu f\circ g\uuu. 
\]
\end{lemma}
\bpf
This follows by unfolding the definition of $g^*$, then using naturality 
of $\lambda_{A^*}$, $\lambda_\II=\rho_\II$, and finally Lemma \ref{lm:CUT}.
\hfill\endproof\newline

In a picture,
\par\vspace{3mm}\par\noindent
\begin{minipage}[b]{1\linewidth}
\centering{\epsfig{figure=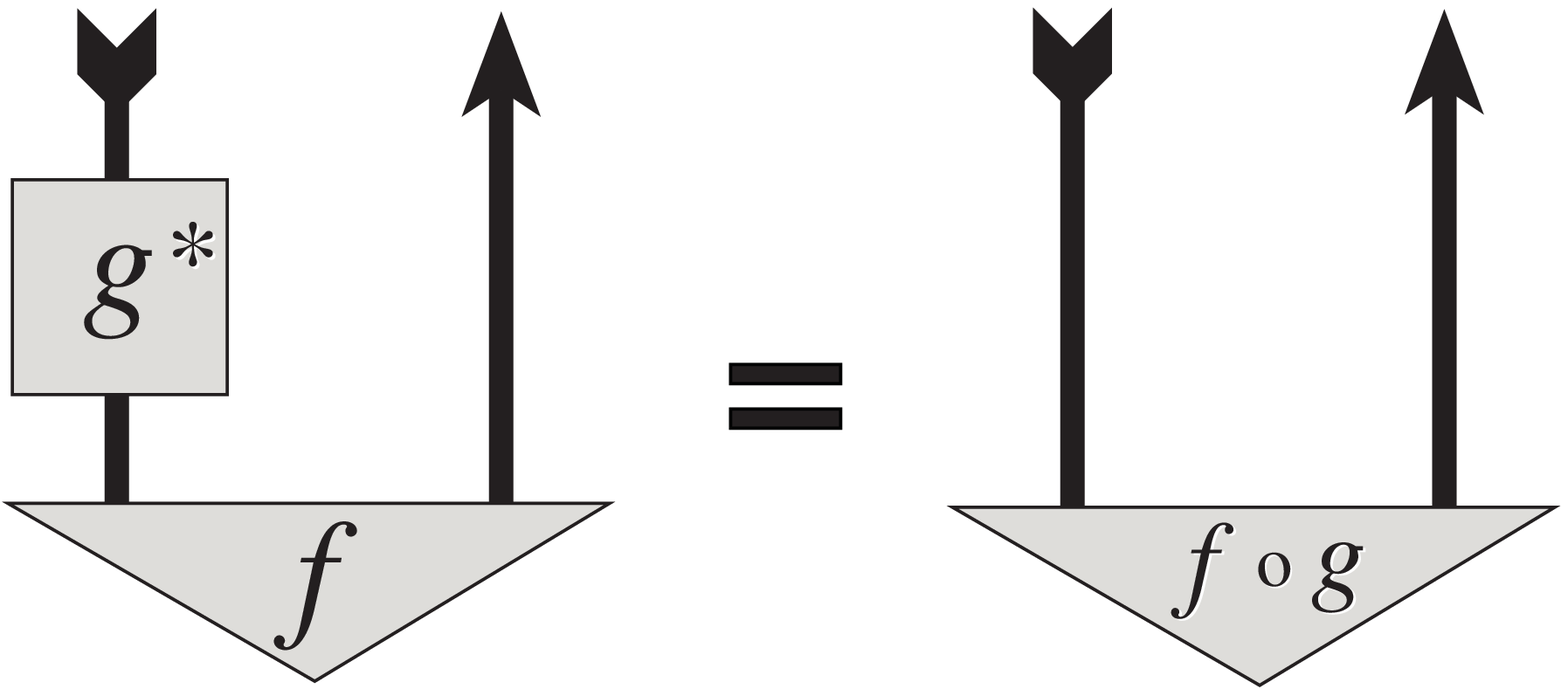,width=145pt}}
\end{minipage}
\par\vspace{3mm}\par\noindent

The obvious analogues of Lemma \ref{lm:precompos} and  \ref{lm:precompos2} for
conames also hold.

\subsection{Quantum information flow in entanglement networks}\label{sec:ABSETNNETW}

We claim that Lemmas \ref{lm:precompos}, \ref{lm:compos} and \ref{lm:CUT} capture the quantum
information flow in the (logic-gate) teleportation\index{quantum teleportation} and entanglement swapping protocols.  
We shall provide a full interpretation of finitary quantum
mechanics in Section \ref{sec:absquantprot} but for now the following rule
suffices:
\bit
\item We interpret \emph{preparation} of an entangled state as a \emph{name}  
and an \emph{observational branch} as a \emph{coname}. 
\eit
For an entanglement network of teleportation-type shape, applying Lemma \ref{lm:compos} yields
\[
U\circ\left(\lambda^{-1}_C\circ (\dd f\ddd\otimes
1)\right)\circ\left((1\otimes\uu g\uuu)\circ\rho_A\right)
=U\circ g\circ f\,.
\]

In a picture, 
\par\vspace{3mm}\par\noindent
\begin{minipage}[b]{1\linewidth}
\centering{\epsfig{figure=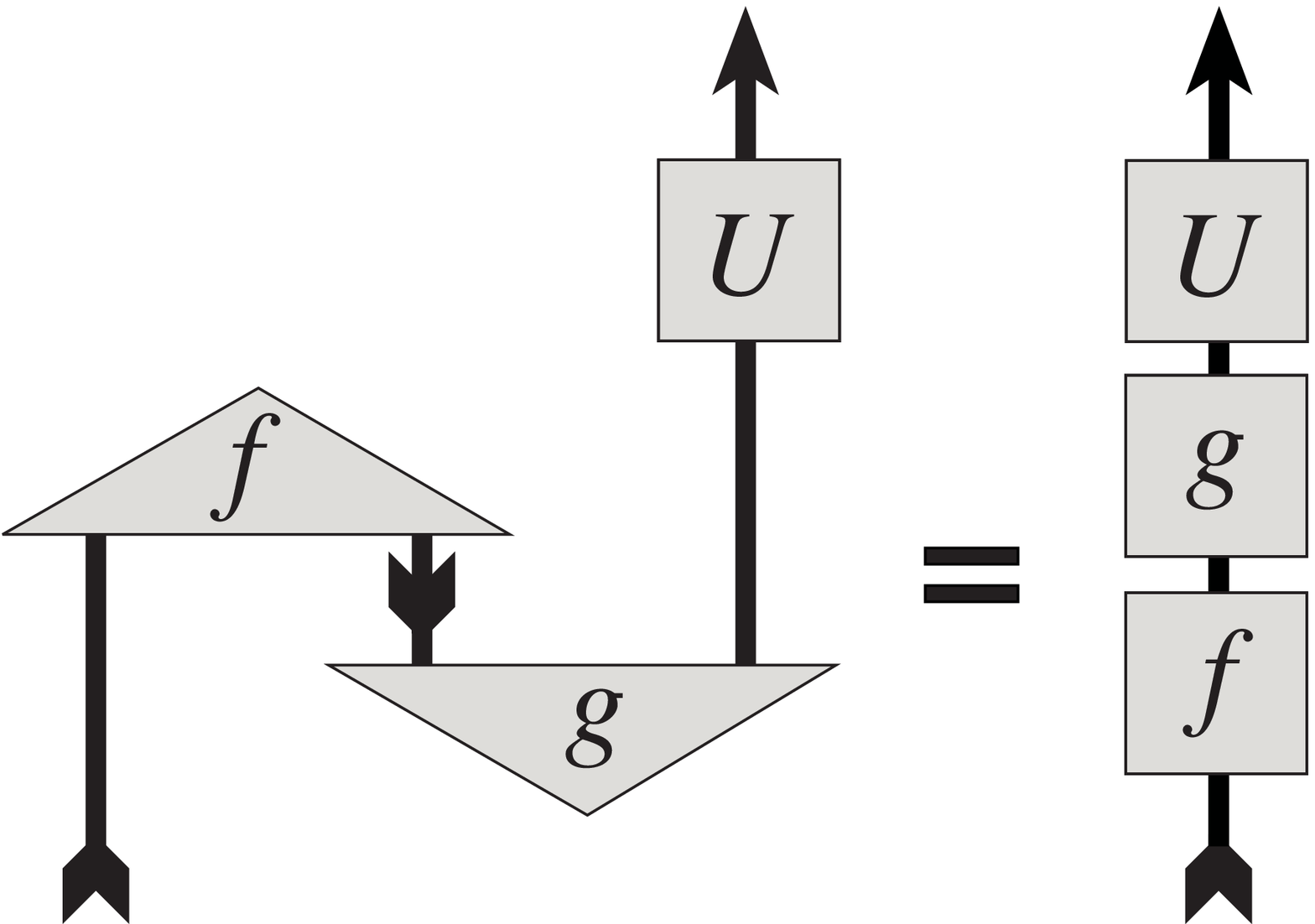,width=145pt}}
\end{minipage}
\par\vspace{3mm}\par\noindent

\noindent We make the information flow more explicit in the following version of the same picture:
\par\vspace{3mm}\par\noindent
\begin{minipage}[b]{1\linewidth}
\centering{\epsfig{figure=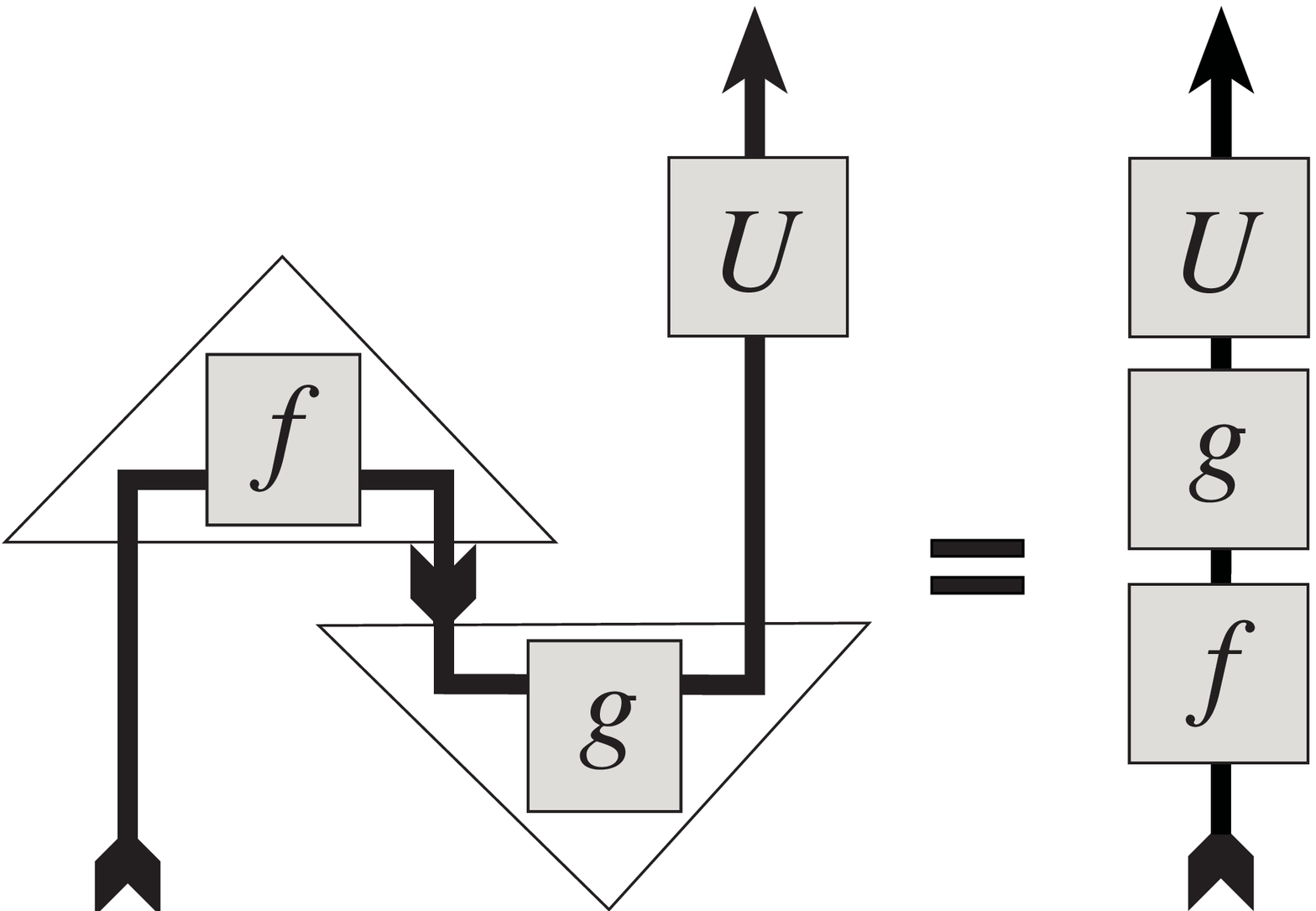,width=146pt}}
\end{minipage}
\par\vspace{3mm}\par\noindent
\noindent
Note that the quantum information seems to flow `following the line' while being
acted on by the functions whose name or coname labels the boxes
(and this fact remains valid for much more complex networks
\cite{Coe1}).

Teleporting the input requires
$U\circ g\circ f=1_A$ --- we assume all functions have type $A\to A$.
Logic-gate teleportation of $h:A\to B$ requires $U\circ g\circ f=h$.

We 
calculate this explicitly in {\bf Rel}\index{category of relations}. For initial state
$x\in X$ after preparing 
${\uu S\uuu\subseteq\{*\}\times(Y\times Z)}$
we obtain 
$\{x\}\times \{(y,z)\mid *\,\uu S\uuu(y,z)\}$
as the state
of the system. For observational branch
$\dd R\ddd\subseteq (X\times Y)\times \{*\}$
we have that $z\in Z$ is the output iff $\dd R\ddd\times 1_Z$
receives  an input $(x,y,z)\in X\times Y\times Z$
such that $(x,y)\dd R\ddd\,*\,$.
Since 
\[
*\,\uu S\uuu(y,z)\Leftrightarrow ySz\qquad {\rm and}\qquad (x,y)\dd
R\ddd\,*\Leftrightarrow xRy
\]
we indeed obtain $x(R;S)z$.
This illustrates that the compositionality
is due to a mechanism of imposing constraints between the components
of the tuples.

In ${\bf FdVec}_\mathbb{C}$   
the vector space of all linear maps of type $V\to W$ is $V\multimap W$ and
hence by 
${V^*\otimes W\simeq V\multimap W}$
we have a bijective
correspondence between linear maps
${f:V\to W}$ and vectors $\Psi\in V^*\otimes W$ (see also
\cite{Coe1,Coe2}):
\[
\Psi_f={1\over\sqrt{2}}\cdot\uu f\uuu(1)\quad\quad{\rm and}\quad\quad\dd
f\ddd=\langle\sqrt{2}\cdot\Psi_f|-\rangle\,.
\]
In particular we have for the Bell base:
\[
b_i={1\over\sqrt{2}}\cdot\uu \beta_i\uuu(1)\quad\quad{\rm and}\quad\quad\dd
\beta_i\ddd=\langle \sqrt{2}\cdot b_i|-\rangle\,.
\]  
Setting $g:=\beta_1=1_V$, $f:=\beta_i$ and $U:=\beta_i^{-1}$
indeed yields 
$\beta_i^{-1}\circ 1_A\circ\beta_i=1_A$, 
which expresses the correctness
of the teleportation protocol along each branch.

Setting $g:=h$ and $f:=\beta_i$ for logic-gate teleportation requires
$U_i$ to satisfy
$U_i\circ h\circ \beta_i=h$ that is 
${h\circ\beta_i=U^\dagger\circ h}$
(since $U$ has to be unitary).
Hence we have derived the laws of logic-gate teleportation --- one should
compare this calculation to the size of the calculation in
Hilbert space.

Deriving the swapping protocol using
Lemma \ref{lm:precompos} and Lemma \ref{lm:CUT} proceeds analogously
to the derivation of the teleportation protocol. 
\par\vspace{3mm}\par\noindent
\begin{minipage}[b]{1\linewidth}
\centering{\epsfig{figure=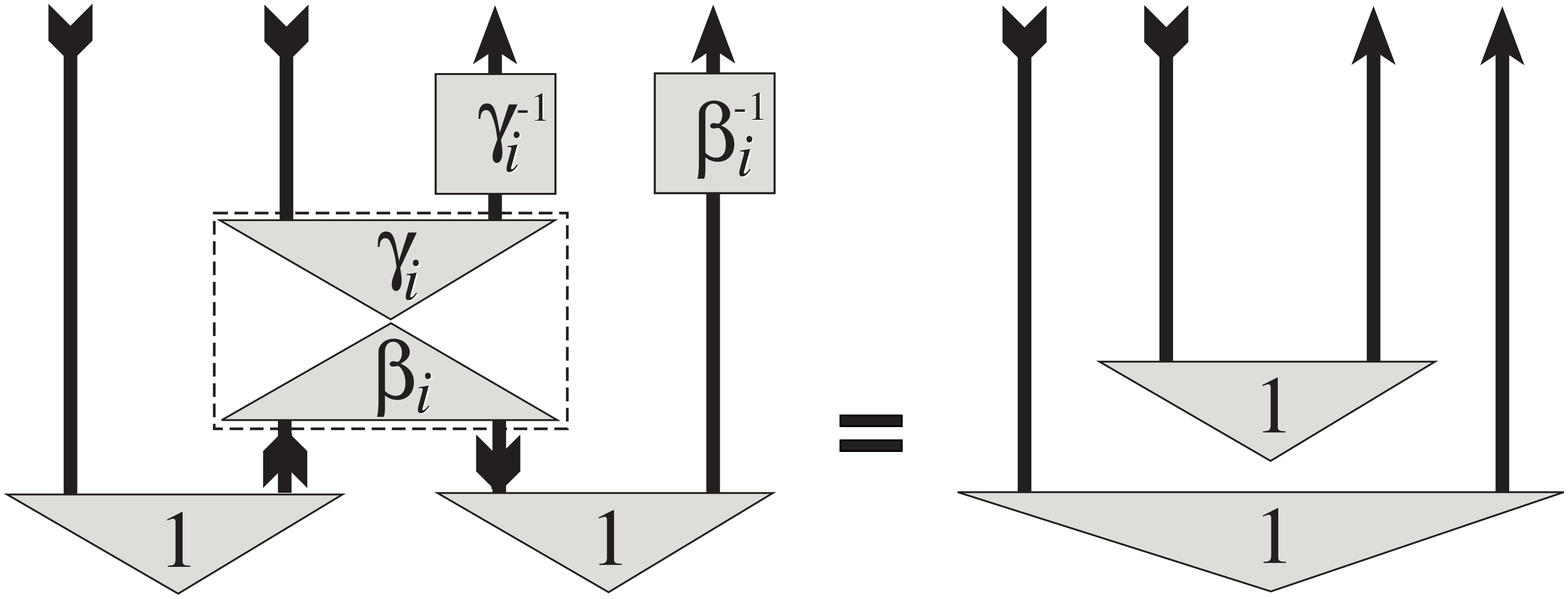,width=258pt}}
\end{minipage}
\par\vspace{3mm}\par\noindent
-- the two triangles within the dashed line stand for $\uu\gamma_i\uuu\circ\dd\beta_i\ddd$.
We obtain two distinct flows due to the fact that a non-destructive
measurement is involved.
\par\vspace{3mm}\par\noindent
\begin{minipage}[b]{1\linewidth}
\centering{\epsfig{figure=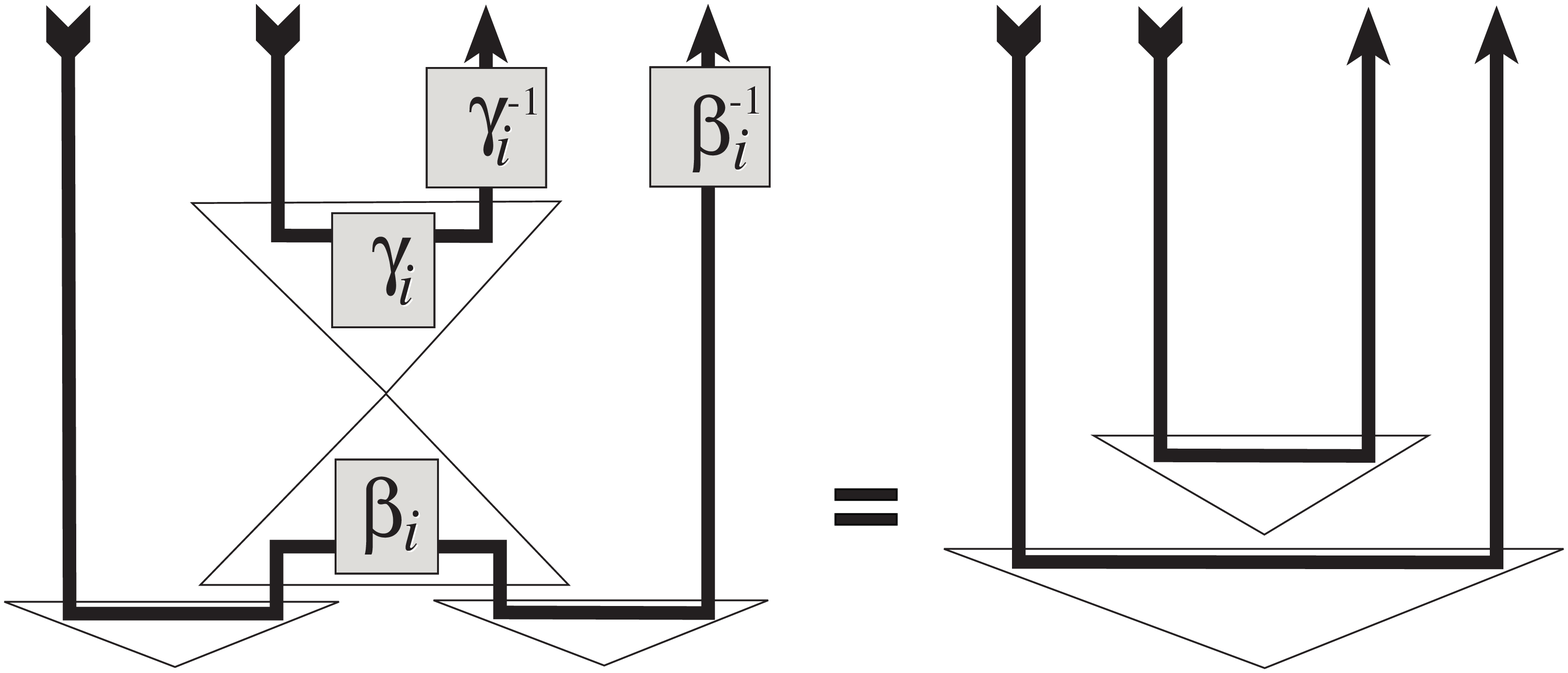,width=258pt}}
\end{minipage}
\par\vspace{3mm}\par\noindent
How $\gamma_i$ has to relate to $\beta_i$ such that they make up a true projector
will be discussed in Section \ref{sec:absquantprot}.

For a general entanglement
network of the swapping-type (without unitary correction and
observational branching) by Lemma
\ref{lm:CUT} we obtain the following `reduction':\vspace{-0mm}

\par\vspace{3mm}\par\noindent
\begin{minipage}[b]{1\linewidth}
\centering{\epsfig{figure=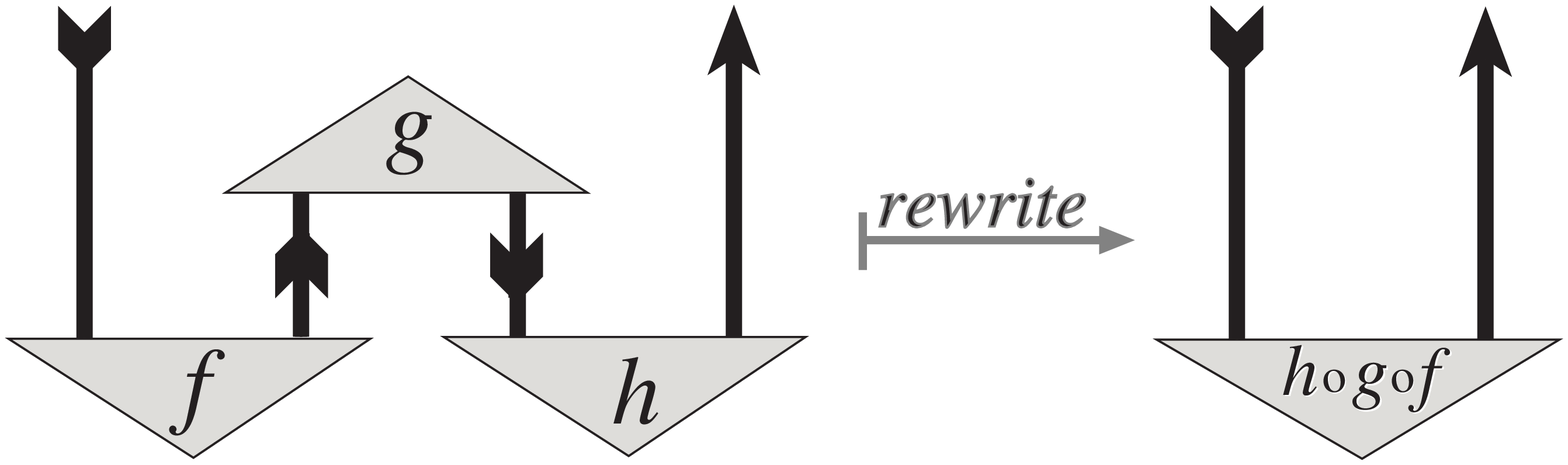,width=258pt}}
\end{minipage}
\par\vspace{3mm}\par\noindent

\vspace{-3mm}\noindent
This picture, and the underlying algebraic property expressed by Lemma 
3.5, is in fact directly related to \emph{Cut-Elimination} in the
logic corresponding to compact-closed categories. If one turns the
above picture upside-down, and interprets names as Axiom-links and
conames as Cut-links, then one has a normalization rule for
proof-nets. This  perspective is 
developed in \cite{AD}.

\section{Strongly Compact Closed Categories and 2-Dimensional Dirac Notation}
\paragraph{The key example} In Section \ref{sec:CCC} we analysed the
compact closed structure of ${\bf FdVec}_{\mathbb{K}}$, where we took
the dual of a vector space $V$ to be the vector space of its linear
functionals ${V^*}$.  In the case that $V$ is
equipped with an inner product we can refine this analysis. We discuss this for the key example of
\emph{finite-dimensional Hilbert spaces}, \ie finite-dimensional complex vector spaces with a \emph{sesquilinear} inner product: the inner product is linear in the second argument, and
\[ \langle \phi \mid \psi \rangle = \overline{ \langle \psi \mid \phi \rangle}  \]
which implies that it is \emph{conjugate-linear} rather than linear in its first argument.

We organize these spaces into a category
 $\FdHilb$, where the morphisms are linear maps. Note that we do \emph{not} require morphisms to
preserve the inner product.

This category provides the basic setting for finite-dimensional quantum mechanics and for quantum information and computation.\footnote{Much of quantum information is concerned with \emph{completely positive maps} acting on \emph{density matrices}. An account of this extended setting in terms of a general categorical construction within our framework is discussed in Section~7.}
 
 In the setting of Hilbert spaces, we can replace the dual space by a more elementary construction.
 In a Hilbert space, each linear functional 
$\bar{\psi}:{\cal H}\to\mathbb{C}$ is witnessed by some $\psi\in{\cal H}$
such that $\bar{\psi} = {\langle \psi\mid\cdot \ \rangle}$. This
however does \emph{not} induce an isomorphism between ${\cal H}$ and
${\cal H}^*$, due to the conjugate-linearity of the inner product in its first
argument. 
This leads us to introduce the \emph{conjugate space} $\bar{\HH}$ of a Hilbert space $\HH$: this has the same additive group of vectors as $\HH$, while the scalar multiplication and inner product are ``twisted'' by complex conjugation:
\[ \alpha \sdot_{\bar{\HH}} \phi := \bar{\alpha} \sdot_{\HH} \phi \qquad 
\langle \phi \mid \psi \rangle_{\bar{\HH}} := \langle \psi \mid \phi
\rangle_{\HH} \]
We can define $\HH^* = \bar{\HH}$, since $\HH$ and $\bar{\HH}$ have the same orthornormal bases, and we can define the counit by
\[ \epsilon_{\HH} : \HH \otimes \bar{\HH} \rightarrow \mathbb{C} :: \phi \otimes \psi\mapsto \langle \psi \mid \phi\rangle_{\HH}
\]
which is indeed (bi)linear rather than sesquilinear! Note that 
\[ \bar{\bar{\HH}} = \HH, \qquad \overline{A \otimes B} = \bar{A} \otimes \bar{B} \, . \]

\subsection{Why compact closure does not suffice}
Note that the categories $\FdHilb$ and $\FdVect_{\mathbb{C}}$ are equivalent! This immediately suggests that some additional categorical structure must be identified to reflect the r\^ole of the inner product.

A further reason for seeking additional categorical structure is to reflect the centrally important notion of \emph{adjoint} in Hilbert spaces:
\[ 
\begin{diagram}[loose,height=.8em,width=0pt]
& A && \rTo^{f} && B \\
\ & && \hLine && & \\
& A && \lTo^{f^{\dagger}} && B \\
\end{diagram} 
\qquad \qquad \langle f \phi \mid \psi \rangle_B = \langle \phi \mid f^{\dagger} \psi \rangle_A \]
This is \emph{not} the same as the dual --- the types are different!
In ``degenerate'' CCC's in which $A^* = A$, e.g. $\mathbf{Rel}$ or real inner-product spaces, we have $f^* = f^\dagger$.
In Hilbert spaces, the  isomorphism $A \simeq A^*$ is not linear, but \emph{conjugate linear}:
\[ \langle \lambda \sdot \phi \mid {-} \rangle \;\; = \;\; \bar{\lambda} \sdot \langle \phi \mid {-} \rangle 
\]
and hence does not live in the category $\mathbf{Hilb}$ at all!

\subsection{Solution: Strong Compact Closure}

A key observation is this: the assignment $\HH \mapsto \HH^*$ on objects has a \emph{covariant} functorial extension $f \mapsto f_*$, which is essentially identity on morphisms; and then we can \emph{define}
\[ f^{\dagger} = (f^*)_* = (f_*)^* . \]
Concretely, in terms of matrices $()^*$ is \emph{transpose}, $()_{*}$ is \emph{complex conjugation}, and the adjoint is the \emph{conjugate transpose}. Each of these three operations can be expressed in terms of the other two. For example, $f^{*} = (f^{\dagger})_{*}$.
All three of these operations are important in articulating the foundational structure of quantum mechanics. All three can be presented at the abstract level as functors, as we shall now show.

\subsection{Axiomatization of Strong Compact Closure}

We shall adopt the most  concise and elegant axiomatization of strongly compact closed categories\index{strong compact closure}, which takes the adjoint as primitive, following \cite{AC2}.

It is convenient to build the definition up in several stages, as in \cite{Selinger}.

\begin{definition}
A \emph{dagger category} is a category $\CC$ equipped with an identity-on-objects, contravariant, strictly involutive functor $f\mapsto f^\dagger$:
\[ 1^{\dagger} = 1, \qquad (g \circ f)^{\dagger} = f^{\dagger} \circ g^{\dagger}, \qquad f^{\dagger\dagger} = f \, . \]
We define an arrow $f : A \rarr B$ in a dagger category to be \emph{unitary} if it is an isomorphism such that $f^{-1} = f^{\dagger}$. An endomorphism $f : A \rarr A$ is \emph{self-adjoint} if $f = f^{\dagger}$.
\end{definition}

\begin{definition}
A \emph{dagger symmetric monoidal category} $(\CC , \otimes , \II, \lambda, \rho, \alpha, \sigma, {\dagger} )$
combines dagger and symmetric monoidal structure, with the requirement that the natural isomorphisms 
$\lambda$, $\rho$, $\alpha$, $\sigma$ are componentwise unitary, and moreover that $\dagger$ is a strong monoidal functor:
\[ (f \otimes g)^{\dagger} = f^{\dagger} \otimes g^{\dagger} \, .
\]
\end{definition}

Finally we come to the main definition.
\begin{definition}
A \emph{strongly compact closed category} is a dagger symmetric monoidal category which is compact closed, and such that the following diagram commutes:
\[ \begin{diagram}
\II & \rTo^{\eta_{A}} & A^{*} \otimes A \\
& \rdTo_{\epsilon_{A}^{\dagger}} & \dTo_{\sigma_{A^{*}, A}} \\
& & A \otimes A^{*}
\end{diagram}
\]
\end{definition}
\noindent This implies that  the counit is \emph{definable} from the unit and the adjoint: 
\[ \epsilon_{A} =  \eta_{A}^{\dagger} \circ \sigma_{A, A^{*}} \]
and similarly the unit can be defined from the counit and the adjoint.
Furthermore, it is in fact possible to replace the two commuting diagrams required in the definition of compact closure by one. We refer to \cite{AC3} for the details.

\begin{definition}
In any strongly compact closed category $\CC$, we can define a covariant monoidal functor
\[ A \mapsto A^{*}, \qquad f : A \rarr B \mapsto f_{*} = (f^{\dagger})^{*} : A^{*} \rarr B^{*} \, . \]
\end{definition}

\paragraph{Examples} Our central example is of course $\FdHilb$.
Any compact closed category such as 
$\mathbf{Rel}$\index{category of relations}, in which $(\ )^*$ is
the identity on objects, is trivially strongly compact closed (we just 
take $f^{\dagger} = f^{*}$). Note that in this case $f_{*} = f^{**} = f$.
Thus in \textbf{Rel} the adjoint  is relational converse. The
category of finite-dimensional  real inner product spaces and linear
maps, with  $A =
A^*$, offers another example of this situation.  
A construction of free strongly compact closed categories over dagger categories is given in \cite{Abr1}.

\paragraph{Scalars}
Self-adjoint scalars $s = s^{\dagger}$ in strongly compact closed categories are of special interest. In the case of $\FdHilb$, these are the \emph{positive reals} $\mathbb{R}^+$. The passage from $s$ to 
$ss^{\dagger}$, which  is self-adjoint, will track the passage in quantum mechanics from \emph{amplitudes} to \emph{probabilities}.

\subsection{Inner Products and Dirac Notation}
With the adjoint available, it is straightforward to interpret Dirac notation\index{Dirac notation} --- the indispensable everyday notation of quantum mechanics and quantum information. A \emph{ket} is simply an arrow $\psi : \II \rarr A$, which we can write as $|\psi \rangle$ for emphasis. We  think of kets as \emph{states}, of a given type of system $A$.  The corresponding \emph{bra} will then be $\psi^{\dagger} : A \rarr \II$, which we can think of as a \emph{costate}.

\paragraph{Example}
In $\FdHilb$, a linear map $f : \mathbb{C} \rarr \HH$ can be identified with the vector $f(1) = \psi  \in \HH$: by linearity, all other values of $f$ are determined by $\psi$. Even better, we can identify $f$ with its image, which is the  \emph{ray} or one-dimensional subspace of $\HH$ spanned by $\psi$ --- the proper notion of (pure) state of a quantum system.

\begin{definition}
Given $\psi,\phi : \II \to A$ we define  their \em
abstract inner product \em $\langle\psi\mid\phi\rangle$  as 
\[ \psi^{\dagger} \circ \phi : \II \lrarr \II \, . \]
\end{definition}
\noindent Note that this is a scalar, as it should be.
In $\FdHilb$, this definition coincides with the usual inner product.
In {\bf Rel}, for $x,y\subseteq\{*\}\times X$:
\[
\langle x\mid y\rangle= 1_{\II}, \ \ x \cap y \neq \varnothing
\qquad{\rm and}\qquad
\langle x\mid y\rangle= 0_{\II}, \ \ x \cap y = \varnothing .
\]

\noindent We now show that two of the basic properties of adjoints in $\FdHilb$ hold in generality in the abstract setting.
\begin{proposition}\label{pr:adjIn}
For
$\psi:\II\to A$, $\phi:\II\to B$ and $f:B\to A$
we have 
\[
\langle f^\dagger\circ\psi\mid\phi\rangle_B=\langle \psi\mid f\circ\phi\rangle_A\,.
\]
\end{proposition}
\bpf
$\langle
f^\dagger\!\circ\psi\mid\phi\rangle=(f^\dagger\!\circ\psi)^\dagger\!\circ\phi=
\psi^\dagger\!\circ f\circ\phi=\langle \psi\mid 
f\circ\phi\rangle$. 
\hfill\endproof

\begin{proposition}\label{pr:UnIn}
Unitary morphisms $U:A\to B$ preserve the inner product, that is  
for all $\psi,\phi:\II\to A$ we have 
\[
\langle U\circ\psi\mid U\circ\phi\rangle_B=
\langle \psi\mid \phi\rangle_A\,.
\]
\end{proposition}
\bpf
By Proposition \ref{pr:adjIn},
$\langle U\circ\psi\mid U\circ\phi\rangle_B=
\langle U^\dagger\!\circ U\circ\psi\mid \phi\rangle_A=
\langle \psi\mid \phi\rangle_A$.\hfill\endproof

Finally, we show how the inner product can be defined in terms of the `complex conjugate' functor $()_{*}$.
\begin{proposition}
For $\psi,\phi : \II \to A$ we have:
\[ \langle \psi\mid \phi\rangle_A = \begin{diagram}
\ \II & \rTo^{\rho_{\II}} & \II \otimes \II& \rTo^{1_\II \!\otimes\! u_{\II}} & \II
\otimes
\II^*\! & \rTo^{\phi \!\otimes\! \psi_*} & A \otimes A^*\! & \rTo^{\epsilon_A} & \II. 
\end{diagram}
\]
\end{proposition}
\bpf
Since $u_\II= \rho^{-1}_{\II^*}\circ \eta_\II$ by naturality of $\rho$ we have
\[
\eta_\II=\rho_{\II^*}\circ \rho^{-1}_{\II^*}\circ \eta_\II=\rho_{\II^*}\circ u_\II=(u_\II\otimes 1_\II)\circ\rho_\II
\]
where we use $\rho^{-1}=\rho^\dagger$ and similarly we obtain $\epsilon_\II=\rho^\dagger_\II\circ(1_\II\otimes u_\II^\dagger)$. Hence by $1_\II=u_\II^\dagger\circ u_\II$ and the analogues to Lemmas \ref{lm:precompos} and  \ref{lm:precompos2} for the counit we obtain
\beqa
\psi^\dagger\circ\phi
=\rho^\dagger_\II\circ( (\psi^\dagger\circ\phi)\otimes 1_\II)\circ\rho_\II\!\!&=&\!\!\epsilon_\II\circ(\psi^\dagger\otimes 1_{\II^*})\circ(\phi\otimes 1_{\II^*})\circ\epsilon_\II^\dagger\\
\!\!&=&\!\!\dd\psi^\dagger\ddd\circ(\phi\otimes 1_{\II^*})\circ\epsilon_\II^\dagger\\
\!\!&=&\!\!\epsilon_\II\circ(1_\II\otimes\psi_*)\circ(\phi\otimes 1_{\II^*})\circ\epsilon_\II^\dagger
\eeqa
which is equal to $\epsilon_\II\circ(\phi\otimes \psi_*)\circ(1_\II\otimes u_\II)\circ\rho_\II$.
\hfill\endproof\newline

\subsection{Dissection of the bipartite projector}
Projectors are a basic building block in the von Neumann-style  foundations of quantum mechanics, and in standard approaches to quantum logic\index{quantum logic}. It is a notable feature of our approach that we are able, at the abstract level of strongly compact closed categories, to delineate a \emph{fine-structure} of bipartite projectors, which can be applied directly to the analysis of information flow in quantum protocols.

We define a \emph{projector} on an object $A$ in a strongly compact closed category to be an arrow $\PP : A \rarr A$ which is idempotent and self-adjoint:
\[ \PP^{2} = \PP, \qquad \PP = \PP^{\dagger} \, . \]

\begin{proposition}
Suppose we have a state $\psi : \II \rarr A$ which is \emph{normalized}, meaning $\langle \psi \mid \psi \rangle = 1_{\II}$. 
Then the  `ket-bra' $|\psi\rangle\langle \psi| = \psi \circ \psi^{\dagger} : A \rarr A$ is a projector.
\end{proposition}
\bpf
Self-adjointness is clear. For idempotence:
\[ \psi \circ \psi^{\dagger} \circ \psi \circ \psi^{\dagger} = \langle \psi \mid \psi \rangle \sdot  \psi \circ \psi^{\dagger} = \psi \circ \psi^{\dagger} \, . \]
\hfill\endproof\newline

\noindent We now want to apply this idea in a more refined form to a state $\psi : \II \rarr A^{*} \otimes B$ of a compound system. Note that, by Map-State duality:
\[ \CC(\II, A^{*} \otimes B) \equiv \CC(A, B) \]
any such state $\psi$ corresponds biuniquely to the name of a map $f : A \rarr B$, \ie $\psi = \uu f \uuu$. This arrow witnesses an information flow from $A$ to $B$, and we will use this to expose the information flow inherent in the corresponding projector. 

Explicitly, we define
\[
\PP_f:=\uu f\uuu\circ(\uu
f\uuu)^\dagger=\uu f\uuu\circ\dd f_*\ddd:A^*\otimes B\to A^*\otimes B\,,
\]
that is, we have an assignment
\[
\PP_{\_}:\CC(\II,A^*\otimes B)\longrightarrow \CC(A^*\otimes B,A^*\otimes
B)::\Psi\mapsto
\Psi\circ \Psi^\dagger
\]
from bipartite elements to bipartite projectors. Note that the strong compact closed structure is essential in order to define  $\PP_f$ as an 
endomorphism.

We can \em normalize \em these projectors $\PP_f$ by considering
$s_f\sdot
\PP_f$ for
$s_f:=(\dd f_*\ddd\circ\uu f\uuu)^{-1}$ (provided this inverse exists
in $\CC (\II ,
\II )$), yielding
\[
(s_f\sdot \PP_f)\circ(s_f\sdot \PP_f)=s_f\sdot(\uu f\uuu\circ
(s_f\sdot(\dd
f_*\ddd\circ\uu f\uuu))\circ\dd f_*\ddd) =s_f\sdot \PP_f\,,
\]
and also
\[
(s_f\sdot \PP_f)\circ\uu f\uuu=\uu f\uuu\qquad{\rm and}\qquad
\dd f_*\ddd\circ(s_f\sdot \PP_f)=\dd f_*\ddd\,.
\]

A  picture corresponding to this decomposed bipartite projector is:
\par\vspace{3mm}\par\noindent
\begin{minipage}[b]{1\linewidth}
\centering{\epsfig{figure=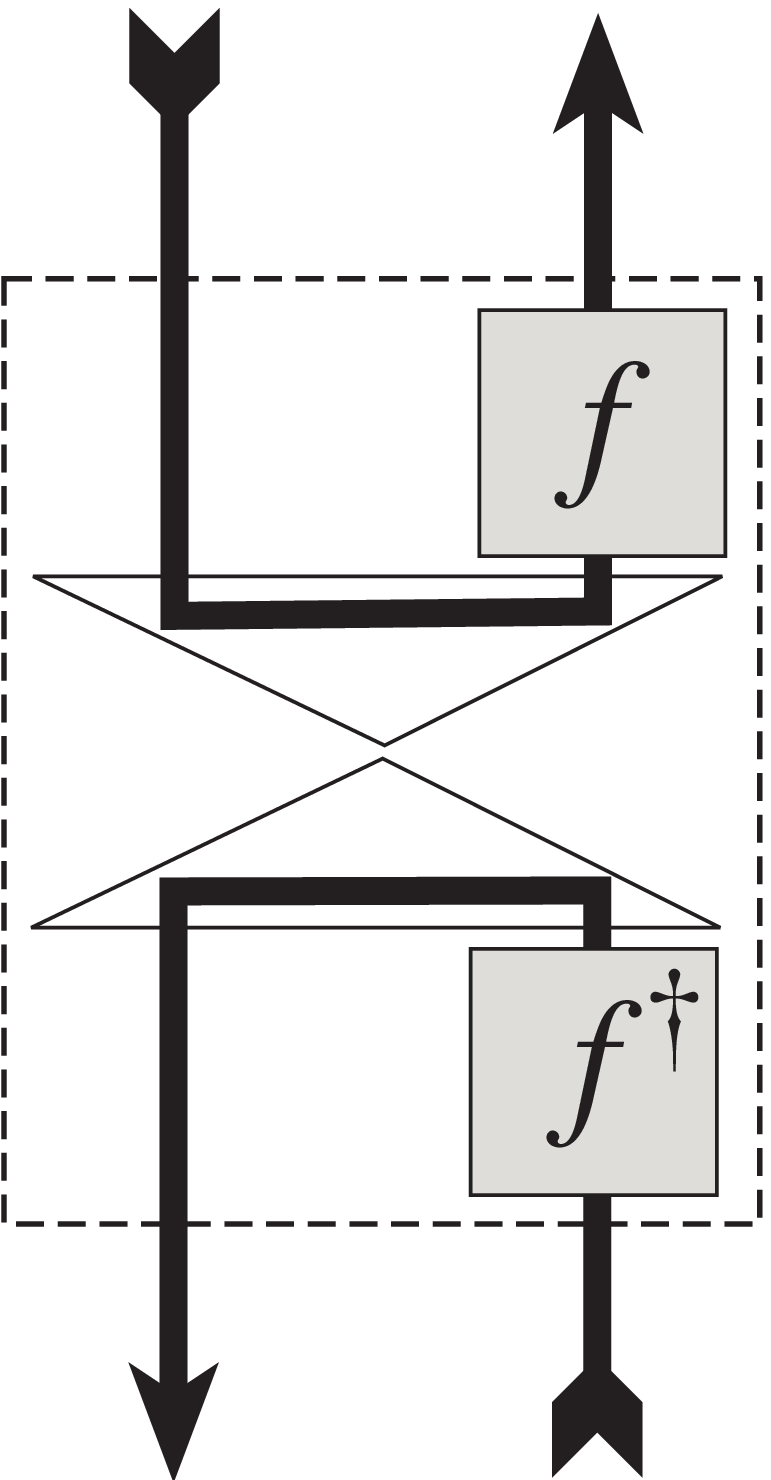,width=62pt}}
\end{minipage}
\par\vspace{3mm}\par\noindent

\subsection{Trace} 

Another essential mathematical instrument in quantum mechanics is the \emph{trace} of a linear map. In quantum information, extensive use is made of the more general notion of \emph{partial trace}, which is used to trace out a subsystem of a compound system. 

A general categorical axiomatization of the notion of partial trace has been given by Joyal, Street and Verity \cite{JSV}. A trace in a symmetric monoidal category $\CC$ is a family of functions 
\[ \Tr_{A,B}^U : \CC (A \otimes U, B \otimes U) \longrightarrow \CC (A, B) \]
for objects $A$, $B$, $U$ of $\CC$, satisfying a number of axioms, for which we refer to \cite{JSV}.
This specializes to yield the total trace for endomorphisms by taking $A = B = \II$. In this case, $\Tr(f) = \Tr_{\II,\II} ^{U}(f): \II \rarr \II$ is a scalar. Expected properties such as the invariance of the trace under cyclic permutations
\[ \Tr(g \circ f) = \Tr(f \circ g) \]
follow from the general axioms.

Any compact closed category carries a canonical (in fact, a unique) trace. The definition can be given slightly more elegantly in the strongly compact closed case. For an endomorphism $f : A \rarr A$, the total trace is defined by
\[ \Tr(f) = \epsilon_{A} \circ (f \otimes 1_{A^{*}}) \circ \epsilon_{A}^{\dagger} \, . \]
More generally, if $f : A \otimes C \rarr B \otimes C$, $\Tr_{A,B}^{C}(f) : A \rarr B$ is defined to be:
\[ \begin{diagram}
A & \rTo^{\rho_{A}} & A\! \otimes\! \II & \rTo^{1\! \otimes\! \epsilon_{C}^{\dagger}}
& A \!\otimes\! C \!\otimes\! C^{*}
& \rTo^{f \!\otimes\! 1_{C^{*}}} & B \!\otimes\! C \!\otimes\! C^{*} & \rTo^{1 \!\otimes\! \epsilon_{C}} & B \!\otimes\! \II &
\rTo ^{\rho_{B}^{-1}} & B \, .
\end{diagram}
\]
These definitions give rise to the standard notions of trace and partial trace in $\FdHilb$.

\section{Biproducts, Branching and Measurements}
\label{sec:biprods} 

As we have seen, many of the basic ingredients of quantum mechanics are present in strongly compact closed categories. What is lacking is the ability to express the probabilistic branching arising from measurements, and the information flows from quantum to classical and back.
We shall find this final piece of expressive power in a rather standard piece of additional categorical structure, namely \emph{biproducts}\index{biproducts}.

\subsection{Biproducts}
Biproducts have been studied as part of the structure of Abelian
categories. For further details, and proofs of the general results
we shall cite in this sub-section, see e.g. \cite{Mitchell,MacLane}.

Recall that  a \emph{zero object}  in a category is one which is both 
initial and terminal. If $\Zero$ is a zero object, there is an 
arrow
\begin{diagram}
\hspace{-8mm}0_{A,B}:A&\rTo&\Zero &\rTo &B
\end{diagram}
between any pair of objects $A$ and $B$. Let $\CC$ be a category with
a zero object and binary products and coproducts.
Any arrow 
\begin{diagram}
\hspace{-8mm}A_1\coprod A_2 &\rTo^f A_1 &\prod A_{2}\hspace{-8mm} 
\end{diagram}
with injections
$q_i:A_i\to A_1 \coprod A_{2}$ and projections $p_j:A_1\prod A_2\to A_j$
can be written uniquely
as a matrix 
\[ \left( \begin{array}{cc}
f_{11} & f_{21} \\
f_{12} & f_{22}
\end{array} \right) \]
 where $f_{ij}:=p_j\circ f\circ q_i : A_{i} \rightarrow A_j$.  If the arrow 
\[ \left( \begin{array}{cc}
1 & 0 \\
0 & 1
\end{array} \right) \]
is an isomorphism for all $A_1$, $A_2$, then we say that $\CC$ has
\emph{biproducts}, and write $A \oplus B$ for the biproduct of $A$ and 
$B$. 

\begin{proposition}[Semi-additivity]
If $\CC$ has biproducts, then we can define an operation of addition
on each hom-set $\CC (A, B)$ by
\begin{diagram} 
A&\rTo^{f+g}&B\\
\dTo^\Delta&&\uTo_\nabla\\
A\oplus A&\rTo_{f\oplus g}&B\oplus B
\end{diagram}
for  $f,g:A\to B$, where 
$\Delta=\langle 1_A,1_A\rangle$ and $\nabla=[1_B,1_B]$
are respectively the diagonal and 
codiagonal.
This operation is associative and commutative, with $0_{AB}$ as an
identity. Moreover, composition is bilinear with respect to this
additive structure. Thus $\CC$ is enriched over abelian monoids.
\end{proposition}
 
 Because of this automatic enrichment of categories with biproducts over abelian monoids, we say that such a category is \emph{semi-additive}.
 
\begin{proposition}
\label{biprodprop}
If $\CC$ has biproducts, we can choose projections $p_1$, $\ldots$, $p_n$ and
injections $q_1$, $\ldots$, $q_n$ for each $\bigoplus_{k=1}^{k=n}A_k$ satisfying\vspace{-3mm}
\[ p_j \circ q_i = \delta_{ij} \qquad{\rm and}\qquad \sum_{k=1}^{k=n}q_k \circ p_k=
1_{\bigoplus_{k}\!A_k} \] 
where $\delta_{ii} = 1_{A_i}$, and $\delta_{ij} = 0_{A_{i}, A_{j}}$, $i \neq j$.
\end{proposition}

\subsection{Strongly compact closed categories with biproducts}
We now come to the full mathematical structure we shall use as a setting for finitary quantum mechanics: namely \emph{strongly compact closed categories with biproducts}.

A first point is that, because of the strongly self-dual nature of compact closed categories, weaker assumptions suffice in order to guarantee the presence of biproducts. The following elegant result is due to Robin Houston \shortcite{Houston}, and was in fact directly motivated by \cite{AC2}, the precursor to the present article.

\begin{theorem}
Let $\CC$ be a monoidal category with finite products and coproducts,
and suppose that for every object $A$ of $\CC$, the functor $A \otimes {-}$  preserves products and the functor ${-} \otimes A$ preserves coproducts. Then C has finite biproducts.
\end{theorem}
Because a compact closed category is closed and self-dual, the existence of products implies that of coproducts, and vice versa, and the functor ${-} \otimes A$ is a left adjoint and hence preserves coproducts. Moreover, since $A^{*} \linimpl B \simeq  A^{**} \otimes B \simeq A \otimes B$, the functor $A \otimes {-}$ is a right adjoint and preserves products.
Hence this result specializes to the following:
\begin{proposition}
If $\CC$ is a compact closed category with either products or coproducts, then it has biproducts, and hence is semiadditive.
\end{proposition}

\paragraph{Examples} 
There are many examples of compact closed categories with biproducts:
 the category of relations for a regular category with
stable disjoint coproducts; the category of finitely generated projective modules
over a commutative ring; the category of finitely generated free semimodules over a
commutative semiring; and the category of free semimodules over a
complete commutative semiring are all semi-additive  compact closed categories.
Examples have also arisen in a Computer Science context in the first
author's work on Interaction Categories \cite{Abramsky}.
Compact closed categories with biproducts with additional
assumptions, in particular that the category is abelian, have been studied in the
mathematical literature on \emph{Tannakian categories} \cite{Deligne}. 

In the case of strongly compact closed categories, we need a coherence condition between the dagger and the biproduct structure. We say that a category is \emph{strongly compact closed with biproducts} if we can choose biproduct structures $p_{i}$, $q_{i}$ as in Proposition~\ref{biprodprop} such that $p_{i}^{\dagger} = q_{i}$ for $i=1,\ldots , n$.

\begin{proposition}
If $\CC$ is strongly compact closed with biproducts, then
\[
\sum_{k=1}^{k=n}p_k^\dagger \circ p_k=\sum_{k=1}^{k=n}q_k \circ q_k^\dagger= 
1_{\bigoplus_{k}\!A_k}\,.
\]
Moreover, there are natural isomorphisms
\[
\nu_{A,B}:(A\oplus B)^*\simeq A^*\oplus B^*\qquad{\rm and}\qquad
\nu_\II:\Zero^*\simeq\Zero\,,
\]
and $(\ )^{\dagger}$ preserves biproducts and hence is additive:
\[ 
(f \oplus g)^{\dagger} = f^{\dagger} \oplus g^{\dagger}\ ,\qquad(f + g)^{\dagger} = f^{\dagger} +
g^{\dagger}
\qquad{\rm and}\qquad 0^{\dagger}_{A,B} = 0_{B,A}\,. 
\]
\end{proposition}

\paragraph{Examples} Examples of semi-additive strongly compact closed
categories are the category $({\bf Rel},\times,+)$, where the biproduct is the
disjoint union\index{category of relations}, and the category
$(\FdHilb,\otimes,\oplus)$, where the biproduct is the
direct sum.  

\paragraph{Distributivity and classical information flow}
As we have already seen, in a compact closed category with biproducts, tensor distributes over the biproduct. This abstract-seeming observation in fact plays a crucial r\^ole in the representation of \emph{classical information flow}.
To understand this, consider a quantum system $A \otimes B$, composed from subsystems A(lice) and B(ob). Now suppose that Alice performs a local measurement, which we will represent as resolving her part of the system into say $A_{1} \oplus A_{2}$. Here the biproduct is used to represent the different \emph{branches} of the measurement. At this point, by the functorial properties of $\oplus$, Alice can perform actions $f_{1} \oplus f_{2}$, which \emph{depend on which branch of the measurement has been taken}. The global state of the system is $(A_{1} \oplus A_{2}) \otimes B$, and as things stand Bob has no access to this measurement outcome. However, under distributivity we have
\[ (A_{1} \oplus A_{2}) \otimes B \simeq (A_{1} \otimes B) \oplus (A_{2} \otimes B) \]
which corresponds to propagating the classical information as to the measurement outcome `outwards', so that it is now accessible to Bob, who can perform an action depending on this outcome, of the form $1_{A} \otimes (g_{1} \oplus g_{2})$.

We shall record distributivity in an explicit form for future use.
\begin{proposition}[Distributivity of $\otimes$ over $\oplus$]\label{distributivity}
In any monoidal closed category there is a right distributivity natural isomorphism
$\tau_{A,B,C}:A\otimes(B\oplus C)\simeq (A\otimes B)\oplus(A\otimes C)$,
which is explicitly defined as
\[
\tau_{A,\cdot,\cdot}:=\langle 1_A\otimes p_1,1_A\otimes  
p_2\rangle\qquad{\rm and}\qquad\tau_{A,\cdot,\cdot}^{-1}:
=[ 1_A\otimes q_1,1_A\otimes q_2]\,.
\]
A left distributivity isomorphism 
$\upsilon_{A, B, C}: (A \oplus B) \otimes C
\simeq (A \otimes C) \oplus (A \otimes C)$ can be defined similarly.
\end{proposition}

\paragraph{Semiring of scalars.} In a strongly compact closed category with  biproducts, 
the scalars  form a commutative semiring. Moreover, scalar multiplication satisfies the usual additive properties 
\[ (s_{1} + s_{2}) \sdot f = s_{1} \sdot f + s_{2} \sdot f, \qquad 0\sdot f = 0 \]
as well as the multiplicative ones. For Hilbert spaces,  this commutative semiring is the field of complex numbers.
In $\mathbf{Rel}$ the commutative semiring of 
scalars is the Boolean semiring $\{0, 1\}$, with disjunction as sum.

\paragraph{Matrix calculus.} 
We can write any arrow of the form
$f:A\oplus B\to C\oplus D$ as a matrix
\[
M_f:=
\left(
\begin{array}{cc}
p_1^{C,D}\!\!\circ f\circ q_1^{A,B} & p_1^{C,D}\circ f\circ q_2^{A,B}\\
p_2^{C,D}\circ f\circ q_1^{A,B} & p_2^{C,D}\circ f\circ q_2^{A,B}\\
\end{array}
\right).
\]
The sum $f+g$ of such morphisms corresponds to the matrix sum $M_f+M_g$ and composition $g\circ f$ corresponds to matrix
multiplication $M_g\cdot M_f$. Hence  categories with biproducts admit a matrix
calculus. 

\subsection{Spectral Decompositions}

We define a \emph{spectral decomposition} of an object $A$ to be a unitary isomorphism 
\[
U : A \to \bigoplus_{i=1}^{i=n} A_i\,.
\]
(Here the `spectrum' is just the set of indices $1, \ldots , n$).
Given a spectral decomposition $U$, we define
morphisms 
\[
\neoiota_j := U^{\dagger} \!\circ q_{j}: A_{j} \to A\qquad{\rm and}\qquad
\pi_j :=  \psi_j^{\dagger} = p_j \circ U : A \to A_{j}\,,
\]
diagramatically  
\begin{diagram}
A_j & \rTo^{\neoiota_j} & A \\
\dTo^{q_j}& \ldTo_{U} & \dTo_{\pi_j}\\
\ \ \bigoplus_{i=1}^{i=n} A_i& \rTo_{p_j} & A_j
\end{diagram}
and finally \emph{projectors} 
\[
{\rm P}_j := \neoiota_j
\circ \pi_j : A \to A\,.
\] 
These projectors are \emph{self-adjoint}
\[ \PP_j^{\dagger} = (\neoiota_j \circ \pi_j )^{\dagger} =
\pi_j^{\dagger} \circ \neoiota_j^{\dagger} = \neoiota_j \circ \pi_j = \PP_j
\]
\emph{idempotent} and \emph{orthogonal}
\[
{\rm P}_i\circ {\rm P}_j=\neoiota_i\circ \pi_i\circ \neoiota_j
\circ \pi_j=\neoiota_i\circ \delta_{ij}\circ \pi_j=  
\delta_{ij}^A\circ{\rm P}_i .
\]
Moreover, they yield a \emph{resolution of the identity}:
\beqa
\sum_{i=1}^{i=n} \PP_i &= & \sum_{i=1}^{i=n} \neoiota_i \circ\pi_i=\sum_{i=1}^{i=n} U^{\dagger} \circ q_i \circ p_i \circ U \\ 
&= & U^{\dagger} \circ (\sum_{i=1}^{i=n} q_i \circ p_i )
\circ U
= U^{-1} \circ 1_{\bigoplus_{i}\! A_i} \circ
U = 1_{A} \, .
\eeqa

\subsection{Bases and dimension}

Writing $n\cdot X$ for types of the shape $\bigoplus_{i=1}^{i=n}X$ it follows by
self-duality of the tensor unit
$\II$ that
\[ 
\nu^{-1}_{{\rm I},\ldots,{\rm
I}}\circ\left(n\cdot{u}_{\rm
I}\right)\ :\,n\cdot{\rm I}\simeq\left(n\cdot{\rm
I}\right)^{*}.
\]
A \em basis \em for an object $A$ is a unitary isomorphism 
\[
{\sf base}:n\cdot\II\to A\,.
\]
Given bases ${\sf base}_A$ and ${\sf base}_B$ for objects $A$ and $B$ respectively we can define the matrix $(m_{ij})$ of any
morphism
$f:A\to B$ in those two bases as the matrix of 
\[
{\sf base}_B^\dagger\circ f\circ{\sf base}_A:n_A\cdot\II\to n_B\cdot\II \, .
\]

\begin{proposition}
Given $f:A\to B$,
${\sf base}_A:n_A\cdot\II\to A$ and ${\sf base}_B:n_B\cdot\II\to A$
the matrix $(m'{\!\!}_{ij})$ of $f^\dagger$ in these bases is the conjugate transpose of the matrix
$(m_{ij})$ of
$f$.
\end{proposition}
\bpf
$m'{\!\!}_{ij}
=p_i\circ {\sf base}_A^{\dagger} \circ f^\dagger \circ{\sf base}_B \circ q_j
=(p_j\circ{\sf base}_B^\dagger\circ f\circ{\sf base}_A\circ q_i)^\dagger
= m_{ji}^\dagger$.
\hfill\endproof\newline

\noindent If in addition to the assumptions of Proposition \ref{pr:adjIn} and Proposition
\ref{pr:UnIn} there exist bases for $A$ and $B$, we can prove converses to both
of them.

\begin{proposition}\label{pro:inprod2}
If there exist bases for $A$ and $B$
then $f:A\to B$ is the adjoint to $g:B\to A$ if and only if
\[
\langle f\circ\psi\mid \phi\rangle_B=\langle \psi\mid g\circ\phi\rangle_A
\]
for all $\psi:\II\to A$ and $\phi:\II\to B$.
\end{proposition}
\bpf
Let $(m_{ij})$ be the matrix of $f^\dagger$ and $(m'{\!\!}_{ij})$ the matrix of $g$ in the given bases.
\[ \begin{array}{lcl}
m_{ij}
& = & p_i\circ{\sf base}_A^\dagger\circ\! f^\dagger\circ {\sf base}_B\circ q_j \\
& = & \langle f\circ {\sf base}_A\!\circ q_i\mid {\sf base}_B\!\circ q_j\rangle_B \\
& = & {\langle f\circ\psi\mid \phi\rangle_B} \\
& = & {\langle \psi\mid g\circ\phi\rangle_A} \\
& = & \langle  {\sf base}_A\!\circ q_i\mid g\!\circ{\sf base}_B\!\circ q_j\rangle_A \\
& = & p_i\circ{\sf base}_A^\dagger\circ\! g\circ {\sf base}_B\circ q_j \\
& = & m'{\!\!}_{ij} \, .
\end{array}
\]
Hence the matrix elements of $g$ and $f^\dagger$ coincide so $g$ and $f^\dagger$ are equal.
The converse is Proposition \ref{pr:adjIn}.
\hfill\endproof

\begin{proposition}\label{pr:UnIn2}
If there exist bases for $A$ and $B$ then a morphism $U:A\to B$ is unitary
if and only if it preserves the inner product, that is 
for all $\psi,\phi:\II\to A$ we have 
\[
\langle U\circ\psi\mid U\circ\phi\rangle_B=
\langle \psi\mid \phi\rangle_A\,.
\]
\end{proposition}
\bpf
We have 
$\langle U^{-1}\!\!\circ\psi\mid \phi\rangle_A=
\langle U\circ U^{-1}\!\!\circ\psi\mid U\circ \phi\rangle_B=
\langle \psi\mid U\circ \phi\rangle_B$
and hence by Proposition \ref{pro:inprod2}, $U^\dagger=U^{-1}$. The converse is
given by Proposition \ref{pr:UnIn}.
\hfill\endproof\newline

\noindent Note also that when a basis is available we can assign to $\psi^\dagger:A\to \II$ and $\phi:\II\to A$
matrices
\[
\left(
\begin{array}{ccc}
\psi_1^\dagger&
\cdots&
\psi_n^\dagger
\end{array}
\right)
\qquad{\rm and}\qquad
\left(
\begin{array}{c}
\phi_1\\
\vdots\\
\phi_n
\end{array}
\right)
\]
respectively, and  the
inner product becomes
\[
\langle \psi \mid \phi \rangle =\left(
\begin{array}{ccc}
\psi_1^\dagger&
\cdots&
\psi_n^\dagger
\end{array}
\right)
\left(
\begin{array}{c}
\phi_1\\
\vdots\\
\phi_n
\end{array}
\right)
=\sum_{i=1}^{i=n}\psi_i^\dagger\circ\phi_i\,. 
\]

\paragraph{Dimension}
Interestingly, two different notions of dimension arise in our setting.  We
assign an \em integer dimension \em ${\sf dim}(A)\in\mathbb{N}$ to an object
$A$ provided there exists a base 
\[
{\sf base}:{\sf dim}(A)\cdot\II\to A\,.
\] 
%and if furthermore any base of $A$ has this form.  
Alternatively, we introduce the \em scalar dimension \em as
\[
{\sf dim}_s(A):= \Tr(1_{A}) = \epsilon_A\circ \epsilon_{A}^{\dagger} \in{\bf C}(\II,\II) .
\]
We also have:
\[
{\sf dim}_s(\II)=1_\II
\qquad\quad
{\sf dim}_s(A^*)={\sf dim}_s(A)
\quad\qquad{\sf dim}_s(A\otimes B)={\sf dim}_s(A){\sf dim}_s(B)
\]
In ${\bf FdVec}_\mathbb{K}$ these notions of dimension coincide, in the sense that ${\sf dim}_s(V)$ is multiplication with the
scalar
${\sf dim}(V)$. In ${\bf Rel}$ the integer dimension corresponds to the cardinality of the set, and is only well-defined for finite sets, while ${\sf
dim}_s(X)$ always exists;  however,  ${\sf
dim}_s(X)$ can only take two values, $0_\II$ and $1_\II$, 
and the two notions of dimension diverge for sets of cardinality greater than 1. 

\subsection{Towards a representation theorem}
As the results in this section have shown, under the assumption of biproducts we can replicate many of the familiar linear-algebraic calculations in Hilbert spaces. One may wonder how far we really are from Hilbert spaces.

The deep results by Deligne \shortcite{Deligne} and Doplicher-Roberts \shortcite{DR} on Tannakian categories, the latter directly motivated by algebraic quantum field theory,  show that under additional assumptions, in particular  that the category is abelian as well as compact closed, we  obtain a representation into finite-dimensional modules over the ring of scalars. One would like to see a similar result in the case of strongly compact closed categories with biproducts, with the conclusion being a representation into inner-product spaces.

\section{Abstract Quantum Mechanics: Axiomatics and Quantum Protocols}
\label{sec:absquantprot}

We can identify the basic ingredients of finitary quantum mechanics in any semi-additive
strongly compact closed category.   
\bit
\item[{\bf 1.}] A \emph{state space} is represented by an object $A$.
\item[{\bf 2.}] A \em basic variable \em (`type of qubits') is a state space $Q$ with a 
given unitary isomorphism 
\[
{\sf base}_Q:{\rm I}\oplus{\rm I}\to Q
\]
which we call the
\em computational basis \em of $Q$. By using the isomorphism $n \cdot \II \simeq
(n\cdot\II )^*$ described in Section~5, we also obtain a computational basis for
$Q^*$.
\item[{\bf 3.}] A \emph{compound system} for which the subsystems are  described by
$A$ and $B$ respectively is
described by $A\otimes B$. If we have computational bases ${\sf base}_A$
and ${\sf base}_B$, then we define\vspace{-1mm} 
\[ 
{\sf base}_{A\otimes B}:=({\sf base}_A\otimes{\sf base}_B)\circ d_{nm}^{-1}
\] 
where 
\[
d_{nm} : n \cdot \II \otimes m \cdot \II \simeq (nm) \cdot \II
\]
is the canonical isomorphism constructed using first the left distributivity
isomorphism $\upsilon$, and then the right distributivity isomorphism
$\tau$, to give the usual lexicographically-ordered computational
basis for the tensor product\index{tensor product}.
\item[{\bf 4.}] Basic data transformations are unitary isomorphisms.
\item[{\bf 5a.}] A \emph{preparation} in a state space $A$ is a morphism 
$\neoiota:{\rm I}\to A$ for which there exists a unitary $U:\II\oplus B \to A$ such that 
\begin{diagram}
{\rm I} & \rTo^{\neoiota} & A \\
\dTo^{q_1}& \ruTo_{U} & \\
{\rm I}\oplus B& 
\end{diagram}
commutes.
\item[{\bf 5b.}] Consider a spectral decomposition 
\[
U : A \to \bigoplus_{i=1}^{i=n} A_i
\]
with associated
projectors $\PP_j$. This gives rise to the \emph{non-destructive measurement}
\[ \langle\PP_i \rangle_{i=1}^{i=n}:A\to n\cdot A .
\]
The projectors 
$\PP_i:A\to A$
for $i=1, \ldots , n$ are called the \emph{measurement branches}.
This measurement is \emph{non-degenerate} if $A_i = \II$ for all $i = 1,
\ldots , n$. In this case we refer to $U$ itself as a \emph{destructive measurement} or  \emph{observation}. The  morphisms 
$\pi_i = p_i \circ
U :A\to\II$
for $i=1, \ldots , n$ are called \emph{observation branches}.
\eit
Note that the type of a non-destructive measurement makes it explicit that it is an operation which
involves  a non-deterministic transition (by contrast with the standard Hilbert space quantum
mechanical formalism). 
\bit
\item[{\bf 6a.}]  Explicit biproducts represent the \emph{branching} arising from the indeterminacy
of measurement outcomes. 
\eit
Hence an operation $f$ acting on an explicit biproduct $A\oplus B$ should itself be an explicit
biproduct, \textit{i.e.}~we want 
\[
f=f_1\oplus f_2:A\oplus B\to C\oplus D\,,
\]
 for $f_1:A\to C$ and $f_2:B\to D$.
The dependency of $f_i$ 
on the branch it is in captures \emph{local} classical communication.
The full force of non-local classical communication is enabled by Proposition \ref{distributivity}.
\bit
\item[{\bf 6b.}]  Distributivity isomorphisms represent \em non-local classical communication\em.
\eit
To see this, suppose e.g. that we have a compound system $Q \otimes A$, and we
(non-destructively) measure the qubit in the first component, obtaining a new system state 
described by
$(Q \oplus Q) \otimes A$. At this point, we know `locally',
\textit{i.e.} at the site of the first component, what the measurement 
outcome is, but we have not propagated this information to the rest
of the system $A$. However, after applying the distributivity
isomorphism
\[ (Q \oplus Q) \otimes A \simeq (Q \otimes A) \oplus (Q \otimes 
A) \]
the information about the outcome of the measurement on the first
qubit has been propagated globally throughout the system, and we can
perform operations on $A$ depending on the measurement outcome, e.g.
$(1_Q \otimes U_0 )  \oplus (1_Q \otimes U_1 )$
where $U_0$, $U_1$ are the  operations we wish to perform on $A$ in
the event that the outcome of the measurement we performed on $Q$ was
0 or 1 respectively.

\subsection{The Born rule} 

We now show how the \emph{Born rule}, which is the key quantitative feature of
quantum mechanics, emerges automatically from our abstract setting.  

For a preparation $\neoiota : \II \to A$ and 
spectral decomposition 
$U : A \to \bigoplus_{i=1}^{i=n} A_i$, 
with
corresponding non-destructive measurement 
\[
\langle\PP_i \rangle_{i=1}^{i=n}:A\to n\cdot A\,,
\]
we can consider the protocol
\begin{diagram}
\II & \rTo^{\neoiota} & A & \rTo^{\langle\PP_i \rangle_{i=1}^{i=n}} & n \cdot
A\,.
\end{diagram}
We define scalars
\[ 
{\sf Prob}({\rm P}_i,\neoiota) := \langle \neoiota \mid \PP_i\mid \neoiota
\rangle =  
\neoiota^{\dagger} \circ \PP_i \circ \neoiota\,. 
\]

\begin{proposition}
With notation as above, 
\[
{\sf Prob}({\rm P}_i,\neoiota) = ({\sf Prob}({\rm
P}_i,\neoiota))^{\dagger}
\]
and
\[
\sum_{i=1}^{i=n}{\sf Prob}({\rm P}_i,\neoiota) = 1\,.
\]
Hence we think of the scalar ${\sf Prob}({\rm P}_j,\neoiota)$ as 
`the probability of obtaining  the $j$'th outcome of the measurement
$\langle\PP_i
\rangle_{i=1}^{i=n}$ on the state $\neoiota$'. 
\end{proposition}

\bpf 
From the definitions of preparation and the projectors, there are
unitaries $U$, $V$ such that 
\[
{\sf Prob}({\rm P}_i,\neoiota) = (V \circ q_{1})^{\dagger} \circ U^\dagger
\circ q_i \circ p_i \circ U
\circ V \circ q_1\]  for each $i$. Hence
\beqa
\sum_{i=1}^{i=n} {\sf Prob}({\rm P}_i,\neoiota)
&=&\sum_{i=1}^{i=n}
p_1
\circ V^\dagger \circ U^\dagger
\circ q_i \circ p_i \circ U \circ V \circ q_1 \\  
&= &p_1 \circ 
V^\dagger
\!\circ U^\dagger \!\circ\! \Bigl(\sum_{i=1}^n \!q_i
\circ  p_i \Bigr)\! \circ
 U \circ V \circ q_1 \\
&= & p_1 \circ  V^{-1}\! \circ U^{-1}\! \circ 1_{n\cdot\II} \circ U 
\circ V
\circ q_1= p_1 \circ q_1 = 1_{\II}\,.
\eeqa 
\hfill\endproof\newline 

\noindent Moreover, since by definition
${\rm
P}_j=\pi_j^\dagger\circ\pi_j$,
we can rewrite the Born rule expression as 
\[ 
{\sf Prob}({\rm P}_j,\neoiota)
=\neoiota^{\dagger} \circ \PP_j\circ \neoiota
=\neoiota^{\dagger} \circ \pi_j^\dagger\circ\pi_j\circ \neoiota
=(\pi_j\circ\neoiota)^\dagger\circ\pi_j\circ \neoiota
=s_j^\dagger\circ s_j
\]
for some scalar $s_j\in{\bf C}(\II,\II)$.  Thus $s_j$ can be thought of as the `probability amplitude' giving rise to the probability $s_j^\dagger\circ s_j$, which is of course self-adjoint. If we consider the protocol 
\begin{diagram}
\II & \rTo^{\neoiota} & A & \rTo^{\langle \pi_i \rangle_{i=1}^{i=n}} & n
\cdot \II\,.
\end{diagram}
which involves an observation $\langle \pi_i \rangle_{i = 1}^{i =n}$, then
these scalars $s_j$ correspond to the branches 
\begin{diagram}
\II & \rTo^{\neoiota} & A & \rTo^{\pi_j } & \II\,.
\end{diagram}

We now turn to the description of the quantum protocols previously discussed in Section~\ref{qmtelepsec} within our framework. In each case, we shall give a complete description of the protocol, including the quantum-to-classical information flows arising from measurements, and the subsequent classical-to-quantum flows corresponding to the classical communications and the actions depending on these performed as steps in the protocols. We shall in each case verify the correctness of the protocol, by proving that a certain diagram commutes. Thus these case studies provide evidence for  the expressiveness and effectiveness of the framework.

Our general axiomatic development allows for considerable generality.
The standard von Neumann axiomatization fits
Quantum Mechanics perfectly, with no room to spare. Our basic setting
of strongly compact closed categories with biproducts is general enough to
allow very different models such as $\mathbf{Rel}$, the category of
sets and relations. 
When we consider specific protocols such as teleportation, a kind of 
`Reverse Arithmetic' (by analogy with Reverse Mathematics \cite{RM})
arises. That is, we can characterize what requirements are placed on
the semiring of scalars  ${\mathbf C} ({\rm I},{\rm I})$ (where ${\rm I}$ is the
tensor unit) in order for the protocol to be realized. This is often
much less than requiring that this be the field of complex numbers,
but in the specific cases which we shall consider, the requirements are
sufficient to exclude $\mathbf{Rel}$.

\subsection{Quantum teleportation} 

\begin{definition}\em
A \em teleportation base \em is a scalar $s$ together with a  
morphism 
\[
{\sf prebase}_{\rm T} : 4\cdot{\rm I}\to Q^*\otimes Q
\]
such that:
\begin{itemize}
\item ${\sf base}_{\rm T} := s \sdot {\sf prebase}_{\rm T}\,$ is unitary.
\item the four maps $\beta_j:Q\to Q$, where $\beta_j$ is defined by 
$\uu\beta_j\uuu:={\sf prebase}_{\rm
T}\circ q_j$,
are unitary. 
\item $2 s^{\dagger} s = 1$.
\end{itemize}
The morphisms $s \sdot \uu\beta_j\uuu$ are the \em base 
vectors \em of the teleportation base.   A teleportation base is a  
\em Bell base \em when
the \em 
Bell base  maps \em 
$\beta_1,\beta_2,\beta_3,\beta_4:Q\to Q$
satisfy\footnote{This choice of
axioms is sufficient for our purposes.  One might prefer to axiomatize a
notion of Bell base such that the
corresponding Bell base maps are exactly the Pauli matrices --- note
that this would introduce a
coefficient $i$ in $\beta_4$.}
\[
\beta_1=1_Q\qquad \beta_2=\sigma_{Q}^{\oplus}\qquad
\beta_3=\beta^\dagger_3\qquad
\beta_4=\sigma_{Q}^{\oplus}\circ\beta_3
\]
where 
\[
\sigma_{Q}^{\oplus}:={\sf base}_Q^{}\circ\sigma^\oplus_{\II,\II}\circ{\sf
base}_Q^{-1}\,.
\]
A teleportation base\index{quantum teleportation} defines a \em teleportation observation \em
\[
\langle s^{\dagger} \sdot 
\dd
\beta_i\ddd\rangle_{i=1}^{i=4}:Q\otimes Q^*\to 4\cdot{\rm
I}\,.
\]
\end{definition}

To emphasize the
identity of the individual qubits we label the three copies of $Q$ we
shall consider as $Q_a$, $Q_b$, $Q_c$. We also use labelled
identities,
e.g.~${1_{bc} : Q_b \rightarrow Q_c}$, and labelled Bell bases.
Finally, we introduce 
\[
\Delta_{ac}^4:=\langle s^{\dagger}s \sdot 1_{ac}\rangle_{i=1}^{i=4}:
Q_a\to 4\cdot Q_c
\]
as the \emph{labelled,
weighted diagonal}. This expresses the intended behaviour of teleportation, namely that the
input qubit is propagated to the output along each branch of the
protocol, with `weight' $s^{\dagger}s$,  corresponding to the probability
amplitude for that branch. Note that the sum of the corresponding
probabilities is 
\[
4(s^{\dagger}s)^{\dagger}s^{\dagger}s = (2s^{\dagger} s)(2 s^{\dagger} s) =
1\,.
\]

\begin{theorem}\label{thm:teleport}
The following diagram commutes.
\begin{diagram}
Q_a&\rIs&Q_a\\ 
&&\dTo^{\rho_a}&\hspace{-1.5cm}{\bf import\ unknown\ state}\\
&&Q_a\otimes{\rm I}\\
&&\dTo^{1_a\otimes (s \sdot \uu 1_{bc}\!\!\uuu)}&\hspace{-1.5cm}{\bf produce\ EPR\mbox{\bf
-}pair}\\ &&Q_a\otimes(Q_b^*\!\otimes Q_c)\\
&&\dTo^{\alpha_{a,b,c}}&\hspace{-1.5cm}{\bf spatial\ delocation}\\
\dTo^{\Delta^4_{ac}}&&(Q_a\otimes Q^*_b)\otimes Q_c\\
&&\dTo^{\quad\quad\langle s^{\dagger} \sdot \dd
\beta_i^{ab}\!\!\ddd\rangle_{i=1}^{i=4}\otimes\! 1_c}&\hspace{-1.5cm}{\bf
teleportation\ observation}\\ &&\left(4\cdot{\rm
I}\right)\otimes Q_c\\
&&\dTo^{\left(4\cdot\lambda^{-1}_c\right)\!\circ\upsilon_c}&\hspace{-1.5cm}{\bf
classical\ communication}\\ &&\ \ \ 4\cdot Q_c\\
&&\dTo^{\bigoplus_{i=1}^{i=4}(\beta_i^c)^{-1}}&\hspace{-1.5cm}{\bf unitary\
correction}\\
\ \ \ \ 4\cdot Q_c&\rIs&4\cdot Q_c\!\!\!\!\!\! 
\end{diagram}
The right-hand-side of the above diagram is our formal description of
the teleportation protocol; the commutativity of the diagram expresses 
the correctness of the protocol.
Hence any strongly compact closed category with biproducts admits  
quantum teleportation provided it contains a teleportation base.
If we do a Bell-base observation then the corresponding
unitary corrections are
\[
\beta^{-1}_i\!\!=\beta_i\ \ {\it for}\ \ i\in\{1,2,3\}
\qquad\mbox{\rm\emph{and}}\qquad
\beta_4^{-1}\!\!=\beta_3\circ\sigma_{Q}^{\oplus}\,.
\]
\end{theorem}
\bpf
For each $j\in\{1,2,3,
4\}$ we have a commutative diagram of the form below. The top trapezoid
is the statement of  the Theorem. We ignore  the scalars  -- which cancel out against each
other -- in this proof.
{\scriptsize
\begin{diagram}
Q_a&&&&&&\rTo^{\ \ \langle 1_{ac}\rangle_i\ \
\!\!\!}&&&&&&4\cdot Q_c\\
&\rdTo^{\,\rho_a}&&&&&\hspace{-2.5cm}\mbox{\bf\normalsize
Quantum\
teleportation}\hspace{-2.5cm}&&&&&\!\!\!\!\!\ruTo^{\!\!\!\!\!\!\!\!\bigoplus_i
(\beta_i^c)_i^{-1}}&\\
&&\hspace{-2mm}Q_a\!\otimes\!{\rm
I}&\rTo^{\!\!1_a\!\otimes\!\uu 1_{bc}\!\!\uuu\!}&\!Q_a\!\otimes\!
Q^*_b\!\!\otimes\! Q_c\!&\rTo^{\hspace{-2mm}\!\langle\dd
\beta^{ab\!\!}_i\ddd\rangle_i\!\otimes\!\!
1_c\!\!\hspace{-1mm}}&\left(4\cdot{\rm
I}\right)\otimes Q_c&\rTo^{\hspace{-2mm}\!\!\!\!\langle p_i^{\rm
I}\!\otimes\!\!  1_c\rangle_i\hspace{-3.0mm}}&
4\cdot\left({\rm I}\!\otimes\!
Q_c\right)&\rTo^{\!\!\!\!4\cdot\lambda^{-1}_c\!\!\!\!\!\!\!\!\!\!\!\!}&4\cdot Q_c\hspace{-2mm}&&\\
\dTo_{1_{Q_a}\hspace{-2mm}}&&&&&\hspace{-6.5mm}\rdTo_{\dd \beta^{ab}_j\!\ddd\!\otimes\!
1_c\hspace{-2mm}}&\dTo_{p^{\rm I}_j\otimes\! 1_c}&&\dTo_{p^{{\rm I}\otimes
Q_c}_j\!\!\!\!\!\!\!}&&\dTo_{p_j^{Q_c}}&&\dTo^{\hspace{-2mm}p_j^{Q_c}}\\ &&Q_c&&\pile{\lTo^{\lambda^{-1}_c\!\!\!}\\
\rTo_{\lambda_c}}&&{\rm I}\otimes Q_c&\rTo_{\!\!\!1_{{\rm I}\otimes Q_c}}&{\rm I}\otimes
Q_c&\rTo_{\!\!\!\lambda^{-1}_c\!\!}&Q_c&&\\ &\ruTo_{\!\beta^{ac}_j}&&&&&&&&&&\rdTo_{(\beta_j^c)^{-1}\!}\\
\hspace{-1mm}Q_a&&&&&&\rTo_{1_{ac}}&&&&&&Q_c
\end{diagram} 
}

\noindent
We use the universal property of the
product, naturality of $\lambda$ and the explicit form of
the natural isomorphism ${\upsilon_c:=\langle p_i^{\rm I}\!\otimes\!
1_c\rangle_{i=1}^{i=4}}$.  In the
specific case of a Bell-base observation we use
$1_Q^\dagger=1_Q$, $(\sigma_{Q}^{\oplus})^\dagger=\sigma_{Q}^{\oplus}$
and
$(\sigma_{Q}^{\oplus}\circ\beta_3)^\dagger=
\beta_3^\dagger\circ(\sigma_{Q}^{\oplus})^\dagger=
\beta_3\circ\sigma_{Q}^{\oplus}$.
\hfill\endproof\newline

\noindent Although in {\bf Rel} teleportation works for `individual observational branches' it fails to
admit the full teleportation protocol since there are only two
automorphisms of $Q$ (which is just a two-element set, \ie the type of 
`classical bits'), and hence
there is no teleportation base.

We now consider sufficient conditions on the ambient category $\CC$
for a teleportation base to exist.
We remark firstly that if ${\bf C}({\rm I},{\rm I})$ contains an
additive inverse for $1$, then it is a ring, and moreover all additive 
inverses exist in each hom-set $\CC (A, B)$, so $\CC$ is enriched over 
Abelian groups.
Suppose then that  ${\bf C}({\rm I},{\rm I})$ is a ring with $1\not= -1$.
We can define a morphism 
\[
{\sf prebase}_{\rm T}= {\sf base}_{Q^*\otimes Q} \circ M :4\cdot{\rm I}\to
Q^*\!\otimes Q
\]
where $M$ is
the endomorphism of $4 \cdot \II$ determined by the matrix
\[
\left(
\begin{array}{cccc}
1&0&1&0\\
0&1&0&1\\
0&1&0&\!\!\!\!-\!1\\
1&0&\!\!\!\!-\!1&0
\end{array}
\right)
\]
The corresponding morphisms $\beta_j$ will have $2 \times 2$ matrices determined
by the columns of this $4 \times 4$ matrix, and will be unitary.
If ${\bf C}({\rm I},{\rm I})$ furthermore contains a scalar $s$
satisfying $2s^{\dagger} s = 1$, then $s \sdot {\sf
prebase}_{\rm T}$ is unitary, and the conditions for a teleportation base are fulfilled.
Suppose we start with a ring $R$ containing an element $s$ satisfying
$2s^2 = 1$. (Examples are plentiful, e.g. any  subring of $\mathbb{C}$, or of
$\mathbb{Q}(\sqrt{2})$, containing $\frac{1}{\sqrt{2}}$). The category 
of finitely generated free $R$-modules and $R$-linear maps is strongly 
compact
closed with biproducts, and admits a teleportation base (in which $s$
will appear as a scalar with $s = s^{\dagger}$), hence
realizes teleportation.

\subsection{Logic-gate teleportation}

Logic gate teleportation of qubits requires only a minor modification as
compared to the teleportation protocol.
\begin{theorem}\label{thm:logicgate}
Let unitary morphism $f:Q\to Q$ be such that for each $i\in\{1,2,3,4\}$ a
morphism
$\varphi_i(f):Q\to Q$ satisfying 
$f\circ\beta_i=\varphi_i(f)\circ f$
exists.
The diagram of Theorem \ref{thm:teleport} with the modifications made
below commutes.   
\begin{diagram}
\dDots&&\dDotsto\\
&&Q_a\otimes{\rm I}\\
&&\dTo^{1_a\otimes(s \sdot \uu f\uuu)}&\hspace{-5mm}{\bf produce}\
f{\bf\mbox{\bf -}state}\\  &&Q_a\otimes(Q^*_b\!\otimes
Q_c)\\ 
\dTo^{\Delta^4_{ac}\!\circ\! f}&&\dDotsto\\
&&\ \ \ 4\cdot Q_c\\
&&\dTo^{\bigoplus_{i=1}^{i=4}(\varphi_i(f))^{-1}}&\hspace{-5mm}{\bf unitary\ correction}\\ 
\ \ \ \ 4\cdot Q_c&\rIs&4\cdot Q_c\!\!\!\!\!\! 
\end{diagram}
The right-hand-side of the diagram is our formal description of
logic-gate teleportation of $f:Q\to Q$; the commutativity of the diagram under
the stated conditions expresses the correctness of logic-gate teleportation for qubits.
\end{theorem}
\bpf
The top trapezoid
is the statement of  the Theorem. The $a$,
$b$ and $c$-labels are the same as in the proof of teleportation. For each $j\in\{1,2,3,4\}$ we have a diagram of the form
below. Again we ignore  the scalars in this proof.
{\scriptsize
\begin{diagram}
Q&&&&&&\rTo^{\langle f\rangle_i\!\!\!}&&&&&&4\cdot Q\\
&\rdTo^{\,\rho_Q}&&&&&\hspace{-2.5cm}\mbox{\bf\normalsize
Logic-gate
teleportation}\hspace{-2.5cm}&&&&&\!\!\!\!\!\!\!\!\!
\ruTo^{\!\!\!\!\bigoplus_i\varphi_i(f)^{-1}\!\!}&\\
&&\hspace{-2mm}Q\!\otimes\!{\rm
I}&\rTo^{\!\!\!\!1_Q\!\otimes\!\uu
f\uuu\!\hspace{-1.3mm}}&\!Q\!\otimes\! Q^*\!\!\otimes\!
Q\!&\rTo^{\!\langle\dd
\beta_i\ddd\rangle_i\!\otimes\!\!
1_Q\!\!}&(4\cdot {\rm
I})\!\otimes\!Q&\rTo^{\hspace{-1mm}\!\langle p_i^{\rm
I}\!\otimes\!\! 1_Q\rangle_i\hspace{-1mm}}&
4\cdot \left({\rm I}\!\otimes\!
Q\right)&\rTo^{\hspace{-2mm}4\cdot
\lambda^{-1}_Q\!\hspace{-5mm}}&4\cdot Q\hspace{-2mm}&&\\
\dTo_{1_Q}&&\mbox{\bf\normalsize Lemma
\ref{lm:compos}}\hspace{-1cm}&&&\hspace{-6.5mm}\rdTo_{\dd \beta_j\ddd\!\otimes\! 1_Q\hspace{-2mm}}&\dTo_{p^{\rm I}_j\otimes\!
1_Q}&&\dTo^{p^{{\rm I}\otimes
Q}_j}&&\dTo^{p_j^Q}&&\dTo^{p_j^Q}\\
\hspace{-1mm}Q&\rTo_{\!\!\!\!f\circ \beta_j}&Q&&
\pile{\lTo^{\lambda^{-1}_Q}\\
\rTo_{\lambda_Q}}&&{\rm I}\otimes Q&\rTo_{1_{{\rm I}\otimes Q}}&{\rm
I}\otimes Q&\rTo_{\lambda^{-1}_Q}&Q&&\\ \dTo_Q&
\ruTo_{\!\!\varphi_j(f)\circ\!\! f\!\!}&&&&&&&&&&
\rdTo_{\varphi_j(f)^{-1}\!}\\
Q&&&&&&\rTo_{f}&&&&&&Q
\end{diagram}
}

\hfill\endproof\newline

This two-dimensional case does not yet provide a universal
computational primitive, which requires teleportation of $Q\otimes
Q$-gates \cite{Gottesman}. We present the example of teleportation of a
$\cnot$ gate \cite{Gottesman} (see also \cite{Coe1} Section 3.3).  

Given a Bell base we define a $\cnot$ gate as one
which acts as follows on tensors of the Bell base maps:
\[
\cnot\circ (\sigma_{Q}^{\oplus}\otimes 1_Q)=(\sigma_{Q}^{\oplus}\otimes
\sigma_{Q}^{\oplus})\circ\cnot\qquad
\cnot\circ (1_Q\otimes \sigma_{Q}^{\oplus})=(1_Q\otimes
\sigma_{Q}^{\oplus})\circ\cnot
\]
\[
\cnot\circ (\beta_3\otimes 1_Q)=(\beta_3\otimes 1_Q)\circ\cnot\qquad
\cnot\circ (1_Q\otimes \beta_3)=(\beta_3\otimes \beta_3)\circ\cnot
\]
It follows from this that  
\[
\cnot\circ (\beta_4\otimes 1_Q)=(\beta_4\otimes
\sigma_{Q}^{\oplus})\circ\cnot\qquad
\cnot\circ (1_Q\otimes \beta_4)=(\beta_3\otimes
\beta_4)\circ\cnot
\]
from which in turn it follows by bifunctoriality of the tensor that the required unitary
corrections factor into single qubit actions, for which we introduce a notation by setting
\[
\cnot\circ (\beta_i\otimes 1_Q)=\varphi_1(\beta_{i})\circ\cnot\qquad
\cnot\circ (1_Q\otimes \beta_i)=\varphi_2(\beta_{i})\circ\cnot
\]
The reader can verify that for
\[
4^2\cdot
(Q_{c_1}\!\!\otimes\! Q_{c_2}):=4\cdot (4\cdot (Q_{c_1}\!\!\otimes\! Q_{c_2}))
\]
and 
\[
\hspace{-1.2mm}\Delta_{ac}^{4^2}\!:=\!\langle s^{\dagger}\!s \sdot \langle s^{\dagger}\!s
\sdot\! 1_{ac}\rangle_{i=1}^{i=4}\rangle_{i=1}^{i=4}\!:\!
Q_{a_1}\!\!\otimes\! Q_{a_2}\!\to 4^2\!\cdot 
(Q_{c_1}\!\!\otimes\! Q_{c_2})\hspace{-1.2mm}
\] 
the following diagram commutes.

{\scriptsize
\begin{diagram}
Q_{a_1}\!\!\otimes\! Q_{a_2}&\rIs&Q_{a_1}\!\!\otimes\! Q_{a_2}\\     
&&\dTo^{\rho_a}&
\hspace{-1.2cm}\mbox{\bf\normalsize import\ unknown\ state}\hspace{0.5cm}\\ 
&&(Q_{a_1}\!\!\otimes\! Q_{a_2})\otimes{\rm I}\\
&&\dTo^{\hspace{-1cm}1_a\otimes (s^2 \sdot \uu
\cnot\uuu)}&\hspace{-1.7cm}
\mbox{\bf\normalsize produce\
\cnot\mbox{\bf -}state}\\ &&\hspace{-2cm}(Q_{a_1}\!\!\otimes\!
Q_{a_2})\otimes((Q_{b_1}\!\!\otimes\! Q_{b_2})^*\!\otimes (Q_{c_1}\!\!\otimes\!
Q_{c_2}))\hspace{-2cm}\\
&&\dTo^{\hspace{-2cm}(\alpha,\sigma)\!\circ(1_a\!\otimes(u_{b}\!\otimes\!
1_c))}&\hspace{-1.7cm}\mbox{\bf\normalsize   spatial\ delocation}\\ &&\hspace{-2.3cm}((Q_{a_1}\!\!\otimes\!
Q_{b_1}^*)\otimes (Q_{c_1}\!\!\otimes\! Q_{c_2}))\otimes(Q_{a_2}\!\!\otimes\!
Q_{b_2}^*)\hspace{-2.3cm}\\ &&\dTo^{\hspace{1cm}(\langle s^{\dagger} \!\!\sdot\!
\dd
\beta_i^{a_1\!b_1\!}\!\!\ddd\rangle_{i=1}^{i=4}\!\otimes\!\! 1_{c})\!\otimes\! 
\!1_{2}}&\hspace{-1.7cm}\mbox{\bf\normalsize 1st\ observation}\\ &&\hspace{-2.3cm}(\left(4\cdot{\rm
I}\right)\otimes (Q_{c_1}\!\!\otimes\!Q_{c_2}))\otimes (Q_{a_2}\!\!\otimes\!
Q_{b_2}^*)\hspace{-2.3cm}\\
\dTo^{\Delta^{4^2}_{ac}\circ\cnot}&&\dTo^{((4\cdot\lambda^{-1}_c)\!\circ\!\upsilon_c)\otimes
1_2}&\hspace{-1.7cm}\mbox{\bf\normalsize 1st\ communication}\\ 
&&\hspace{-2.3cm}(4\cdot(Q_{c_1}\!\!\otimes\!Q_{c_2}))\otimes
(Q_{a_2}\!\!\otimes\! Q_{b_2}^*)\hspace{-2.3cm}\\
&&\dTo^{\hspace{-2.3cm}\left(\bigoplus_{i=1}^{i=4}(\varphi_1^c(\beta_{i}))^{-1}\right)\!\otimes
\! 1_2}&\hspace{-1.7cm}\mbox{\bf\normalsize 1st\ correction}\\
&&\hspace{-2.3cm}(4\cdot(Q_{c_1}\!\!\otimes\! Q_{c_2}))\otimes
(Q_{a_2}\!\!\otimes\! Q_{b_2}^*)\hspace{-2.3cm}\\
&&\dTo^{\hspace{-2cm}(4\cdot 1_c)\!\otimes\!\langle s^{\dagger} \!\sdot\!
\dd
\beta_i^{a_2\!b_2\!}\!\!\ddd\rangle_{i=1}^{i=4}}
&\hspace{-1.7cm}\mbox{\bf\normalsize 2nd\ observation}\\
&&(4\cdot(Q_{c_1}\!\!\otimes\!Q_{c_2}))\otimes (4\cdot{\rm
I})\\ 
&&\dTo^{(4\cdot\rho^{-1}_{4c})\!\circ\!\tau_{4c}}&\hspace{-1.7cm}\mbox{\bf\normalsize
2nd\ communication}\\ 
&&(4\cdot(4\cdot(Q_{c_1}\!\!\otimes\!Q_{c_2})))\\
&&\dTo^{\hspace{-2cm}\bigoplus_{i=1}^{i=4}(4\cdot
\varphi_2^c(\beta_{i}))^{-1}}&\hspace{-1.7cm}\mbox{\bf\normalsize 2nd\ correction}\\
\ \ \ \ 4^2\cdot
(Q_{c_1}\!\!\otimes\! Q_{c_2})&\rIs&4^2\cdot
(Q_{c_1}\!\!\otimes\! Q_{c_2})\!\!\!\!\!\! 
\end{diagram}
}

\subsection{Entanglement swapping}

\begin{theorem}\label{thm:swap}
Setting\hspace{-2.5mm}
\[
\begin{array}{lcl}
\gamma_i&\!\!:=\!\!&(\beta_i)_*\vspace{1.5mm}\\
{\rm
P}_{i}&\!\!:=\!\!&s^{\dagger}s\sdot(\uu\gamma_i\uuu\circ\dd\beta_i\ddd)\vspace{1.5mm}\\ 
\zeta_i^{ac\!\!}&\!\!:=\!\!&\bigoplus_{i=1}^{i=4}\left((1_b^*\otimes
\gamma_i^{-1})\otimes(1_d^*\otimes\beta_i^{-1})\right)\vspace{1.5mm}\\ 
\Theta_{ab\!\!}&\!\!:=\!\!&1_d^*\otimes\langle {\rm
P}_{i}\rangle_{i=1}^{i=4}\otimes\! 1_c\vspace{1.5mm}\\
\Omega_{ab\!\!}&\!\!:=\!\!&\langle s^{\dagger}s^3\sdot(\uu 1_{ba}\!\uuu\!\otimes\! \uu
1_{dc}\!\uuu)\rangle_{i=1}^{i=4}
\end{array}
\]   

\hspace{-2.5mm}\noindent
the following diagram commutes.  
\begin{diagram}
{\rm I}\otimes{\rm I}&\rIs&{\rm I}\otimes{\rm I}\\ 
&&\dTo^{s^2\sdot(\uu 1_{da}\!\uuu\!\otimes \!\uu 1_{bc}\!\uuu)}&\hspace{-1.8cm}{\bf produce\ 
EPR\mbox{\bf -}pairs}\\ &&(Q_d^*\otimes Q_a)\otimes(Q_b^*\!\otimes Q_c)\\ 
&&\dTo^{\alpha}&\hspace{-1.8cm}{\bf spatial\ delocation}\\
&&Q_d^*\otimes (Q_a\otimes Q^*_b)\otimes Q_c\\
\dTo^{\Omega_{ab\!\!}}&&\dTo^{\Theta_{ab}}&\hspace{-1.8cm}{\bf
Bell\mbox{\bf -}base\ measurement}\\ &&\hspace{-3mm}Q_d^*\otimes \left(4\cdot (Q_a\!\otimes
Q^*_b)\right)\otimes Q_c\\
&&\dTo^{(4\cdot(\alpha,\sigma))\circ(\tau,\upsilon)}&\hspace{-1.8cm}{\bf
classical\ communication}\\ &&\hspace{-3mm}4\cdot (\left(Q^*_b\!\otimes Q_a
\right)\!\otimes\!(Q_d^*\otimes Q_c))\\
&&\dTo^{\hspace{-3cm}
\zeta_i^{ac}} &\hspace{-1.8cm}{\bf unitary\  
correction}\\
4\cdot (\left(Q^*_b\!\otimes Q_a
\right)\hspace{-1.15cm}&&\hspace{-1cm}\hspace{-1.0mm}\otimes(Q_d^*\otimes Q_c))\hspace{3mm}
\end{diagram}
The right-hand-side of the above diagram is our formal description of 
the entanglement swapping protocol.
\end{theorem}
\bpf
The top trapezoid
is the statement of the Theorem. We have a diagram of the form below for each $j\in\{1,2,3,
4\}$. To simplify the notation of the types we set
$(a^*\!,b,c^*\!,d)$ for 
${Q_a^*\otimes Q_b\otimes Q_c^*\otimes Q_d}$ etc. Again we ignore the
scalars in this proof. 
{\footnotesize
\begin{diagram}
{\rm I}\!\otimes\!{\rm I}&&&&&\rTo^{\hspace{-16mm}\langle\uu
1_{ba}\!\uuu\otimes\uu
1_{dc}\!\uuu\rangle_i\hspace{-14mm}}&&&&&4\!\cdot\!
(b^*\!\!,a,d^*\!\!,c)\\  
%1
&\rdTo^{\!\!\!\uu 1_{da}\!\uuu\!\otimes \!\uu
1_{bc}\!\uuu\!\!\!}&&&&\hspace{-5cm}\mbox{\bf\normalsize Entanglement
swapping}\hspace{-5cm}&&&&\ruTo~{\sharp}&\\ 
%2
&&(d^*\!\!,a,b^*\!\!,c)&\rTo^{\Theta_{ab}}&(d^*\!\!,4\!\cdot\!
(a,b^*),c)&\rTo^{(\tau,\upsilon)}&4\!\cdot\!(d^*\!\!,a,b^*\!\!,c)&
\rTo^{\!\!4\!\cdot\!\sigma\!\!}&
4\!\cdot\!(b^*\!\!,a,d^*\!\!,c)&&&\\ 
%3
\uTo_{\rho_{\rm I}}&\hspace{-0.2cm}
\mbox{\bf \normalsize  Lemma \ref{lm:CUT}}\hspace{-1.6cm}&&
\rdTo~{\!1^*_d\otimes\dd\beta_j\ddd\otimes 1_c}&&
\rdTo~{1_d^*\!\otimes p_j^{(a,b^*)}\!\otimes\!1_c}&
\dTo_{p_j}&&
\dTo_{p_j}&&\dTo_{p_j\hspace{-5mm}}\\ 
%4
&&(d^*\!\!,c)&\lTo_{\!\!\rho^{-1}_{d^*}\!\!\otimes\! 1_c\!\!}&(d^*\!\!,{\rm
I}\,,c)&\rTo_{\hspace{-6mm}{\!1^*_d\!\otimes\!\uu\gamma_j
\!\uuu\!\otimes\!
1_c}\hspace{-5mm}}&(d^*\!\!,a,b^*\!\!,c)&\rTo_{\sigma}&(b^*\!\!,a,d^*\!\!,c)&&\\ 
%5
&\ruTo^{\uu\beta_j\uuu\!\!\!}&
\dTo_{\lambda_{d^*}\!\otimes 1_c}&
\hspace{13mm}\ldTo_{\!\!\!\!\sigma^{-1}}&
\dTo~{{\hspace{-7mm}1_{d}^*\!\otimes\!\uu
1_{ab\!}\!\uuu\!\otimes\!\beta_j^{-1}\hspace{-5mm}}}&
\mbox{\bf\normalsize
L.~\ref{lm:precompos}}\hspace{-0mm}&
\dTo~{\diamond}&&&\rdTo~{\ddagger}&\\ 
%6
{\rm I}&&({\rm
I}\,,d^*\!\!,c)&&(d^*\!\!,a,b^*\!\!,c)&\rIs&(d^*\!\!,a,b^*\!\!,c)&&
\rTo_{\hspace{-8mm}\sigma}&&(b^*\!\!,a,d^*\!\!,c)\\
%7
\dTo_{\lambda_{\rm I}}&\hspace{-0.2cm}\mbox{\bf\normalsize Lemma
\ref{lm:precompos}}\hspace{-1.6cm}&&
\rdTo~{\uu 1_{ba\!}\!\uuu\!\otimes\!1_{d}^*\!\otimes\!\beta_j^{-1}}&
\dTo_{\sigma^{-1}}\\
%8
{\rm I}\!\otimes\!{\rm I}&&\rTo_{\uu 1_{ba}\!\uuu\otimes\uu
1_{dc}\!\uuu}&&(b^*\!\!,a,d^*\!\!,c)
\end{diagram}
}
\par\medskip\par\noindent where\[
\sharp:= \bigoplus_i(1_b^*\!\otimes\!\gamma_i^{\!-1}\!\!\otimes\!\!1_d^*\!\otimes\!\beta_i^{-1}\!)\quad,\quad\ddagger:=1_b^*\!\otimes\!\gamma_j^{\!-1}\!\!\otimes\!1_d^*\!\otimes\!\beta_j^{-1}
\quad\mbox{\rm and}\quad
\diamond:=1_d^*\!\otimes\!
\gamma_j^{\!-1}\!\!\otimes\!1_b^*\!\otimes\!\beta_j^{-1}\]

\hfill\endproof\newline 

We use $\gamma_i=(\beta_i)_*$ rather than $\beta_i$ to make ${\rm
 P}_i$ an endomorphism and hence a projector. 
The general definition of a `bipartite entanglement projector' is
\[
\hspace{-1.5mm}{\rm P}_f:=\uu f\uuu\circ\dd f_*\ddd=\uu
f\uuu\circ\dd f^\dagger\!\!\ddd\circ \sigma_{A^*\!, B}:A^*\otimes B\to A^*\otimes
B\hspace{-1.5mm}
\]
for $f:A\to B$, so in fact ${\rm P}_i={\rm P}_{(\beta_i)_*}$.

%%%%%%%%%%%%%%%%%%%%%%%%
%%% EXTRA SECTIONS START HERE
\section{Extensions and Further Developments}%%%%%%%%%
%%%%%%%%%%%%%%%%%%%%%%%%

Since its first publication in 2004, a number of   elaborations on the categorical quantum axiomatics described above have been proposed, by
ourselves in collaboration with members of our group, Ross Duncan, Dusko
Pavlovic, Eric Oliver Paquette, Simon Perdrix and Bill Edwards, and also
by others elsewhere, most notably Peter Selinger and Jamie Vicary.  We shall
present some of the main developments.

\subsection{Projective structure}  
We shall discuss our first topic at considerably greater length than the others we shall cover in this survey.
The main reason for this is that it concerns the passage to a projective point of view, which makes for an evident comparison with the standard approaches to quantum logic\index{quantum logic} going back to \cite{BvN}.
Thus it seems appropriate to go into some detail in our coverage of this topic, in the context of the Handbook in which this article will appear.

The axiomatics we have given corresponds to the pure state picture of quantum mechanics. The very fact that we can faithfully carry out linear-algebraic calculations using the semi-additive structure provided by biproducts means that states will 
 typically carry redundant global phases, as  is the case for vectors in Hilbert spaces.   Eliminating these means `going projective'.    The \em quantum logic \em tradition provides one way of doing so \cite{BvN}. Given a Hilbert space one eliminates global scalars by passing to the projection lattice. The non-Boolean nature of the resulting lattice is then taken to be  characteristic for quantum behaviour.  This leads one then to consider certain classes of non-distributive lattices as `quantum structures'.

{\small
\begin{diagram}
\hspace{-2.5cm}\mbox{\bf Hilbert space }&&\\
\hspace{-1cm}\dTo^{\hspace{-2.5cm}\mbox{\sf kill redundant global scalars}} &
\rdTo^{\sf Birkhoff\mbox{\rm -}von\ Neumann}&\\
\hspace{-3.5cm}\mbox{\bf lattice of subspaces }
&
\rTo_{\hspace{0.5cm}\mbox{\sf go\ abstract}\hspace{0.5cm}} & {\bf 
non\mbox{\rm -}distributive\ lattices}\hspace{-2.5cm}
\end{diagram} }

It is well-known that there is no obvious counterpart for the Hilbert space tensor product\index{tensor product} when passing to these more general classes of lattices.  This is one reason why Birkhoff-von Neumann style quantum logic never penetrated the mainstream 
physics community, and is particularly unfortunate in the light of the important role that the tensor product plays in quantum information and computation.  

But one can also  start from the whole category of
finite dimensional Hilbert spaces and linear maps {\bf FdHilb}. Then we can consider
`strongly compact closed categories + some additive structure' as
its appropriate abstraction, and hope to find some abstractly valid counterpart to
`elimination of redundant global scalars'.

{\small
\begin{diagram}
\hspace{-0.7cm}\mbox{\bf FdHilb} &
\rTo^{\hspace{0.7cm}\mbox{\sf go\ abstract}\hspace{0.3cm}} &
\mbox{\bf `vectorial' strong compact
closure}\hspace{-5.3cm}&\hspace{3.6cm}
\\   &
\rdTo_{\sf our\ approach} &
\dTo_{\mbox{\sf \ kill redundant global scalars}\hspace{-2.5cm}}\\ &&
\ \,\mbox{\bf `projective' strong compact closure}\!\!\!\hspace{-5.55cm}
\end{diagram} }

The major advantage which such a construction has is that the tensor product is now part of the mathematical object under consideration, and hence will not be lost in the passage from vectorial spaces to projective ones.

This passage was realised in \cite{deLL} as follows.  For morphisms in {\bf FdHilb}, i.e.~linear maps, if
we have that $f=e^{i\theta}\cdot g$ with $\theta\in[0,2\pi[$ for $f,g:{\cal
H}_1\to{\cal H}_2$, then
\[ 
f\otimes f^\dagger = e^{i\theta}\!\!\cdot g\otimes (e^{i\theta}\!\!\cdot g)^\dagger=e^{i\theta}\!\!\cdot g\otimes e^{-i\theta}\!\!\cdot g^\dagger=g\otimes g^\dagger\,.
\]   
Now in abstract generality, given a strongly compact closed category ${\bf C}$, we can define a new category $\WProj({\bf C})$ with the same objects as those of ${\bf C}$, but with
\[\label{eq:WPROJconst} 
\WProj({\bf C})(A,B):=\left\{f\otimes f^\dagger\bigm|  f\in{\bf C}(A,B)\right\}
\]
as hom-sets and in which composition is given by
\[ 
(f\otimes f^\dagger)\,\bar{\circ}\,(g\otimes g^\dagger):=  (f\circ
g)\otimes(f\circ g)^\dagger\!.
\]
One easily shows that $\WProj({\bf C})$ is again a strongly compact closed category.  The abstract counterpart to elimination of global phases is expressed by the following propositions.

\begin{proposition}{\rm\cite{deLL}}\label{Pr:phase1} 
For morphisms $f$, $g$ and
scalars $s$, $t$ in a strongly compact closed category, we have
\[ s\bullet f=t\bullet g \; \wedge \; s\circ s^\dagger=t\circ t^\dagger=1_\II\
\;\;
\Longrightarrow \;\;  f\otimes f^\dagger\!= g\otimes g^\dagger.
\]
\end{proposition}

\begin{proposition}{\rm\cite{deLL}}\label{Pr:phase2} For   morphisms $f$ and $g$ in a strongly compact closed category with scalars $S$ we have
\[  f\otimes f^\dagger\!=g\otimes g^\dagger\ \ \ \Longrightarrow\ \ \
\exists s,t\in S.\, s\sdot f=t\bullet g\; \wedge \; s\circ s^\dagger=t\circ
t^\dagger.
\] 
In particular we can set
\[ \label{eq:Hsscalers} s:=(\uu f\uuu)^\dagger\circ\uu f\uuu
\qquad{\rm and}\qquad t:=(\uu g\uuu)^\dagger\circ\uu f\uuu.
\]
\end{proposition}

While the first proposition is straightforward, the second one is somewhat more surprising.  It admits a simple graphical proof.  We represent
units by dark triangles and their adjoints by the same triangle but
depicted upside down. %We take a from bottom to top reading convention. 
Other morphisms are depicted by square boxes as before, with the exception of scalars which are depicted by `diamonds'.  The scalar $s:=(\uu f\uuu)^\dagger\circ\uu f\uuu$ is depicted as
\par\vspace{3mm}\par\noindent
\begin{minipage}[b]{1\linewidth}
\centering{\epsfig{figure=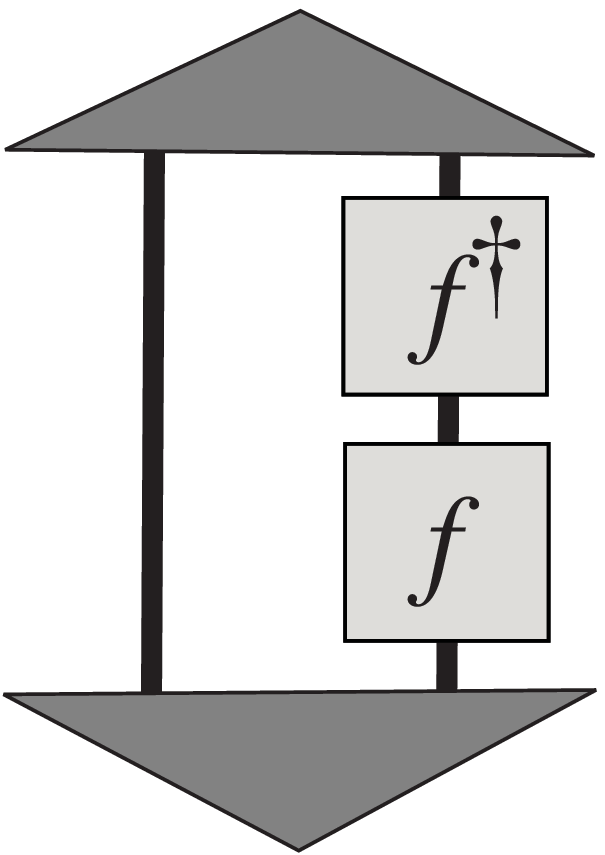,width=60pt}}
\end{minipage}
\par\vspace{3mm}\par\noindent
Bifunctoriality means that we can move these boxes upward and
downward, and naturality provides additional modes of movement, e.g.~scalars admit arbitrary movements. Now, given that $f\otimes f^\dagger\!=g\otimes g^\dagger$, that is, in a picture,
\par\vspace{3mm}\par\noindent
\begin{minipage}[b]{1\linewidth}
\centering{\epsfig{figure=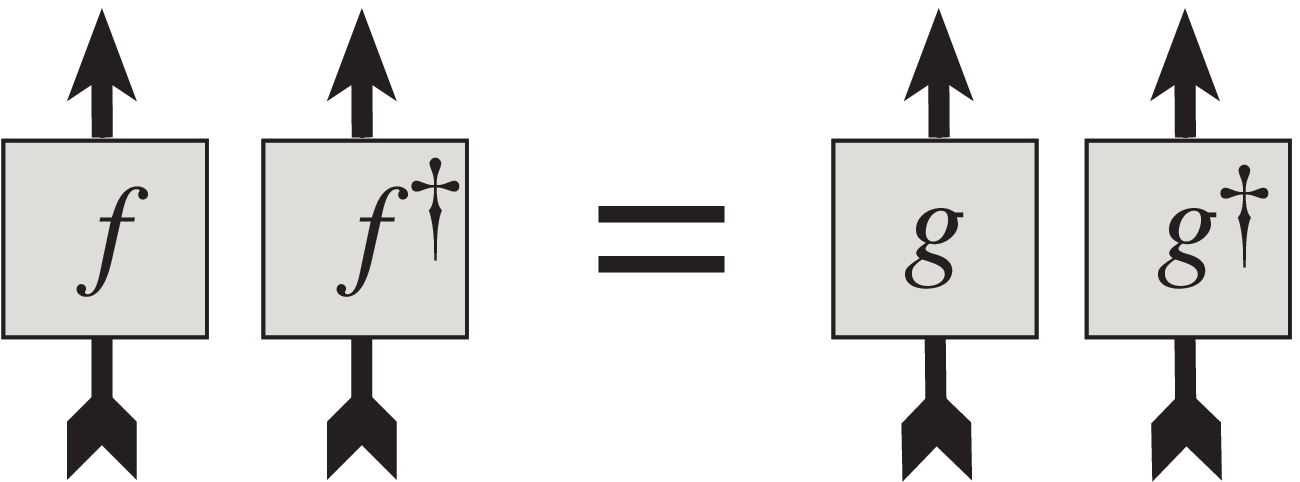,width=128pt}}
\end{minipage}
\par\vspace{3mm}\par\noindent
we need to show that $s\bullet f=t\bullet g$ and $s\circ s^\dagger=t\circ
t^\dagger$ for some choice of scalars $s$ and $t$, that is, in a picture,
\par\vspace{3mm}\par\noindent
\begin{minipage}[b]{1\linewidth}
\centering{\epsfig{figure=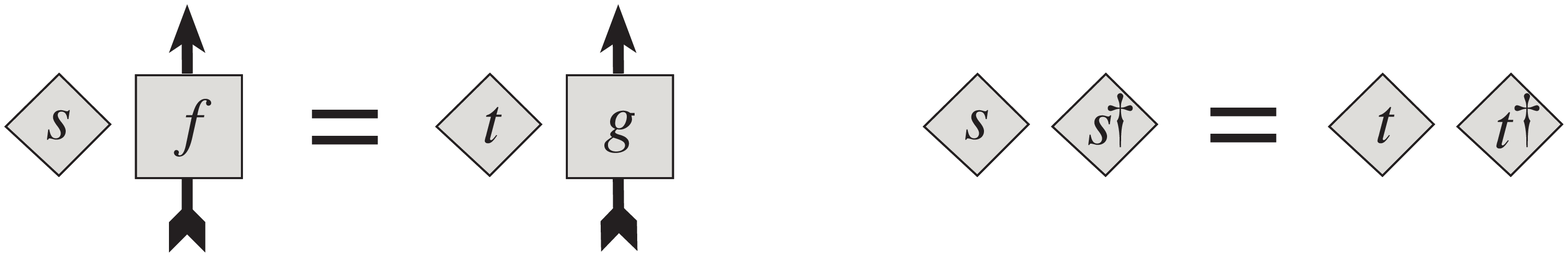,width=300pt}}
\end{minipage}
\par\vspace{3mm}\par\noindent
The choice that we will make for $s$ and $t$ is
\par\vspace{3mm}\par\noindent
\begin{minipage}[b]{1\linewidth}
\centering{\epsfig{figure=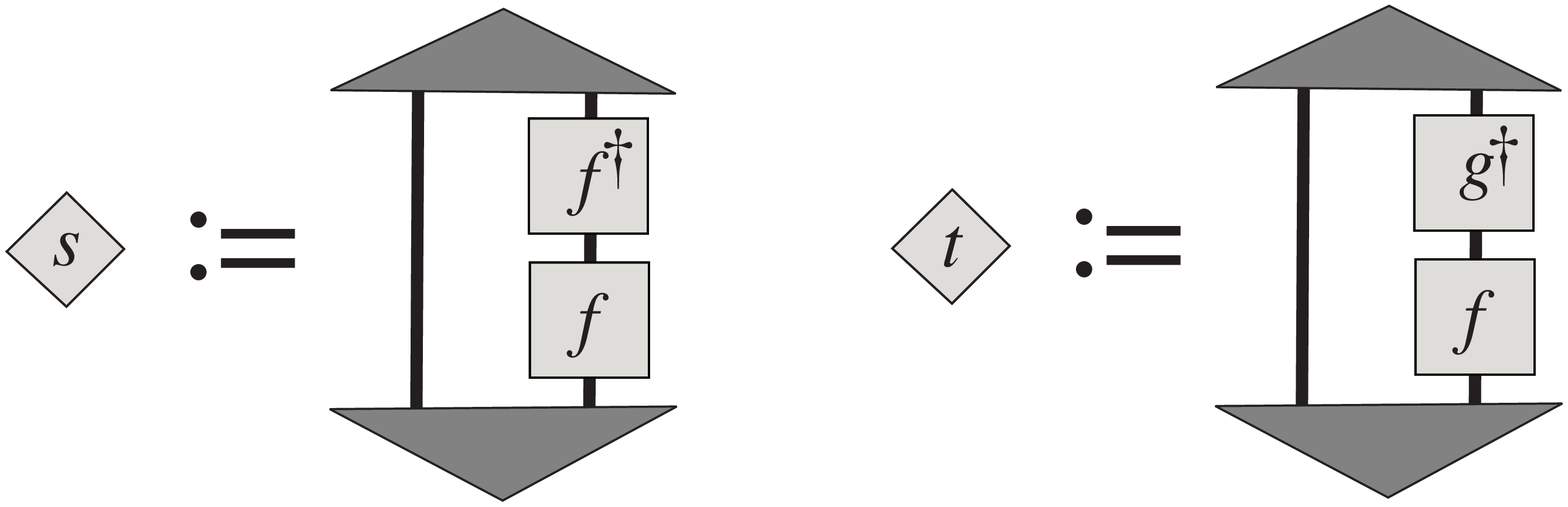,width=265pt}}
\end{minipage}
\par\vspace{3mm}\par\noindent
Then we indeed have  $s\bullet f=t\bullet g$ since in
\par\vspace{3mm}\par\noindent
\begin{minipage}[b]{1\linewidth}
\centering{\epsfig{figure=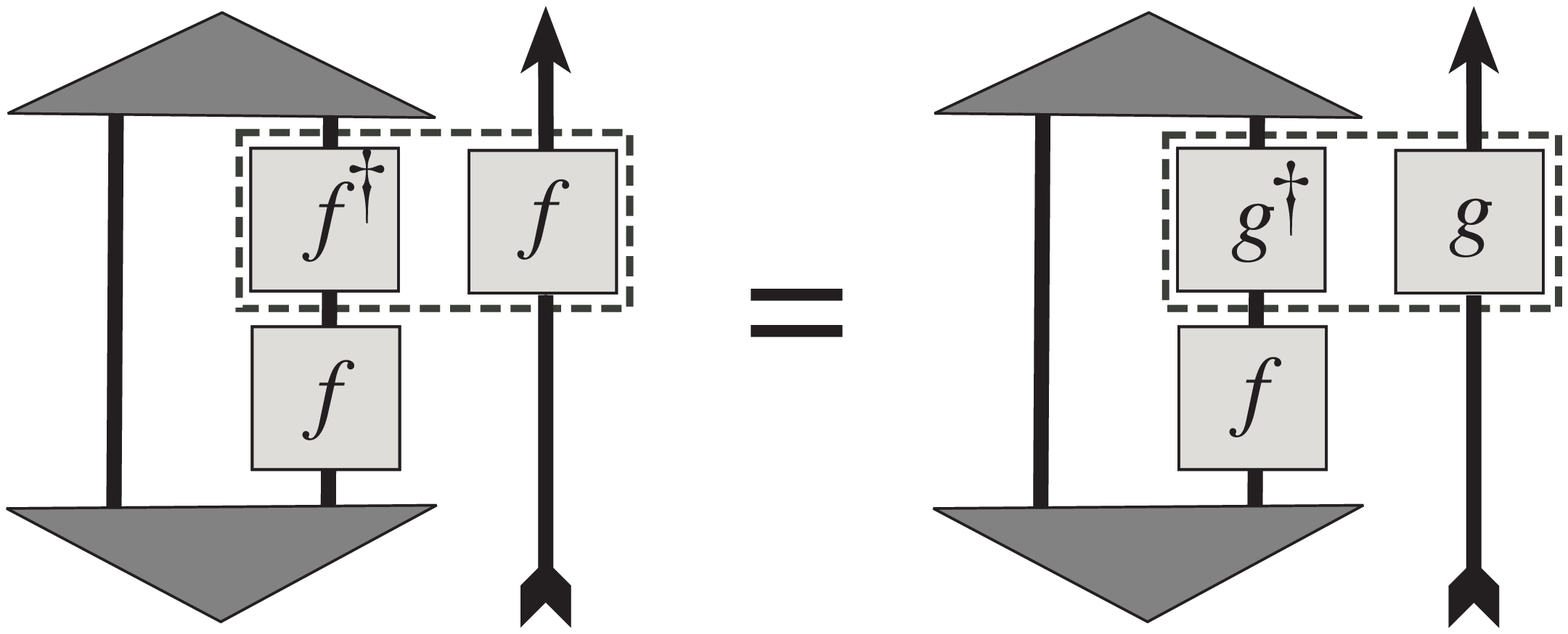,width=215pt}}
\end{minipage}
\par\vspace{3mm}\par\noindent
the areas within the dotted line are equal by assumption. We also have that $s\circ s^\dagger=t\circ t^\dagger$ since
\par\vspace{3mm}\par\noindent
\begin{minipage}[b]{1\linewidth}
\centering{\epsfig{figure=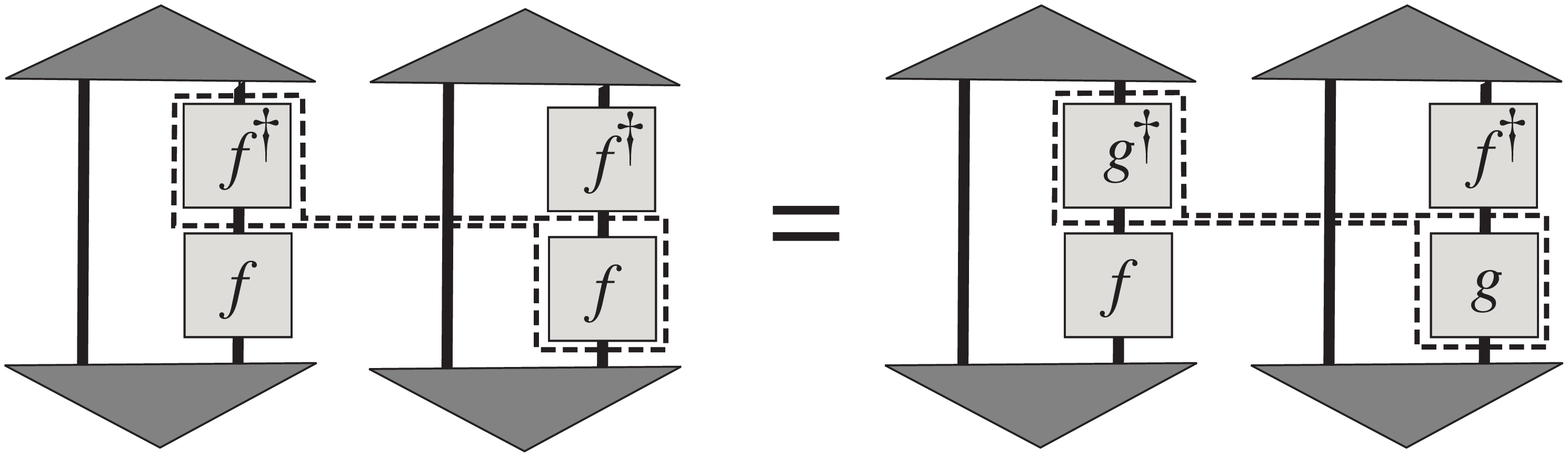,width=300pt}}
\end{minipage}
\par\vspace{3mm}\par\noindent
which  completes the proof.

As expected, biproducts do not survive the passage from ${\bf C}$ to $\WProj({\bf C})$ but the  weaker structure which results still suffices for a comprehensive description of the protocols we have discussed in this article.  In particular, the distributivity natural isomorphisms
\[
{\rm dist}_{0,l}:A\otimes 0\simeq 0\qquad\qquad {\rm dist}_l:  A\otimes(B\oplus
C)\simeq(A\otimes B)\oplus(A\otimes C)\,\,
\]
\[
{\rm dist}_{0,r}:0\otimes A\simeq 0\qquad\qquad {\rm dist}_r: (B\oplus C)\otimes
A\simeq(B\otimes A)\oplus(C\otimes A)\,.
\] 
carry over to $\WProj({\bf C})$.  Details can found in \cite{deLL}.

Our framework also allows a precise general statement of the incompatibility of biproducts with projective structure. 

Call a strongly compact closed category \em projective \em iff  equality of projections implies equality of the corresponding states, that is, 
\beq\label{eq:idempQL4}
\forall\ \psi,\phi:\II\to A\,.\ \ \
\psi\circ\psi^\dagger=\phi\circ\phi^\dagger\ \Longrightarrow\
\psi=\phi\,.
\eeq

\begin{proposition}{\rm\cite{deLL}}\label{biprodincop} 
If a strongly compact closed category with biproducts  is projective and the semiring of scalars admits negatives, \ie is a ring, then we have $1=-1$, that is, there are no non-trivial negatives. 
\end{proposition}

\noindent Having no non-trivial negatives of course obstructs the description of interference.

\subsection{Mixed states and Completely Positive Maps}

The categorical axiomatics set out in this article primarily refers to the pure-state picture of quantum mechanics. However, for many purposes, in particular those of quantum information, it is \emph{mixed states}, acted on by \emph{completely positive maps}, which provide the most appropriate setting.
Peter Selinger \shortcite{Selinger} proposed a general categorical construction, directly in the framework of the categorical axiomatics of \cite{AC2} which has been described in this article, to capture the passage from the pure states to the mixed states picture.

The construction proceeds as follows.
Given any strongly compact closed category ${\bf C}$ we define a new category $\CPM({\bf C})$ with the same objects as ${\bf C}$ but with 
morphisms given by
\[
\CPM({\bf C})(A,B):=\hspace{9.2cm}\vspace{-2mm}
\]
\[\label{eq:CPMconst} 
\ \ \ \ \ \ \ \ \ \left\{(1_B\otimes\epsilon_C\otimes 1_{B^*})\circ(1_{B\otimes C}\otimes\sigma_{B^*,C^*})\circ(f\otimes f_*)\bigm| f\in{\bf C}(A,B\otimes C)\right\}
\]
where for simplicity we assume that the monoidal structure is strict. Composition in $\CPM({\bf C})$ is  inherited pointwise from ${\bf C}$.  The morphisms of the category $\CPM({\bf FdHilb})$ are exactly the \emph{completely positive maps}, and the morphisms in the hom-set $\CPM({\bf FdHilb})(\mathbb{C}, {\cal H})$ are exactly the self-adjoint  operators with positive trace on ${\cal H}$.  The category $\WProj({\bf C})$ faithfully embeds in $\CPM({\bf C})$ by setting
\[
f\otimes f^\dagger \mapsto f\otimes f_*.
\]
Metaphorically, we have
\[ 
{\CPM({\bf C})\over\WProj({\bf C})} 
=
{\mbox{\rm density\ operators} 
\over \mbox{\rm projectors}}\ .
\]
For more details on the CPM-construction we refer the reader to \cite{Selinger}.  

Recently it was shown that the CPM-construction  does not require strong compact closure, but only dagger symmetric monoidal structure.  Details are in \cite{RR08}.  An axiomatic presentation of categories of completely positive maps is given in \cite{Bob_Selinger}.

\subsection{Generalised No-Cloning and No-Deleting theorems}

The No-Cloning theorem \cite{Dieks,WZ} is a basic limitative result for quantum mechanics, with particular significance for quantum information. It says that there is no unitary operation which makes perfect copies of an unknown (pure) quantum state. A stronger form of this result is the No-Broadcasting theorem  \cite{Broadcast}, which applies to mixed states. There is also a No-Deleting theorem \cite{Pati}.

The categorical and logical framework which we have described provides new possibilities for exploring the structure, scope and limits of of quantum information processing, and the features which distinguish it from its classical counterpart.
One area where some striking progress has already been made  is the axiomatics of No-Cloning and No-Deleting. It is possible to delimit the classical-quantum boundary here in quite a subtle way. On the one hand, we have the strongly compact closed structure which is present in the usual Hilbert space setting for QIC, and which we have shown accounts in generality for the phenomena of entanglement. Suppose we were to assume that \emph{either} copying \emph{or} deleting were available in a strongly compact closed category as \emph{uniform operations}. Mathematically, a uniform copying operation means a \emph{natural diagonal} 
\[ \Delta_{A} : A \to A \otimes A \]
\ie a monoidal natural transformation, which moreover is co-associative and co-commutative:
\[
\begin{diagram}
A & \rTo^{\Delta} & A \otimes A & \rTo^{1 \otimes \Delta} & A \otimes (A \otimes A) \\
\deq & & & & \dTo_{\alpha_{A,A,A}} \\
A & \rTo_{\Delta} &  A \otimes A & \rTo_{\Delta \otimes 1} & (A \otimes A) \otimes A
\end{diagram}
\qquad \qquad
\begin{diagram}
A & \rTo^{\Delta} & A \otimes A \\
& \rdTo_{\Delta} & \dTo_{\sigma_{A,A}} \\
& & A \otimes A
\end{diagram}
\]
Thinking of the diagonal associated with the usual cartesian product, one sees immediately that co-commutativity and co-associativity are basic requirements for a copying operation: if I have two copies of the same thing, it does not matter which order they come in, and if I produce three copies by iterating the copying operation, which copy I choose to perform the second copying operation on is immaterial.
Naturality, on the other hand, corresponds essentially to \emph{basis-independence} in the Hilbert space setting; it says that the operation exists `for logical reasons', in a representation-independent form.

We have shown recently that under these assumptions \emph{the category trivializes}; in other words, that this combination of quantum and classical features is \emph{inconsistent}, leading to a collapse of the structure. The precise form of the result is that under these hypotheses every endomorphism in the category is a scalar multiple of the identity. 

Similar generalizations of the No-Deleting theorem \cite{Pati} and  the No-Broadcasting theorem \cite{Broadcast} also hold.   Papers on these results are in preparation. 

One striking feature of these results is that they are visibly in the same genre as a well-known result by Joyal in categorical logic \cite{LS} showing that a `Boolean cartesian closed category' trivializes, which provides a major road-block to the computational interpretation of classical logic. In fact, they strengthen Joyal's result, insofar as the assumption of a full categorical product (diagonals \emph{and} projections) in the presence of  a classical duality is weakened. This shows a heretofore unsuspected connection between limitative results in proof theory and No-Go theorems in quantum mechanics.

Another interesting point is the way that this result is delicately poised. The basis structures to be discussed in the next sub-section do assume commutative comonoid structures existing in strongly compact closed categories---indeed with considerable additional properties, such as the Frobenius identity. Not only is this consistent, such structures correspond to a major feature of Hilbert spaces, namely orthonormal bases. The point is that there are many such bases for a given Hilbert space, and none are canonical. Indeed, the 
\emph{choice} of basis corresponds to  the \emph{choice} of measurement set-up, to be made by a `classical observer'. The key ingredient which leads to inconsistency, and which basis structures lack, is \emph{naturality}, which, as we have already suggested, stands as an abstract proxy for basis-independence.

\subsection{Basis Structures and Classical Information}\label{secClassOb}

In this article, an approach to measurements and classical information has been developed based on biproducts. This emphasizes the \emph{branching structure} of measurements due to their probabilistic outcomes.

One may distinguish the `multiplicative' from the `additive' levels of our axiomatization (using the terminology of Linear logic \cite{Girard}). The multiplicative, purely tensorial level of strongly compact closed categories shows, among other things, how a remarkable amount of multilinear algebra, encompassing much of the structure needed for quantum mechanics and quantum information, can be done without any substrate of linear algebra. Moreover, this level of the axiomatization carries a very nice diagrammatic calculus, which we have sampled informally. In general, the return on structural insights gained from the axiomatization seems very good. The additive level of biproducts reinstates a linear (or `semilinear') level of structure, albeit with fairly weak assumptions, and there is more of a sense of recapitulating familiar definitions and calculations. While a diagrammatic calculus is still available here (see \cite{AD}), it is subject to a combinatorial unwieldiness familiar from process algebra in Computer Science \cite{Milner} (cf. the `Expansion Theorem').

An alternative approach to measurements and classical information has been developed in a series of papers \cite{CPav,CPaq,CPV,CPaqPer} under various names, the best of which is probably `basis structure'. Starting from the standard idea that a measurement set-up corresponds to a choice of orthonormal basis, the aim is to achieve an axiomatization of the notion of basis as an additional structure. Of course, the notion of basis developed in Section~\ref{sec:biprods} has all the right properties, but it is defined in terms of biproducts, while the aim here is to achieve an axiomatization purely at the multiplicative level.

This is done in an interesting way, bringing the informatic perspective to the fore.
One can see the choice of a basis as determining a notion of `classical data', namely the basis vectors. These vectors are subject to the classical operations of \emph{copying} and \emph{deleting}, so in a sense classical data, defined with respect to a particular choice of basis, stands
 as a contrapositive to the No-Cloning and No-Deleting theorems.
 Concretely, having chosen a basis $\{ |i\rangle \}$ on a Hilbert space $\HH$, we can define linear maps
 \[ \HH \lrarr \HH \otimes \HH :: |i\rangle \mapsto |ii\rangle , \qquad \HH \lrarr \mathbb{C} :: |i\rangle \mapsto 1 \]
 which do correctly copy and delete the basis vectors (the `classical data'), although not of course the other vectors.
 
These considerations  lead to the following definition.  A \emph{basis structure} on an object $A$ in a strongly compact closed category is a commutative comonoid structure on $A$
\[
Copy:A\to A\otimes A,  \qquad Delete: A\to \II
\]
subject to a number of additional axioms, the most notable of which is
 the \emph{Frobenius identity} \cite{CarboniWalters}.  In {\bf FdHilb} these structures exactly correspond to orthonormal bases \cite{CPV},  which justifies their name and interpretation.  

Quantum measurements can be defined relative to these structures, as self-adjoint Eilenberg-Moore coalgebras for the comonads induced by the above comonoids \cite{CPav}.  In {\bf FdHilb} these indeed correspond exactly to  projective spectra.  The Eilenberg-Moore coalgebra square
\begin{diagram} 
A&\rTo^{Measure}&X\otimes A\\
\dTo^{\qquad Measure}&&\dTo_{1_X\otimes Measure}\\
X\otimes A&\rTo_{Copy\otimes 1_A}&X\otimes X\otimes A
\end{diagram}
can be seen as an operational expression of von Neumann's projection postulate in a resource-sensitive setting: measuring twice is the same as measuring once and then copying the measurement outcome.  
This abstract notion of measurement admits generalisation to POVMs and PMVMs, for which a generalised  Naimark dilatation theorem can be proved at the abstract level \cite{CPaq}.   

Within $\CPM({\bf C})$ the \em decoherence \em aspect of quantum measurement, which, concretely in ${\bf FdHilb}$, is the completely positive map which eliminates non-diagonal elements relative to the measurement basis, arises as
\[
Copy\circ Copy^\dagger: X\otimes X\to X\otimes X
\]
where $X$ is now taken to be self-dual, that is, $X=X^*$.

These basis structures not only allow for classical data, measurement and control operations to be described  but also provide useful expressiveness when discussing multipartite states and unitaries.  For example, they capture GHZ-states in a canonical fashion \cite{CPav}, and enable an elegant description of the state-transfer protocol \cite{CPaqPer}.

Vicary showed that if one drops the co-commutativity requirement of basis structures in ${\bf FdHilb}$, then, rather than all orthonormal bases, one finds exactly all finite dimensional C*-algebras \cite{Vicary_Cstar}\index{C*-algebra}.

The fact that these multiplicative basis structures allow measurements to be expressed without any explicit account of branching may be compared to the way that the pure $\lambda$-calculus can be used to encode booleans and conditionals \cite{Bar}.
From this perspective, explicit branching can be seen to have its merits, while \emph{model-checking} \cite{CGP} has done much to ameliorate the combinatorial unwieldiness mentioned above.
It is likely that further insights will be gained by a deeper understanding of the relationships between the additive and multiplicative levels.

\subsection{Complementary observables and phases}\label{secCompPhases}

A further step is taken in \cite{CD}, where  abstract counterparts to (relative) phases are defined.   Given a basis structure on $X$ and a point $\psi:\II\to X$  its  action on $X$ is defined to be the morphism 
\[
\Delta(\psi):=Copy^\dagger\circ(\psi\otimes 1_X):X\to X.
\]
From the axioms of basis structures it  follows that the set of all these actions on the hom-set ${\bf C}(\II,X)$ is a commutative monoid.  Defining \em unbiassed points \em as those $\psi\in {\bf C}(\II,X)$ for which $\Delta(\psi)$ is unitary, the corresponding set of \em unbiassed actions \em is always  an abelian group, which we call the  \em phase group\em.  In the case of the qubit in ${\bf FdHilb}$ the phase group corresponds to  the equator of the Bloch sphere, that is, indeed, to relative phase data.

Also in \cite{CD} an axiomatics is proposed for \emph{complementary observables}\index{complementary observables}. It is shown that for all known constructions of complementary quantum observables, the corresponding basis structures obey a `scaled' variant of the bialgebra laws.  This scaled bialgebra structure together with the phase group is sufficiently expressive to describe all linear maps, hence all mutipartite states and unitary operators, in ${\bf FdHilb}$. It provides an abstract means to reason about quantum circuits and to translate between quantum computational models, such as the circuit model and the measurement-based model.

As an application, a description is given of the quantum Fourier transform, the key ingredient of Shor's factoring algorithm \cite{ShorAlg}, the best-known example of a quantum algorithm.

\subsection{The quantum harmonic oscillator}

Jamie Vicary  \shortcite{Vicary} gave a  purely categorical treatment of the quantum harmonic oscillator, directly in the setting described in this article, of strongly compact closed categories with biproducts. 
In Linear logic terminology, he introduced an `exponential level' of structure, corresponding to Fock space. This provides a monoidal adjunction that encodes the raising and lowering operators into a co-commutative comonoid.  Generalised coherent states arise through the hom-set isomorphisms defining the adjunction, and it is shown that they are eigenstates of the lowering operators. Similar results were independently obtained in \cite{Fiore} in an abstract `formal power series' context, with a motivation stemming from Joyal's theory of species.

\subsection{Automated quantum reasoning}

The structures uncovered by the research programme we have described provide a basis for  the design of software tools for automated reasoning about quantum phenomena, protocols and algorithms.  Several MSc students at Oxford University Computing Laboratory have designed and implemented such tools for their Masters Thesis projects.  An ongoing  high-level comprehensive approach has recently be initiated  by Lucas Dixon and Ross Duncan \shortcite{DixonDuncan2008}.

\subsection{Diagrammatic reasoning}\label{sec:Diagrammatic}

We have used a diagrammatic notation for tensor categories in an informal fashion.
In fact, this diagrammatic notation, which can be traced back at least to  Penrose \shortcite{Penrose}, was made fully formal by Joyal and Street \shortcite{JoyalStreet}; topological applications can be found in \cite{Turaev}.

The various structures which have arisen in the above discussion, such as strong compact closure, biproducts, dagger Frobenius comonoids, phase groups, scaled bialgebras, and the exponential structures used in the description of the quantum harmonic oscilator, all admit  intuitive diagrammatic presentations in this tradition. References on these include \cite{AD,CPaq,CD,Vicary}.
Tutorial introductions to these diagrammatic calculi are given in \cite{LNPCoecke,SelingerTutorial}

These diagrammatic calculi provide very effective tools for the communication of the structural ideas. The software tools mentioned in the previous sub-section all support the presentation and manipulation of such diagrams as their interface to the user.

\subsection{Free constructions} 

In \cite{Abr1}  a number of free constructions are described in  a simple, synthetic and conceptual manner, including the free strongly compact closed category over a dagger category, and the free traced monoidal category. The Kelly-Lapalaza \shortcite{KL} construction of the free compact closed category  is recovered in a structured and conceptual fashion.

These descriptions of free categories in simple combinatorial terms provide a basis for the use of diagrammatic calculi as discussed in the previous sub-section.

\subsection{Temperley-Lieb algebra and connections to knot theory and topological quantum field theory}

Our basic categorical setting has been that of symmetric monoidal categories. If we weaken the assumption of symmetry, to \emph{braided} or \emph{pivotal} categories, we come into immediate contact with a wide swathe of developments relating to knot theory, topology, topological quantum field theories, quantum groups, etc. We refer to \cite{FreydYetter,Kauffbook,Turaev,YetterBook,Kock,StreetBook} for a panorama of some of the related literature.

In \cite{Abr2}, connections are made between the categorical axiomatics for quantum mechanics developed in this article, and the Temperley-Lieb algebra, which plays a central r\^ole in the Jones polynomial and ensuing developments. 
For illustration, we show the defining relations of the Temperley-Lieb algebra, in the diagrammatic form introduced by Kauffman:
%\begin{center}
%\input{TLRel123}
%\end{center}
\par\vspace{3mm}\par\noindent
\begin{minipage}[b]{1\linewidth}
\centering{\epsfig{figure=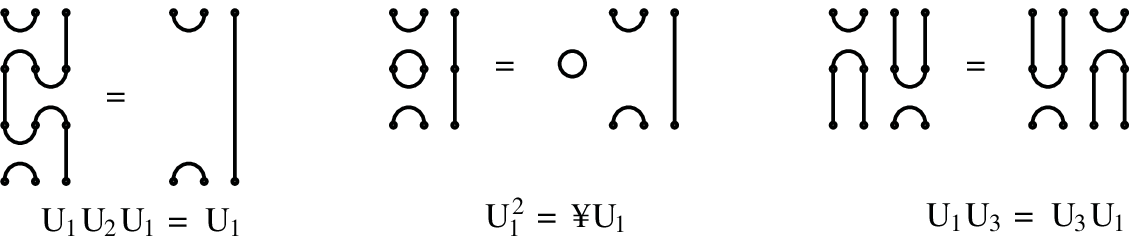,width=280pt}}
\end{minipage}
\par\vspace{0mm}\par\noindent
The relationship with the diagrammatic notation we have been using should be reasonably clear, The `cups' and `caps'  in the above diagrams correspond to the triangles we have used to depict units and counits.

An important mediating r\^ole is played by the \emph{geometry of interaction} \cite{Gi89,Abramsky96}, which provides a mathematical model of information flow in logic (cut-elimination of proofs) and computation (normalization of $\lambda$-terms). 

The Temperley-Lieb algebra is essentially the (free) \emph{planar} version of our quantum setting; and new connections are made between logic and geometry in \cite{Abr2}. For example, a simple, direct description of the Temperley-Lieb algebra, with no use of quotients, is given in \cite{Abr2}. This leads in turn to full completeness results for various non-commutative logics. Moreover, planarity is shown to be an invariant of the information flow analysis of cut elimination. 

This leads to a number of interesting new kinds of questions: 
\begin{itemize}
\item It seems in practice that few naturally occurring quantum protocols require the use of the symmetry maps. (For example, none of those described in this paper do). How much of Quantum Informatics can be done `in the plane'? What is the significance of this constraint?
\item Beyond the planar world we have \emph{braiding}, which carries 3-dimensional geometric information. Does this information have some computational significance? Some ideas in this direction have been explored by Kauffman and Lomonaco \shortcite{KaufLom}, but no clear understanding has yet been achieved.
\item Beyond this, we have the general setting of Topological Quantum Field Theories \cite{Wit,Atiyah} and related notions. This may be relevant to Quantum Informatic concerns in (at least) two ways:
\begin{enumerate}
\item A novel and promising paradigm of \emph{Topological Quantum Computing} has recently been proposed \cite{Kitaev}.
\item The issues arising from \emph{distributed quantum computing}, \emph{quantum security protocols} etc. mean that the interactions between quantum informatics and spatio-temporal structure will  need to be considered.
\end{enumerate}
\end{itemize}

\subsection{Logical syntax}

In \cite{AD}  a strongly normalising proof-net calculus corresponding to the logic of strongly compact closed categories with biproducts is presented. The calculus is a full and faithful representation of the free strongly compact closed category with biproducts on a given category with an involution. This syntax can be used to represent and reason about quantum processes. 

In \cite{RossThesis} this is extended to a description of the free strongly compact category generated by a monoidal category. This is applied to the description of the measurement calculus of \cite{DKP}.

\subsection{Completeness}

In  \cite{Selinger2} Selinger  showed that finite-dimensional Hilbert spaces are equationally complete for strongly compact closed categories.  This  result shows that  if we want to verify an equation expressed purely in the language of strongly compact closed categories, then it suffices to verify that it holds for Hilbert spaces.  

\subsection{Toy quantum categories}

In \cite{Spek} it is shown that  Spekkens'  well-known `toy model' of quantum mechanics described in  \cite{Spekkens} can be regarded as an instance of the categorical quantum axiomatics.   The category ${\bf Spek}$ is defined to be the dagger symmetric monoidal subcategory of ${\bf Rel}$ generated by those objects whose cardinality is a   power of 4, the symmetry group on 4 elements, and a well-chosen copying-deleting pair for the 4 element set.

%%%%%%%%%%%%%=

\end{document}